\documentclass[aps,prd,reprint, superscriptaddress]{revtex4-1}

\usepackage{amsmath,amssymb, amsthm,amstext}
\usepackage{natbib}
\usepackage{graphicx}
\usepackage{color}
\usepackage{array, enumerate}
\usepackage{bm}
\usepackage{multirow}
\usepackage[breaklinks,colorlinks,citecolor=blue]{hyperref}
\usepackage{braket}
\usepackage{txfonts}

\def\nn{\nonumber}
\def\be{\begin{equation}}
\def\ee{\end{equation}}
\def\dotJI{\dot J_{\rm mig, I}}

\def\mnras{MNRAS}
\def\aap{A\&A}
\def\apjl{Astrophys.J.Lett.}
\def\aj{Astronomical Journal}

\begin{document}
\title{Formation Rate of Extreme Mass Ratio Inspirals in Active Galactic Nuclei  }
\author{Zhen Pan}
\email{zpan@perimeterinstitute.ca}
\affiliation{Perimeter Institute for Theoretical Physics, Ontario, N2L 2Y5, Canada}
\author{Huan Yang}
\email{hyang@perimeterinstitute.ca}
\affiliation{Perimeter Institute for Theoretical Physics, Ontario, N2L 2Y5, Canada}
\affiliation{University of Guelph, Guelph, Ontario N1G 2W1, Canada}
\begin{abstract}
  Extreme Mass Ratio Inspirals (EMRIs) are important sources for space-borne gravitational wave detectors, such as LISA (Laser Interferometer Space Antenna) and TianQin. Previous EMRI rate studies have focused on the ``loss cone" scenario, where stellar-mass black holes (sBHs) are scattered into highly eccentric orbits near the central massive black hole (MBH) via multi-body interaction. In this work, we calculate the rate of EMRIs  of an alternative formation channel: EMRI formation assisted by the accretion flow around accreting massive black holes. In this scenario, sBHs and stars on inclined orbits are captured by the accretion disk, and then subsequently migrate towards the MBH, under the influence of density wave generation and head wind. By solving the Fokker-Planck equation incorporating both  sBH-sBH/sBH-star scatterings and sBH/star-disk interactions, we find that an accretion disk usually boosts the EMRI formation rate per individual MBH by  $\mathcal O(10^1-10^3)$ compared with the canonical ``loss cone" formation channel. Taking into account that the fraction of active galactic nuclei (AGNs) is  $\sim \mathcal O(10^{-2}-10^{-1})$, where the MBHs are expected to be rapidly accreting, we expect EMRI formation assisted by AGN disks to be an important   channel for all EMRIs observed by space-borne gravitational wave detectors. These two channels also predict distinct distributions of EMRI eccentricities and orbit inclinations  with respect to the MBH spin equatorial plane, which can be tested by future gravitational wave observations.
\end{abstract}
\maketitle

\section{Introduction}
With the Laser Interferometer Space Antenna (LISA) and TianQin planned for
launch in early 2030s \cite{LISA2019,TianQin2020},
the mHz band will be available for gravitational-wave (GW) observation.
One primary target source of space-borne GW detectors is extreme mass ratio inspiral (EMRI),
which usually consists of a stellar-mass compact object, e.g. a black hole or a neutron star, and a massive black hole (MBH).
The stellar-mass object may stay in the LISA band for years and complete $10^4\sim10^5$ circles around the MBH before their final mergers \cite{LISA2017}.
Because of such large number of cycles, small modification of the metric of the EMRI system \cite{Babak:2017tow} and possible environmental/astrophysical effects \cite{Bonga:2019ycj,Yang:2019iqa,Barausse:2014tra} may accumulate over the duration of the waveform and generate detectable phase shift. With a population of events, the distributions of masses and spins of the host MBHs may be measured as a way to infer the growth history of galactic centre MBHs \cite{Berti:2008af,Pan:2020dze}.

One important problem  related to EMRIs is to evaluate their event rate.
The ``canonical" EMRI formation is expected to be a stellar-mass black hole (sBH)
captured  by  a MBH via multi-body scatterings in the core of a galaxy \cite{Amaro2018},
which has been the main assumption for previous rate calculations \cite{Gair:2017ynp,Babak2017},
(while other formation channels involving tidal disruption or tidal capture of binary sBHs
may also contribute a fraction of EMRIs \cite{Miller2005,Chen2018}).
Given the MBH mass and the initial distributions of surrounding stars and sBHs,
the EMRI rate per MBH can be obtained by solving the Fokker-Planck equation or by N-body simulations
\cite{Preto2010,Amaro2011}. In addition to the generic rate per MBH, the LISA detectable EMRI rate also depends on
the mass function of MBHs at different redshifts, the fraction of MBH living in star clusters, and the
relative abundance of sBHs in star clusters. Taking account of all these uncertainties with semi-analytic models,
Babak et al. \cite{Babak2017} forecasted that there will be tens to thousands EMRIs detected by LISA per year,
and in a similar analysis, Fan et al. \cite{Fan2020} forecasted a slightly lower EMRI detection rate by TianQin.

In this work, we consider another possible EMRI formation channel,
where a sBH in the core of a galaxy is captured by the accretion disk around an accreting MBH.
As extensively studied in the context of star-disk-satellite systems \cite{Goldreich1979,Goldreich1980,Ward1989,Tanaka2002,Tanaka2004},
a planet within inclined orbit with repesct to the disk excites density waves that
drive the planet's inward migration, circularize the planet's orbit and
drive the planet orbit toward the disk plane.
 In a MBH-disk-sBH system, similar processes also work and assist the EMRI formation. A sBH initially resides on a inclined orbit crossing the accretion disk generally moves towards lower incliantion orbits and eventually get captured by the accretion disk. sBHs within the disk generally interact with the disk through wind effects, density wave genration and dynamic friction. As we can find in the discussions in Sec.~\ref{sec:interaction}, sBHs migrate within the disk under these effects.

The timescale of sBH capture and  migration apparently depends on the disk profile. In particular, certain disk models \cite{Sirko2003,Thompson2005} predict local density maxima at distance $\sim \mathcal{O}(10^2)-\mathcal{O}(10^3)$ times the MBH size away. As a result, the migration torque due to density wave generation changes sign at these locations where the migrating object are trapped. This trapping mechanism has been extensively discussed in the context of investigating binary black hole (BBH) mergers in AGN disks  \cite{McKernan2012,McKernan2014,Stone2017,Bartos2017,McKernan2018, Leigh2018,Yang2019prl,Yang2019,
Secunda2019,Secunda2020,McKernan2020,McKernan2020b,Graham:2020gwr,Tagawa2020,Tagawa2020b,Tagawa2021,Tagawa2020arXiv} as a possile way to generate hierarchical stellar-mass BBH  mergers that are observable by ground-based gravitational-wave detectors. However, we notice that these studies have made improper assumptions in calculating
the disk structure and overlooked an important component of the disk force: the head wind, which orginates from the accretion of disk materials onto the moving sBH \cite{Kocsis2011}. With the improper assumptions corrected and the wind influence included, we find that the total torque is always positive for realistic disk parameters, so that the disk trap is unlikely giving rise to hierarchical stellar-mass binary mergers, nor does it stop the sBHs migrating towards the MBH to become EMRIs.

We incorprate all the relevant disk-star/sBH interactions into a Fokker-Planck code, and then compute the EMRI formation rate with varying MBH mass and disk profiles. The initial distribution of the star cluster is specified according to Tremaine's stellar cluster model \cite{Tremaine1994}. In the limit of zero disk effects, the code reproduces known results from previous ``loss-cone" calculations. With the presence of an ``realistic" accretion disk (Sec.~\ref{sec:emri_disk}), we find that the disk-assisted EMRI formation is generically  $\mathcal O(10^1-10^3)$ times faster than the ``loss cone" mechanism for the same MBH. Taking into account the fraction  of active galactic nuclei (AGNs) observed $\mathcal O(10^{-2}-10^{-1})$ \cite{Galametz2009,Macuga2019}, we conclude that disk-assisted EMRI formation may be an important or even dominant channel for LISA EMRI observation.

Interestingly, disk-assisted EMRIs tend to have low eccentricities ($e\simeq 0$) and low inclinations ($\iota\lesssim 0.1$) if the MBH spin is aligned with angular momentum direction of the accretion disk when they enter the LISA band. These distributions are very different from the ones predicted by ``loss cone" formation
(0--0.2 and $0$--$\pi/2$ at plunge, respectively). Environmental effects of AGN disks may also induce detectable phase shifts to EMRI gravitational waveforms  \cite{Kocsis2011,Yunes2011,Derdzinski2021}.
This means that it is possible to seperate out these two channels with a population of events. With the rate inferred for each channel from observations, one can then further constrain the distribution of stars around MBHs and AGN/disk physics. In particular, EMRIs within AGNs may produce both gravitational wave and electromagnetic signals for multi-messenger observations \cite{Kocsis2011, McGee2020}.

The paper is organized as follows. In Section~\ref{sec:review}, we first briefly review
the canonical EMRI formation channel via loss cone and numerically calculate the EMRI rate for
a fiducial MBH + star/sBH cluster system. In Section~\ref{sec:interaction},
we introduce a few commonly used AGN accretion disk models, explore different interactions
between sBHs/stars with accretion disks, and discuss the existence problem of migration traps
in AGN disks. In Section~\ref{sec:emri_disk}, we incorporate the sBH/star-disk interactions into
the Fokker-Planck equation and numerically calculate the accretion disk-assisted EMRI rate.
Summary and discussion are given in Section~\ref{sec:discussion}. Some numerical details are
placed in Appendix~\ref{apa} and \ref{apb}.

Throughout this paper, we use geometrical units $G=c=1$.

\section{Review of EMRI Formation via Loss Cone}\label{sec:review}
In this section, we  briefly review how stars fall into a MBH via the loss cone mechanism, following
Refs.~\cite{Shapiro1978,Hopman2005,Bar-Or2016}, and then compute the EMRI formation rate for a given
MBH-stellar cluster system by numerically solving the Fokker-Planck equation. Many technical details discussed are also
useful for the rate calculation with the presence of AGN disks.

\subsection{Basics}

Let us consider a stellar mass BH (sBH) orbiting around a MBH
with mass $M_\bullet$, which locates in the center of a galaxy being surrounded by a stellar cluster
with velocity dispersion $\sigma$.
Assuming the sBH is on an eccentric orbit with eccentricity $e$ and semi-major axis length $a$,
its specific orbit energy and specific orbital angular momentum are
\be\label{eq:EJ} E:= \phi(r) - \frac{v^2}{2}=\frac{M_\bullet}{2a}, \quad J =M_\bullet\sqrt{\frac{1-e^2}{2E}}\ ,
\ee
where $\phi(r)=M_\bullet/r$ is the (positive) gravitational potential.
For later convenience, we also define $R\equiv J^2/J^2_c(E)$, where $J_c(E)$ is
the specific orbital angular momentum of a sBH with specific energy $E$ on a circular orbit. For the point-mass gravitational potential, we have $R= 2EJ^2/M_\bullet^2 = 1-e^2$.
The sBH gradually spirals inward as GW emission takes away energy and angular momentum on a timescale $t_{\rm gw}$.
On the other hand, stars and sBHs in the cluster  continously scatter the sBH via mutual gravitational interactions, changing its orbit angular momentum
on a relaxation timescale $t_J$. As a result, the inspiral is susceptible to scatterings
if $t_{\rm gw} > t_J$ and the sBH will either be scattered to an wider orbit or plunges into the MBH.
Therefore a stable EMRI forms only if the GW dissipation dominates over scatterings, i.e., the orbit must be
tight and highly eccentric ($e\rightarrow1$) to enable efficient GW emission, $t_{\rm gw} < t_J$.

In the phase space, there is an region bounded by the energy-dependent angular momentum, $ J_{\rm lc}(E)$, where  stellar mass BHs initially populating the region
promptly fall into the MBH within one orbital period \cite{Cutler1994}. This part of phase space is usually referred as the ``loss cone".
The overall infall rate is set by the rate of diffusion/relaxation processes which drive sBHs to the loss cone by successive two-body scatterings, and the condition of infall is written as  $P(E) < t_J$ \cite{Lightman1977},
where $P(E)$ is the orbital period.
For comparison, the condition of stable
inspirals  $t_{\rm gw} < t_J$ is much stronger, because the orbital period $P(E)$
is usually much shorter than the GW dissipation timescale $t_{\rm gw}$.
For the problem we are discussing, the relevant orbits are nearly zero-energy (with $a\gg M_\bullet$ and $E\simeq 0$), and the boundary of the loss cone is defined by \cite{Cutler1994}
\be
J_{\rm lc, bh}(E\simeq 0) = 4M_\bullet \ .
\ee
The above discussion equally applies to stars around a MBH, except the star loss cone is determined by tidal disruption
and $J_{\rm lc, star}(E)$ is slightly larger than $J_{\rm lc, bh}(E)$
\footnote{A star orbiting around a MBH will be disrupted as long as its periapsis is
$r_p \lesssim r_{\rm star}(M_\bullet/m_{\rm star})^{1/3}$,
where $r_{\rm star}\sim 10^6$ km is the typical star radius.
As a result, we find $J_{\rm lc, star}(E\simeq 0) = M_\bullet\sqrt{a(1-e^2)/M_\bullet} =M_\bullet\sqrt{r_p(1+e)/M_\bullet} \simeq M_\bullet\sqrt{2r_p/M_\bullet}$, where we have used Eq.~(\ref{eq:EJ}) in the first equality,
the relation $r_p=a(1-e)$ in the second and $e\simeq 1$ in the third.}.
For numerical convenience, we simply take $J_{\rm lc, star}(E\simeq 0) = 4M_\bullet$.

\subsection{EMRI rate via loss cone}

\subsubsection{Initial condition}\label{subsub:ini}
For a given MBH, to accurately compute the EMRI rate via the loss-cone mechanism,
we need to know the distribution functions
of the surrounding stars and sBHs,  $f_i(t, \vec x, \vec v)$ \ ($i= {\rm star, bh}$).
As argued in Refs.~\cite{Cohn1978,Cohn1979}, these distribution functions are approximately functions of
the action variables: $f_i \approx f_i(t,E,R)$. In alignment with previous studies \cite{Preto2010,Amaro2011}, we use Tremaine's MBH+stellar cluster model \cite{Tremaine1994,Dehnen1993}
as the initial condition for the Fokker-Planck evolution. Assuming there are two components in the stellar cluster:
light stars with mass $m_{\rm star}$ and heavy sBH with mass $m_{\rm bh}$, and the total star/sBH mass in the cluster
are $M_{\rm star}$ and $M_{\rm bh}$, respectively,
the number densities of stars and sBHs in the Tremaine's cluster model are given by
\be\label{eq:numden}
\begin{aligned}
  n_{\rm star}(r)& = \frac{M_{\rm star}}{m_{\rm star}}\frac{3-\gamma}{4\pi}\frac{r_a}{r^\gamma (r+r_a)^{4-\gamma}}, \\
 n_{\rm bh}(r) &= \delta \times n_{\rm star}(r)\ ,
\end{aligned}
\ee
with $r_a$ being the density transition radius, $\gamma$ being the density scaling power index, and $\delta$ being the relative abundance of sBHs. From Eq.~(\ref{eq:numden}), we see $n_i(r)\sim r^{-\gamma}$ ($i=$ star, bh) for $r\ll r_a$,
$n_i(r)\sim r^{-2-\gamma/2}$ for $r=r_a$, and $n_i(r)\sim r^{-4}$ for $r\gg r_a$. Different combinations of model parameters $\gamma$ and $r_a$ produce rich cluster profiles. For example, the Galactic nuclear stellar cluster is
approximately described by the Tremaine's model with $\gamma=1.8$ and $r_a=4 r_{\rm h}=4M_\bullet/\sigma^2$, where
the star density profile is $n_{\rm star}(r)\sim r^{-1.8}$ within the influence radius $r_{\rm h}$ and
becomes as steep as $n_{\rm star}(r)\sim r^{-3}$ at a distance a few times larger than $r_{\rm h}$ \cite{Launhardt2002,Schodel2014}

From the density profiles (\ref{eq:numden}), one can obtain the magnitude of the gravitational potential
\be\label{eq:phi}
\phi(r) = \frac{M_{\bullet}}{r} + \frac{M_{\rm star}+M_{\rm bh}}{r_a} \frac{1}{2-\gamma}
\left[1 - \left(\frac{r}{r+r_a}\right)^{2-\gamma}\right]\ ,
\ee
where the second term is contributed by the stars and sBHs. In the case that the initial distribution functions $f_i(t=0,E,R)$ only depend on the energy $E$, they are related to the position
space number density by \cite{Tremaine1994}
\be
f_i(t=0,E,R) = \frac{\sqrt{2}}{(2\pi)^2}  \frac{d}{dE}\int_0^E \frac{dn_i}{d\phi} \frac{d\phi}{\sqrt{E-\phi}}\ ,
\ee
where $n_i(r)$ has been written as an implicit function of $\phi(r)$.

In the more general case, $f_i$ depends on both $E$ and $R$.
In order to invert the distribution function $f_i(E,R)$ to find the number density $n_i(r)$, we first
list the  properties of star orbits in given potential field $\phi(r)$
\cite{Cohn1979}. From the energy definition $E=\phi-v^2/2$,
we have
\be
2(\phi-E) = v^2 = \frac{J^2}{r^2} + v_r^2\ ,
\ee
where $v_r$ is the radial velocity. For a circular orbit of energy $E$, its orbit radius and angular momentum $J_c$
are determined by
\be
\begin{aligned}
  J_c^2(E) = -r_c^3\phi'(r_c)\ , \\
  2(\phi(r_c)-E) = \frac{J_c^2}{r_c^2}\ .
\end{aligned}
\ee
For a general non-circular orbit with parameters $(E,R)$,
its turning points (apsis/periapsis) $r_\pm$ are determined by
\be
2(\phi(r_\pm) -E) = \frac{J^2}{r_\pm^2}\ ,
\ee
and its orbit period $P(E,R)$ is defined as
\be
P(E,R) = 2\int_{r_-}^{r_+} \frac{dr}{v_r}\ .
\ee
Defining the number density in the $(E,R)$ phase space as $N_i(E,R)dEdR:=\int_{r_-}^{r_+} d^3rd^3v f_i(E,R)$,
we have \cite{Cohn1978,Cohn1979}
\be\label{eq:C}
\begin{aligned}
  N_i(E,R)
  &= 4\pi^2 P(E,R) J_c^2(E)f_i(E,R) \\
  :&=\mathcal C(E,R) f_i(E,R)\ .
\end{aligned}
\ee

With these listed properties, one can show that the position-space number density $n_i(r)$ is related to
the distribution function $f_i(E,R)$ by \cite{Cohn1979}
\be
n_i(r) = \frac{2\pi}{r^2}\int_0^{\phi(r)} dE J_c^2(E) \int_0^{R_{\rm max}} \frac{dR}{v_r} f_i(E,R)\ ,
\ee
where $R_{\rm max}(r, E)= 2r^2(\phi(r)-E)/J_c^2(E)$, and $v_r(r, E,R) = 2(\phi-E)-J^2/r^2 = (R_{\rm max}-R)J_c^2(E)/r^2$.
In the case of isotropic distribution $f_i=f_i(E)$, the above equation simplifies as \cite{Chernoff1990}
\be
n_i(r) = 4\pi\int_0^{\phi(r)} dE\sqrt{2(\phi(r)-E)} f_i(E)\ .
\ee

\subsubsection{Fokker-Planck equation}
Given initial distributions of stars and sBHs, $f_i(t=0, E, R)$, their evolution is governed by the
orbit-averaged Fokker-Planck equation \cite{Cohn1979}
\be\label{eq:FP}
  \mathcal C\frac{\partial f}{\partial t}
  = - \frac{\partial}{\partial E} F_E
  - \frac{\partial}{\partial R}F_R \ ,
\ee
with $\mathcal C$ the weight function defined in Eq.~(\ref{eq:C}) and $F_{E,R}$ the flux in the $E/R$ direction:
\be\label{eq:flux}
\begin{aligned}
  -F_E &= \mathcal D_{EE}\frac{\partial f}{\partial E} + \mathcal D_{ER}\frac{\partial f}{\partial R} + \mathcal D_E f\ ,\\
  -F_R &= \mathcal D_{RR}\frac{\partial f}{\partial R} + \mathcal D_{ER}\frac{\partial f}{\partial E} + \mathcal D_R f\ ,
\end{aligned}
\ee
where the diffusion coefficients $\{\mathcal D_{EE}, \mathcal D_{ER}, \mathcal D_{RR}\}_i$ and the advection coefficients $\{\mathcal D_E, \mathcal D_R\}_i$ are functions of $f_i(t,E,R)$  \cite{Cohn1979,Binney1987} and we detail their calculation  in  Appendix~\ref{apa}.
In particular, the local relaxation timescale of the system is approximately \cite{Spitzer1971, Binney1987,Bahcall1976,Cohn1978}
\be\label{eq:trlx}
t_{\rm rlx}(r) = 0.34\frac{\sigma^3}{\sum_{i}n_{i}(r)m_{i}^2 \ln\Lambda}\ ,
\ee
where the Coulomb's logarithm $\ln\Lambda$ weakly depends the total number of stars within the
influence radius and we take  $ \ln\Lambda=10$ in this work.

\begin{figure*}
\includegraphics[scale=0.6]{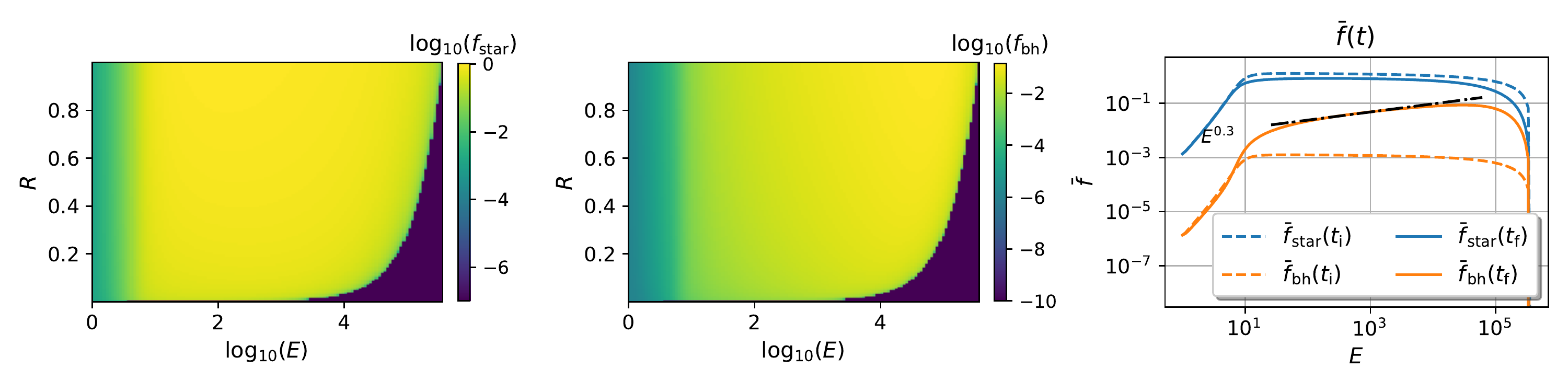}
\caption{\label{fig:lc}  Final distribution functions $f_{\rm star}(t_{\rm f},E,R)$ (left panel),
$f_{\rm bh}(t_{\rm f},E,R)$ (middle panel). Inital and final $R$-integrated distribution functions (right panel). All the distribution functions are shown in units of $10^5{\rm pc}^{-3}/(2\pi\sigma^2)^{3/2}$ and energies $E$ are shown in units of $\sigma^2$.}
\end{figure*}

We aim to evolve $f_{i}(t, E, R)$ according to Eq.~(\ref{eq:FP}) and subject to following boundary conditions
\footnote{In principle, we should evolve both the distributions $f_i$ and the potential field $\phi(r)$ self-consistently.
For the problem we are discussing, the potential field $\phi(r)$ barely changes \cite{Preto2010}
because during the evolution time range the distributions evolve mainly
within the influence radius where the potential field is dominated by the MBH.}.
On the $E\rightarrow 0$ boundary,
\be
f_i(t,E,R)|_{E\rightarrow 0} = f_i(t=0,E,R)|_{E\rightarrow 0}\ ,
\ee
i.e., the distributions far away from the central MBH barely evolve due to its long relaxation timescale.
On the $R=1$ boundary, the flux in the $R$ direction should vanish for both stars and sBHs,
\be
F_{R}|_{R \rightarrow 1} = 0 \ .
\ee
On the loss cone boundary $R=R_{\rm lc}(E):=J_{\rm lc}^2/J_c^2(E)$, the flux in the $R$ direction has been derived in Ref.~\cite{Cohn1978} as
\be\label{eq:bd3}
-\frac{F_R}{\mathcal C} \simeq \left(\frac{D_{RR}}{R}\right)_{R\rightarrow0} \frac{f(R_0)}{\ln(R_0/R_{\rm lc}) +  \mathcal F(y_{\rm lc})}\ ,
\ee
where $D_{RR}:=\mathcal D_{RR}/\mathcal C$, $y_{\rm lc}:=R_{\rm lc}/[(D_{RR}/R)_{R\rightarrow0} P]$, $P=P(E,R)$ is the orbital period, $R_0$ is any small $R$
in the range of $R_{\rm lc}\leq R \ll 1$,
$\mathcal F(y_{\rm lc})\simeq 1/y_{\rm lc}$ for $y_{\rm lc}\lesssim 1$ and $\mathcal F(y_{\rm lc})\simeq 0.824y_{\rm lc}^{-1/2}$ for $y_{\rm lc}\gtrsim 1$.  At $R_0=R_{\rm lc}$, Eq.~(\ref{eq:bd3}) is simplified as $f(R_{\rm lc})\simeq 0$ (empty loss cone) for $y_{\rm lc}\gg 1$ and $F_R\simeq 0$ for $y_{\rm lc}\ll 1$
(full loss cone).

\subsubsection{EMRI rate with the loss-cone mechanism}

We consider a fiducial model of MBH+star/sBH cluster with $M_\bullet=4\times 10^6 M_\odot$, and  two components in the cluster: light stars $m_{\rm star}=1 M_\odot$, and heavy BHs with mass $m_{\rm bh}=10 M_\odot$.
We assume that stellar velocity dispersion $\sigma$ follows the $M_\bullet-\sigma$ relation \cite{Tremaine2002,Gultekin2009}
\be\label{eq:M_sigma}
M_\bullet=1.53\times 10^6M_\odot\left(\frac{\sigma}{70 {\rm km/s}}\right)^{4.24}\ .
\ee
The initial star/sBH distributions are specified according to the Tremaine's model
outlined in Section.~\ref{subsub:ini}
with the total star mass $M_{\rm star}=20 M_\bullet$,
the density transition radius $r_a = 4 r_{\rm h}=4M_\bullet/\sigma^2$,
the density power index $\gamma=1.5$ and
the relative abundance of sBHs $\delta=10^{-3}$. For reference, the initial total number of stars
within the influence sphere is $N_{\rm star}(r<r_{\rm h})=5.6\times 10^6$ in this stellar cluster.

To numerically solve the Eq.~(\ref{eq:FP}), we introduce a new dimensionless
variable $Z=\ln(1+5E/\sigma^2)$ following Ref.~\cite{Cohn1978} and implement
an uniform $128\times 128$ grid on the $(R,Z)$ space.
From the initial distribution functions $f_i(t=0, E,R)$, we
compute all the diffusion coefficients and the advection coefficients,
with which we evolve a discretized version of Eq.~(\ref{eq:FP}) with time step
\be
\delta t = 0.25\mathcal C_{j,k} {\rm min.}\left\{ \frac{\Delta R}{\mathcal D_{R}},
\frac{(\Delta R)^2}{\mathcal D_{RR}}, \frac{\Delta Z}{\mathcal D_{E}}\frac{dE}{dZ},
\frac{(\Delta Z)^2}{\mathcal D_{EE}}\left(\frac{dE}{dZ}\right)^2\right\}_{i,j,k}\ , \nn
\ee
where $i=\{\rm star, sBH\}$, $j$ and $k$ are the grid indices.
We update the coefficients according to the new distribution functions every $100$ steps,
and we stop the simulation at $t_{\rm f}=5$ Gyr.
For convergence test, we also run a low-resolution ($64\times128$) and a high-resolution ($256\times256$)
simulation. We find the simulation results agree with each other to
a good precision.

In Fig.~\ref{fig:lc}, we show the final distributions $f_i(t_{\rm f},E,R)$ in the first two panels and
the $R$-integrated distribution functions
\be
\bar f_i (t,E) := \int_0^1 f_i(t,E,R) dR\ ,
\ee
in the third panel,
where  $\bar f_{\rm star}(t_{\rm f},E)\sim E^{0}$ and  $\bar f_{\rm bh}(t_{\rm f},E)\sim E^{0.3}$ for $E/\sigma^2\gtrsim 10$. The sBH density profile $\bar f_{\rm bh}$ here turns out to be shallower than that obtained from solving 1-d Fokker-Planck equations \cite{Preto2010}. The steeper profile from 1-d calculation is expected because sBHs' leaving the cluster and falling into the MBH via the loss cone cannot be incorporated in the 1-d Fokker-Planck equation, and consequently sBHs which are supposed to fall into the MBH,
are instead accumulated in the large $E$ regime.
The phase space distribution function being $\bar f(E)\sim E^\xi$ is approximately equivalent
to the position space density being $n(r)\sim r^{-(3/2+\xi)}$ \cite{Lightman1977}.
As a result, the star/sBH number density profile in the final state are
$n_{\rm star}(r)\sim r^{-1.5}$ and  $n_{\rm bh}(r)\sim r^{-1.8}$ for $r\lesssim 0.1r_{\rm h}$.
We can also find that the  density profile of the more massive, sBH component is steeper than that of the star component,  a  phenonmenon known as mass segregation,
which has been shown to enhance the EMRI rate \cite{Alexander2009, Preto2010, Amaro2011}.

\begin{figure}
\includegraphics[scale=0.8]{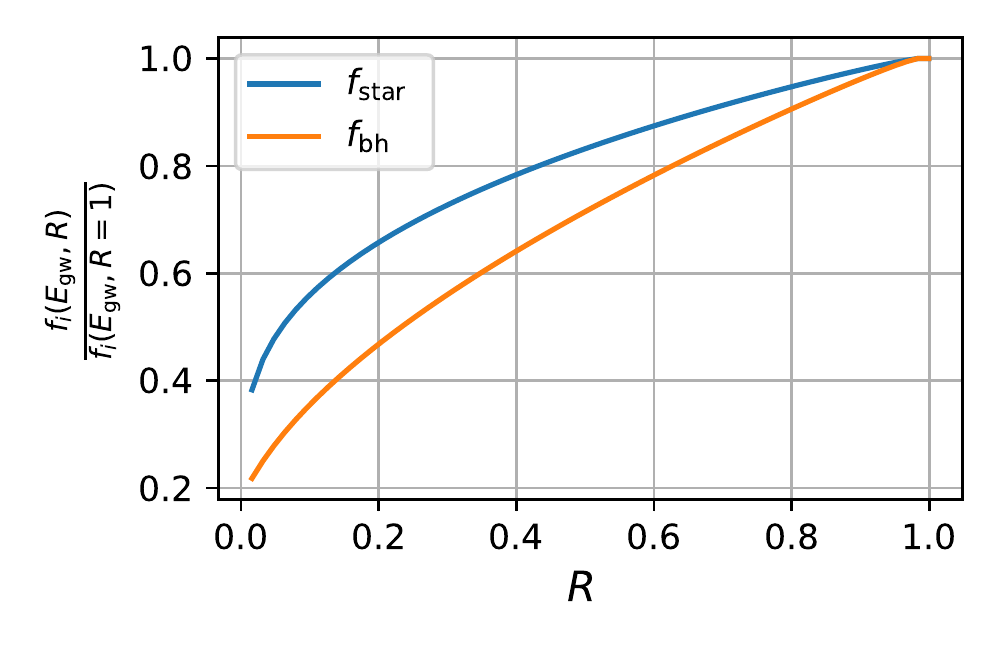}
\caption{\label{fig:R}  Mass segregation in the $R$-direction, where the heavier component
is more concentrated on circular orbits.}
\end{figure}

In fact, mass segregation not only occurs in the $E$- dimension, but also in the $R$-dimension.
In Fig.~\ref{fig:R}, we show the final distributions of the two components at $E=E_{\rm gw}$ [see Eq.~(\ref{eq:emri})] in the $R$-direction,
where the sBH component is more concentrated on circular orbits. As a result, the mass segregation in the
$R$-direction is expected to mildly reduce the EMRI rate, which is mainly contributed by sBHs on highly eccentric orbits.

\begin{figure}
\includegraphics[scale=0.8]{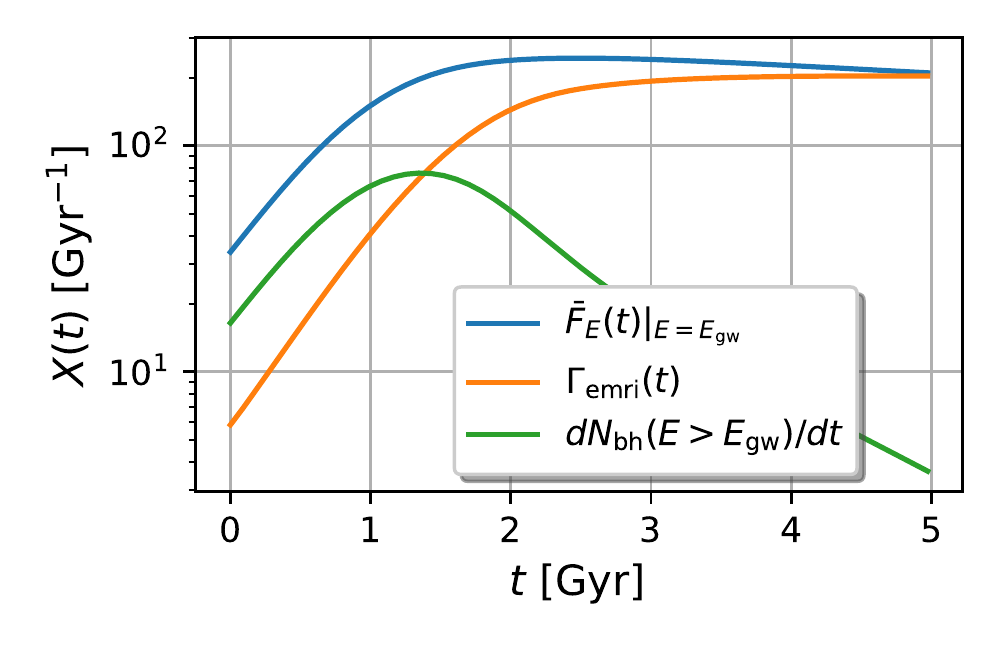}
\caption{\label{fig:rates}  The time dependence of the inflow rate $\bar F_E|_{E=E_{\rm gw}}$,
the EMRI rate $\Gamma_{\rm emri}$ and the sBH number growth rate $dN_{\rm bh}(E>E_{\rm gw})/dt$.}
\end{figure}

With the distribution function $f_{\rm bh}(t,E,R)$ (and all the diffusion and advection coefficients)
obtained, we are ready to calculate the EMRI rate for the loss cone mechanism.
As discussed in Refs.~\cite{Hopman2005,Amaro2011}, the EMRI condition $t_{\rm gw} < t_J$ is approximately formulated as
 $a<r_{\rm gw} = 0.01 r_{\rm h}$ . Therefore the EMRI rate per MBH via loss cone is given by
\be\label{eq:emri}
\Gamma_{\rm emri} = \int_{E>E_{\rm gw}}  \vec F \cdot d\vec l\ ,
\ee
where $E_{\rm gw}= M_\bullet/(2r_{\rm gw})$, $\vec F = (F_E, F_R)$,
and $d\vec l = (dE, dR)$ is the line element along the boundary of the loss cone.
According to the flux conservation in the steady state, the EMRI rate should be equal to the inflow rate
$\bar F_E$ at $E=E_{\rm gw}$
\be
\bar F_E|_{E=E_{\rm gw}} = \int_0^1 F_E(E,R) |_{E=E_{\rm gw}} dR\ .
\ee
In Fig.~\ref{fig:rates}, we plot the time dependence of three different rates: $\bar F_E|_{E=E_{\rm gw}}$,
$\Gamma_{\rm emri}$, and the sBH number growth rate $dN_{\rm bh}(E>E_{\rm gw})/dt$.
We find that the EMRI rate has reached an quasi-steady state at $t\sim 3$ Gyr,
when $\Gamma_{\rm emri}\simeq 210 \ {\rm Gyr}^{-1}$. In addition,  the inflow rate $\bar F_E|_{E=E_{\rm gw}}$
initially increases up to $t\sim 2$ Gyr and slowly decreases until $\bar F_E|_{E=E_{\rm gw}}\simeq\Gamma_{\rm emri}$ at $t_{\rm f}=5$ Gyr.
The total number of sBHs  $N_{\rm bh}|_{E>E_{\rm gw}}$ increases from its initial value
$\simeq 10$ to the final value $\simeq 160$. After $5$ Gyr (not shown in plot),
$N_{\rm bh}(E>E_{\rm gw})$ starts to slowly decrease because the consumption rate via loss cone $\Gamma_{\rm emri}$
becomes larger than the inflow supply rate $\bar F_E|_{E=E_{\rm gw}}$.

\begin{figure}
\includegraphics[scale=0.8]{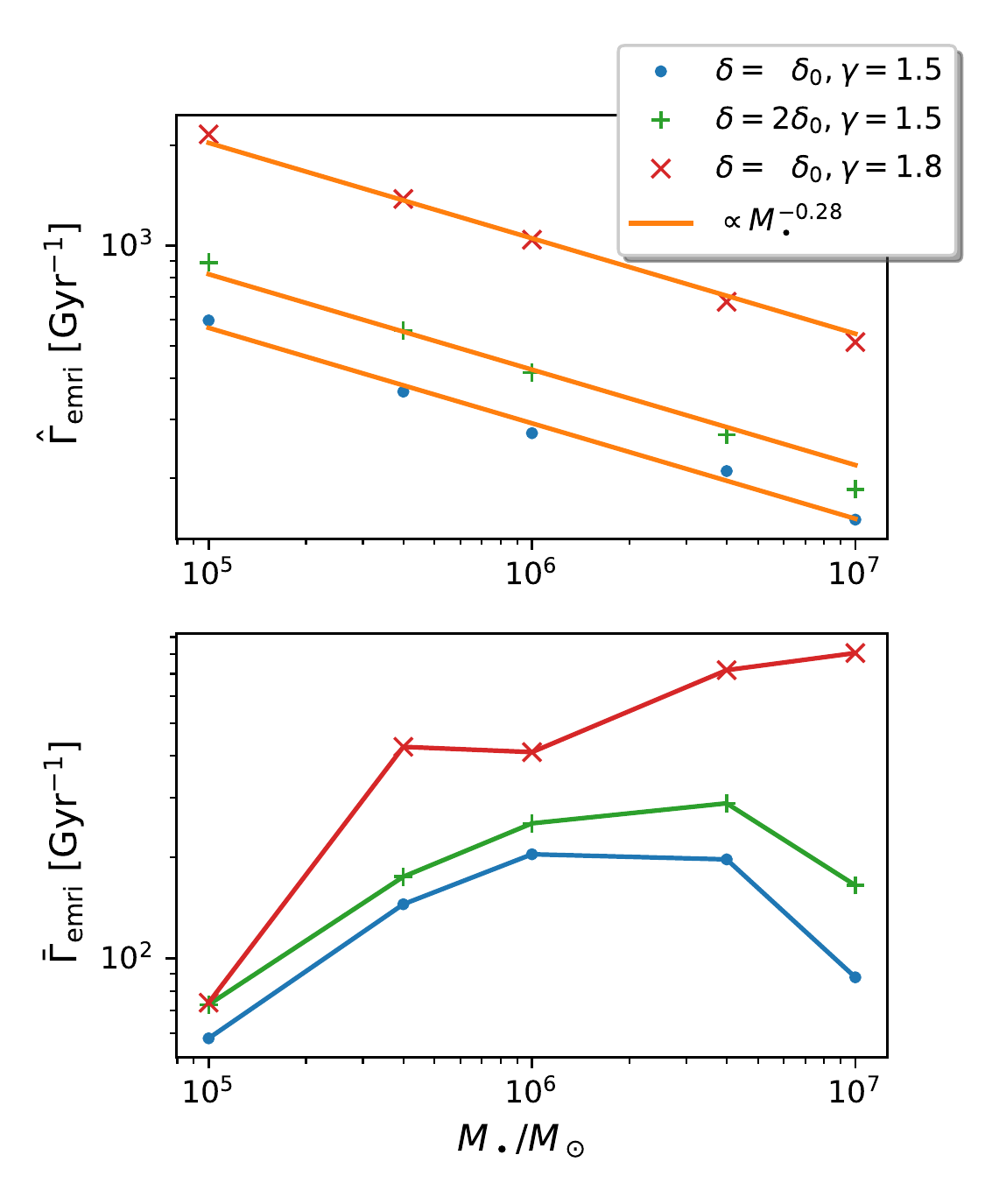}
\caption{\label{fig:Gamma_M} Upper panel:the dependence of the characteristic EMRI rate per MBH $\hat\Gamma_{\rm emri}$
on the MBH mass $M_\bullet$ for different initial cluster models, where $(\delta=\delta_0=10^{-3},\gamma=1.5)$ is our fiducial model, $(\delta=2\delta_0, \gamma=1.5)$ is  a similar cluster model with higher sBH fraction,
and $(\delta=\delta_0, \gamma=1.8)$ is a Galactic nuclear stellar cluster like model.
Lower panel: the average  rate  $\bar\Gamma_{\rm emri}$.}
\end{figure}

To explore the dependence of EMRI rate on the MBH mass, we have performed  additional 4 simulations
similar to the fiducial model case, except with different MBH mass $M_\bullet$.
To quantify the EMRI rate via loss cone, we define the average rate
\be
\bar \Gamma_{\rm emri} :=\frac{1}{t_{\rm f}}\int_0^{t_{\rm f}} \Gamma_{\rm emri}(t) dt\ ,
\ee
and  a characteristic rate
\be
\hat\Gamma_{\rm emri}:= \Gamma_{\rm emri}(t)|_{{\rm when}\ N_{\rm bh}(t,E>E_{\rm gw})\ {\rm maximizes} }\ ,
\ee
i.e., the EMRI rate when the total sBH number $N_{\rm bh}(E>E_{\rm gw})$ is maximal in the range of $t\in[0,t_{\rm f}]$. For the fiducial model we see that
$\hat\Gamma_{\rm emri}=\Gamma_{\rm emri}(t=t_{\rm f})$ (see Fig.~\ref{fig:rates}), while $\hat\Gamma_{\rm emri}$
turns out to be the EMRI rate at some earlier time $\Gamma_{\rm emri}(t<t_{\rm f})$ for cases with lighter MBHs, for which
the relaxation timescales are shorter and the peak $N_{\rm sBH}(E>E_{\rm gw})$ comes earlier. In the upper panel of Fig.~\ref{fig:Gamma_M}, we show the characteristic rate $\hat\Gamma_{\rm emri}$ as a function of $M_\bullet$. We find that the MBH mass dependence is mild, which can be approximated as $\propto M_\bullet^{-0.28}$.
This scaling can be  qualitatively understood as follows \cite{Hopman2005}:
\be \hat\Gamma_{\rm emri}\propto \frac{N_{\rm bh}(E>E_{\rm gw})}{t_{\rm rlx}(r_{\rm gw})}
\propto \frac{\sigma^3}{M_\bullet}\ ,
\ee
where we have used $N_{\rm bh}(E>E_{\rm gw})\propto M_\bullet$,
$t_{\rm rlx}(r_{\rm gw})\propto \sigma^3/n_{\rm star}(r_{\rm gw})\propto
\sigma^3 r_{\rm gw}^3/N_{\rm star}(E>E_{\rm gw})\propto \sigma^3 r_{\rm gw}^3/M_\bullet$,
and $r_{\rm gw}\propto M_\bullet/\sigma^2$.
In combination with the $M_\bullet-\sigma$ relation [Eq.~(\ref{eq:M_sigma})], we have $\hat\Gamma_{\rm emri}\propto M_\bullet^{-0.29}$,
which is  close to the scaling fitted from our numerical results.

To explore the effect of the initial cluster profile, we also run two sets of simulations with two different
initial stellar cluster models: one with higher sBH fraction $(\delta=2\times10^{-3},\gamma=1.5)$,
and another with steeper density profile $(\delta=10^{-3},\gamma=1.8)$, where the initial total number of stars
within the influence sphere is about $50\%$ higher with $N_{\rm star}(r<r_{\rm h})=8.1\times10^6$. We find the dependence of $\hat\Gamma_{\rm emri}$ on $M_\bullet$ can be approximately fitted by the same scaling
$\hat\Gamma_{\rm emri}\propto M_\bullet^{-0.28}$ in both cases. As expected, both higher sBH fraction
and steeper density profile enhance the characteristic rate $\hat\Gamma_{\rm emri}$. In more details,
the EMRI rate increases by less than a factor of 2 when we double the sBH fraction,  because
the sBH-sBH coupling becomes stronger which tends to flatten the density profile of sBHs.
The EMRI rate $\hat\Gamma_{\rm emri}$ in the stellar cluster with a steeper density profile
is higher by a factor of $\sim 3$, which is mainly contributed by higher sBH numbers and shorter relaxation
timescale (higher density of stars) within the influence sphere. The normalized EMRI rate $\hat\Gamma_{\rm emri}/N_{\rm star}^2(r<r_{\rm h})$ in the two stellar clusters differs by only $\sim 50\%$.

To compare future LISA detectable EMRI rate with the model prediction,
the average rate $\bar\Gamma_{\rm emri}$ is more relevant,
whose dependence on the MBH mass $M_\bullet$ and on the initial cluster model are shown in the lower panel of Fig.~\ref{fig:Gamma_M}. We see no simple scaling between the average rate
and the MBH mass for any initial cluster model. The overall behavior is the ratio
$\bar\Gamma_{\rm emri}/\hat\Gamma_{\rm emri}$ is lower for lower MBH mass, while the ratio depends more on the initial condition for MBHs on the high mass end.  For a MBH on the low mass end, sBHs in the
vicinity of the MBH are rapidly depleted and the average rate is limited by the number of sBHs available.
For  a MBH on the high mass end, the relaxation timescale of the system is long and the two rates are different by
no more than a factor of $2$.

In Ref.~\cite{Amaro2011}, stars and sBHs in the cluster are evolved following 1-d ($E$-direction) Fokker-Planck equations,
in which no star/sBH loss is takend account of.
Therefore the numbers of stars and sBHs in the cluster are conserved and there
exists a steady state, base on which the EMRI rate was calculated assuming
the standard logarithmic distribution in the $R$-direction and the relaxation timescale (\ref{eq:trlx})
in the $R$-direction \cite{Hopman2005}.
Though there is no strictly steady state in our approach because the cluster continually
loses stars/sBHs via the loss cone. The characteristic rate $\hat \Gamma_{\rm emri}$ is roughly comparable with
the steady-state EMRI rate in Ref.~\cite{Amaro2011}, in terms of either magnitude or the scaling with $M_\bullet$.

\section{sBH-accretion disk interactions}\label{sec:interaction}

About $1\%$ low-redshift ($z \lesssim 1$) galaxies and as high as $10\%$
high-redshift ($1\lesssim z \lesssim 3$) galaxies are active \cite{Galametz2009,Macuga2019}
in which MBHs are expected to be rapidly accreting gas in a disk configuration.
The interaction with disk could completely reshape the distribution of sBHs.
In this section, we will first discuss models of AGN accretion disks,
then introduce relevant disk-sBH interactions, including density waves and wind.
Other possible interactions, e.g., dynamic friction \cite{Chandrasekhar1943,Ostriker1999}
and heating torque \cite{Masset2017,Hankla2020}, are negligible as we will explain in Section~\ref{subsec:wind}.

\subsection{AGN disk models }\label{subsec:disk}

\emph{$\alpha/\beta$-disk:} With the $\alpha$-viscosity prescription \cite{Shakura1973} and the thin disk assumption,
the 1-d disk structure is governed by the following equations \cite{Sirko2003}:
\begin{subequations}
  \be\label{eq:Teff}
    \sigma_{\rm SB} T_{\rm eff}^4 = \frac{3}{8\pi} \dot M_\bullet \Omega^2\ ,
  \ee
  \be
  T_{\rm mid}^4 =\left(\frac{3}{8}\tau + \frac{1}{2}+\frac{1}{4\tau} \right) T_{\rm eff}^4\ ,
  \ee
  \be
  \tau = \frac{\kappa\Sigma}{2}\ ,
  \ee
  \be
  \beta^b H^2\Omega^2\Sigma = \frac{\dot M_\bullet \Omega}{3\pi\alpha}\ ,
  \ee
  \be
  p_{\rm rad} =\frac{\tau}{2}\sigma_{\rm SB}T_{\rm eff}^4\ ,
  \ee
  \be
  p_{\rm gas} = \frac{\rho k T}{m_{\rm H}}\ ,
  \ee
  \be
  \beta   =  \frac{p_{\rm gas}}{p_{\rm rad}+   p_{\rm gas}}\,
  \ee
  \be
  \Sigma =2\rho H\ ,
  \ee
  \be
  c_s = H\Omega = \sqrt{p_{\rm tot}/\rho} \ ,
  \ee
  \be
  \kappa = \kappa(\rho, T_{\rm mid})\ ,
  \ee
\end{subequations}
with $T_{\rm mid}$ being the middle plane temperature, $T_{\rm eff}$ being the effective radiation temperature,
$\tau$ being the disk optical depth, $H=rh$ being the scale height of the disk, $c_s$ being the local sound speed and
$\kappa$ being the gas opacity \cite{Alexander1994,Iglesias1996}. The parameter $b$ can be either $0$ or $1$ depending on whether the viscosity is proportional
to the toal pressure ($\alpha$-disk) or the gas pressure ($\beta$-disk).

In the outer region, the viscosity dissipation heating becomes less efficient and
the disk will be prone to the disk self-gravity if the Toomre's stability parameter
\be \label{eq:Toomre}
Q = \frac{c_s\Omega}{\pi \Sigma}\simeq \frac{\Omega^2}{2\pi\rho}
\ee
is less than unity. Following Ref.~\cite{Sirko2003}, we assume some external feeding back mechanism heats the disk and maintains a minimum value of the Toomre's parameter as $Q_{\rm min}\simeq 1$.
In outer parts where the equilibrium between the local viscosity dissipation heating
and the radiation cooling no longer holds, Eq.~(\ref{eq:Teff}) is replaced by Eq.~(\ref{eq:Toomre})
with $Q=Q_{\rm min}$.

Given the value of accretion rate $\dot M_{\bullet}$ and the gas opacity function $\kappa(\rho,T_{\rm mid})$, all the disk variables
$\{T_{\rm eff}, T_{\rm mid}, \tau, \Sigma, \rho, H, p_{\rm rad}, p_{\rm gas}, \beta, c_s, \kappa\}$ can be numerically
solved as functions of the angular velocity $\Omega$, which is approximated by the Kepler angular velocity
$\Omega = \sqrt{-\phi'(r)/r}$. In this work, we assume the potential $\phi(r)$ [Eq.(\ref{eq:phi})] of the fiducial
MBH+cluster model.

As two fiducial disk models, we caculate the structure of an $\alpha$-disk and an $\beta$-disk both with
$M_\bullet=4\times10^6 M_\odot$,
$\alpha=0.1$, $\dot M_\bullet = 0.1 \dot M_{\bullet}^{\rm Edd}$, where the Eddington accretion rate is related to the Eddington luminosity by $\dot M_{\bullet}^{\rm Edd}:= L_{\bullet}^{\rm Edd}/0.1$.
We plot the fiducial $\alpha/\beta$-disk structure in left panels of Fig.~\ref{fig:alpha}:
the disk surface density $\Sigma$, the middle plane temperature $T_{\rm mid}$, the disk
optical depth $\tau$ and the disk aspect ratio $h$ as functions of radius $r$.
The two disks only differ within radius $\sim 10^3 M_\bullet$, beyond which the gas pressure dominates over the radiation
pressure, so that the difference in the viscosity prescriptions of the two disks is negligible.
In the inner region, the radiation pressure dominates, so that the viscosity in the $\alpha$-disk is larger than
in the $\beta$-disk, which results in a larger radial gas velocity and a lower gas surface density.

\emph{TQM disk:} In the TQM disk model \cite{Thompson2005}, the disk angular momentum is
assumed to be carried away by global torques instead of local viscosity,
and the gas inflow velocity is parameterized as a constant fraction of local sound speed:
$v_{{\rm gas},r} = X c_s$. In outer parts of the disk, star formation is assumed to heat the disk and maintain its stability
against disk self gravity. In addition to the gas pressure and the radiation pressure, a turbulence pressure
driven by supernova explosion in the disk is also incorporated. The 1-d disk structure is governed by following equations:
\begin{subequations}
  \be\label{eq:Teff_TQM}
    \sigma_{\rm SB} T_{\rm eff}^4 = \frac{3}{8\pi} \dot M_\bullet \Omega^2 + \frac{1}{2}\epsilon_\star\dot\Sigma_\star\ ,
  \ee
  \be
  T_{\rm mid}^4 =\left(\frac{3}{8}\tau + \frac{1}{2}+\frac{1}{4\tau} \right) T_{\rm eff}^4\ ,
  \ee
  \be
  \tau = \frac{\kappa\Sigma}{2}\ ,
  \ee
  \be
  \dot M(r) = X c_s (2\pi r \Sigma)\ ,
  \ee
   \be
  c_s = H\Omega = \sqrt{p_{\rm tot}/\rho} \ ,
  \ee
  \be
  p_{\rm gas} = \frac{\rho k T_{\rm mid}}{m_{\rm H}}\ ,
  \ee
    \be\label{eq:p_rad}
  p_{\rm rad} =\frac{\tau}{2}\sigma_{\rm SB}T_{\rm eff}^4\ ,
  \ee
  \be
  p_{\rm tb} = \epsilon_\star \dot\Sigma_\star\ ,
  \ee
  \be
  \Sigma =2\rho H\ ,
  \ee
  \be
  \dot M(r) = \dot M_\bullet + \int_{r_{\rm min}}^r 2\pi r \dot\Sigma_\star dr \ ,
  \ee
  \be
  \kappa = \kappa(\rho, T_{\rm mid})\ .
  \ee
\end{subequations}
In inner parts where $Q>1$,
the star formation ceases ($\dot\Sigma_\star = 0$), the accretion rate is radius-independent $\dot M\equiv \dot M_\bullet$ and the turbulence pressure $p_{\rm tb}$ vanishes. In outer parts, the Toomre's stability parameter is assumed to be
$Q=1$ and the density is specified by
\be
\rho = \frac{\Omega^2}{2\pi}\ .
\ee

In Fig.~\ref{fig:TQM}, we show the structure of an example TQM disk with $M_\bullet=4\times10^6M_\odot$, $\dot M_\bullet=0.1 \dot M_\bullet^{\rm Edd}$ , $X=0.1$ and $\epsilon_\star=10^{-3}$. A salient feature in the disk is a opacity gap at $r\sim 10^4 M_\bullet$, inside which the disk is optically thin $\tau < 1$.
In Ref.~\cite{Thompson2005}, a sharp density increase was found
on the inner edge of the opacity gap, while the density increase in our solution is rather mild.
This difference is traced back to different equations of radiation pressure assumed:  in Ref.~\cite{Thompson2005}, $p_{\rm rad}=\frac{4}{3}\sigma_{\rm SB}T_{\rm mid}^4$ was assumed, which should hold only in the optically thick regime and break down inside the opacity gap, while our equation of radiation pressure [Eq.~(\ref{eq:p_rad})] is more general.

The $\alpha$-viscosity prescription is consistent with the turbulence viscosity driven by magnetorotational instability
in inner parts of accretion disks where the gas is fully ionized \cite{Balbus1991,Balbus1998,Martin2019}. In outer parts,
the physical mechanisms of the angular momentum transport and the external heating processes (in addition to the disk viscosity heating) maintaining the disk stability are still open issues. In this work, we follow Ref.~\cite{Sirko2003} to consider $\alpha/\beta$ disks
assuming $\alpha$-viscosity prescription throughout the disk and certain implicit heating process in outer parts of the disk.
In consistent with  Ref.~\cite{Thompson2005}, we also consider TQM disks where a more efficient angular momentum transport
mechanism is assumed. In addition, star formation in outer parts of AGN disk is explictly taken into account  as the external heating process. To our best knowledge, we expect $\alpha/\beta$ disk models to be a closer description to inner parts of AGN disks in nature, while it is not clear which disk model works better or whether any of them accurately describes the nature in outer regions.

\subsection{Density waves}\label{subsec:den_wav}
As extensively studied in the context of star-disk-satellite systems \cite{Goldreich1979,Goldreich1980,Ward1989,Tanaka2002,Tanaka2004},
a planet excites density waves consisting of three components:  regular density waves excited excited by the circular motion of
the planet, eccentricity waves excited by the non-circular motion and bending waves excited by the motion normal to the disk.
The regular density waves exert a negative torque on the planet
and drive an inward migration (commonly called type-I migration) on a timescale $t_{\rm mig,I}$;
the eccentricity waves work to damp the eccentricity $e$ of the planet orbits
on a timescale $t_{\rm wav}$ and the bending waves work to drive the planet onto the disk on a same timescale $t_{\rm wav}$.
Similar processes should also work in the MBH-disk-sBH system, with torque arising from density waves \cite{Tanaka2002,Tanaka2004}
\be\label{eq:JI}
\dotJI = C_{\rm I}\frac{m_{\rm bh}}{M} \frac{\Sigma}{M }\frac{r^4\Omega^2}{h^2}\ ,
\ee
where  $M=M(<r)$ is the total mass within radius $r$, $C_{\rm I} = -0.85 + d\ln \Sigma/d\ln r + 0.9\ d\ln T_{\rm mid}/d\ln r$ \cite{Paardekooper2010}. In some special cases we will show later, the surface density $\Sigma(r)$ is a fast increasing function of radius $r$, and the type-I torque becomes positive.
The corresponding migration timescale $t_{\rm mig,I}$ and eccentricity/inclination damping timescale $t_{\rm wav}$ are
\be\label{eq:twav}
\begin{aligned}
  t_{\rm mig,I} &= \frac{J}{|\dotJI|} = \frac{r^2\Omega}{|\dotJI|}\sim
\frac{M}{m_{\rm bh}}\frac{M}{\Sigma r^2}\frac{h^2}{\Omega} , \\
  t_{\rm wav} &= \frac{M}{m_{\rm bh}}\frac{M}{\Sigma r^2}\frac{h^4}{\Omega}\ ,
\end{aligned}
\ee
where in the ``$\sim$" sign we have take $|C_{\rm I}|=1$ for order of magnitude estimate.
In the context of turbulent protoplanet disks, planets are also subject to
stochastic migration due to gravitational interaction with turbulent density fluctuations in the disk \cite{Nelson2005}.
It is not clear under what condition the stochastic migration of sBHs in AGN disks prevails the type-I migration \cite{Johnson2006,Yang2009}. We do not include the possible contribution from stochastic migration in this paper, and
should be straightforward to do so by adding an diffusion term in the Fokker-Planck equation as long as
it is better quantified.

A gap in the disk might be opened if the sBH is so massive that its tidal torque moves
gas away faster than the viscosity replenishing rate. The gap width can be estimated
as \cite{Lin1986,Bryden1999,Crida2006,Kocsis2011}
\be
\Delta \simeq \left(0.02 \frac{r^2\Omega}{\nu}\frac{m_{\rm bh}^2}{M^2}\right)^{1/3} r\ ,
\ee
where $\nu=\dot M/(3\pi\Sigma)$ is the kinetic viscosity coefficient.
The gap opening requires \cite[see also][]{Duffell2013}
\be\label{eq:gap}
H < \Delta, \quad r_{\rm Hill} < \Delta \ ,
\ee
where $H$ is the disk thickness and $r_{\rm Hill}=(m_{\rm bh}/3M)^{1/3}r$ is the Hill radius of the sBH inside which the tidal field of the sBH dominates. As long as a gap opens, type-I migration turns off and the sBH is subject to type-II
migration. Following Ref.~\cite{Syer1995} \cite[see also][]{Duffell2014,Durmann2015}, the type-II torque on the sBH can be estimated as
\be
\dot J_{\rm mig,II} = -\frac{2\pi r^2\Sigma}{m_{\rm bh}} r\Omega |v_{{\rm gas},r}|\ ,
\ee
where $v_{{\rm gas},r} = -\dot M/(2\pi r\Sigma)$ is the gas inflow velocity.
The corresponding  timescale of type-II migration is defined as
\be
t_{\rm mig,II} = \frac{r^2\Omega}{|\dot J_{\rm mig,II}|}\ .
\ee
The above analysis equally applies to stars except with a lower mass $m_{\rm star}$.

\begin{figure*}
\includegraphics[scale=0.65]{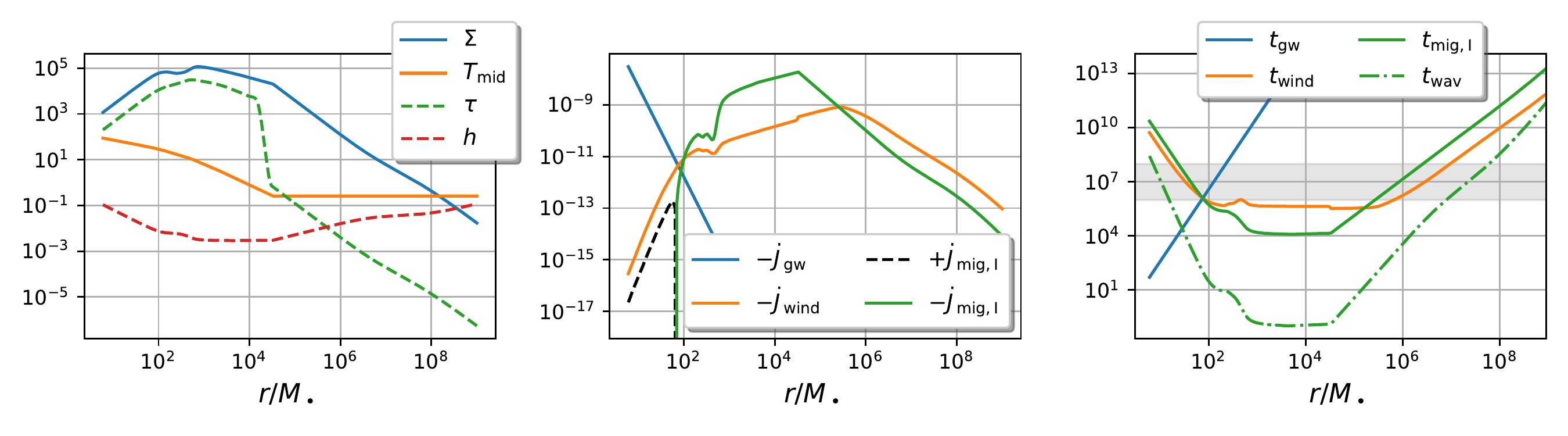}
\includegraphics[scale=0.65]{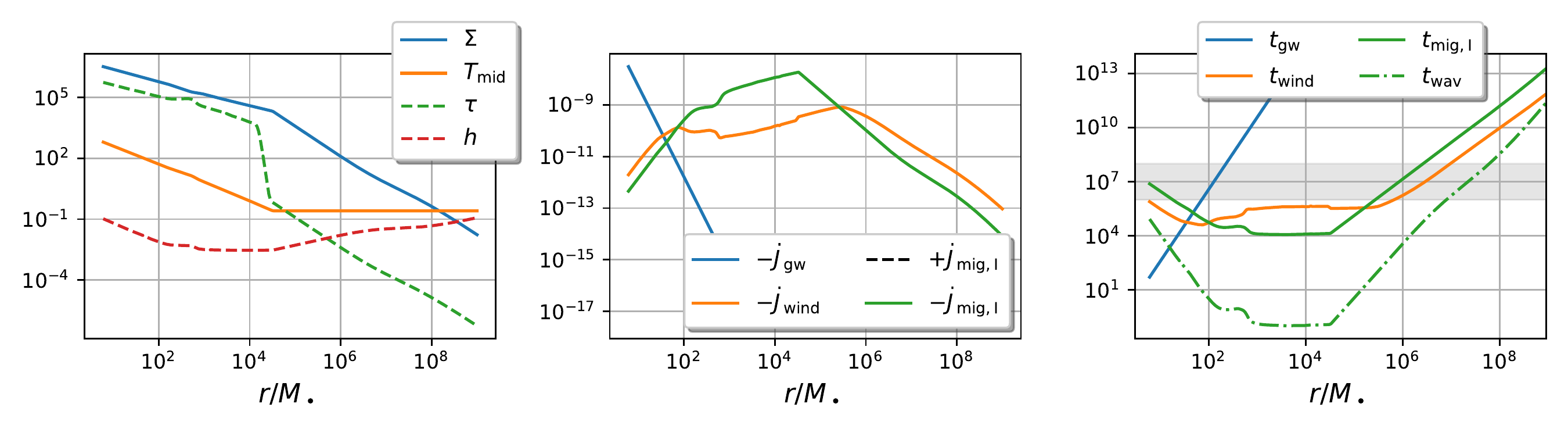}
\caption{\label{fig:alpha} Fiducial $\alpha/\beta$-disk in the upper/lower row.
Left panel: disk structure
with $\Sigma\ [{\rm g/cm}^3]$ the surface density, $T_{\rm}\ [\rm eV]$ the middle plane temperature,
$\tau$ the disk optical depth and $h:=H/r$ the disk aspect ratio.
Middle panel: the torques (in units of $c^2=1$) exerted on the sBH from GW emission $\dot J_{\rm gw}$ [\ref{eq:Jgw}],
disk wind $\dot J_{\rm wind}$ [\ref{eq:Jwind}]
and density waves $\dot J_{\rm mig, I}$ [\ref{eq:JI}], where  $\dot J_{\rm mig, I}$ changes its sign close to the local density maxima.
Right panel: the corresponding timescales (in units of yr) on which different torque change the sBH orbital angular momentum by order of unity $t_i:= J/|\dot J_i|$. with $i={\rm gw/ wind/ mig,I}$. For comparison, the typical disk life span $\sim (10^6,10^8)$ yrs is plotted as a horizontal gray band.}
\end{figure*}

\begin{figure*}
\includegraphics[scale=0.65]{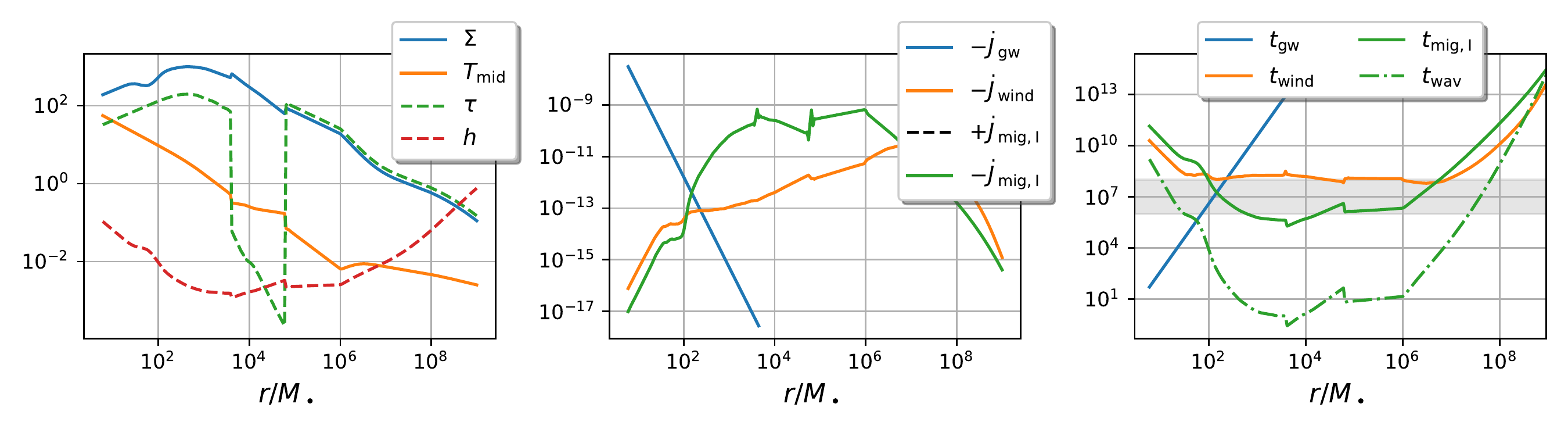}
\caption{\label{fig:TQM} Same to Fig.~\ref{fig:alpha} except for a fiducial TQM disk.}
\end{figure*}

\begin{figure*}
\includegraphics[scale=0.65]{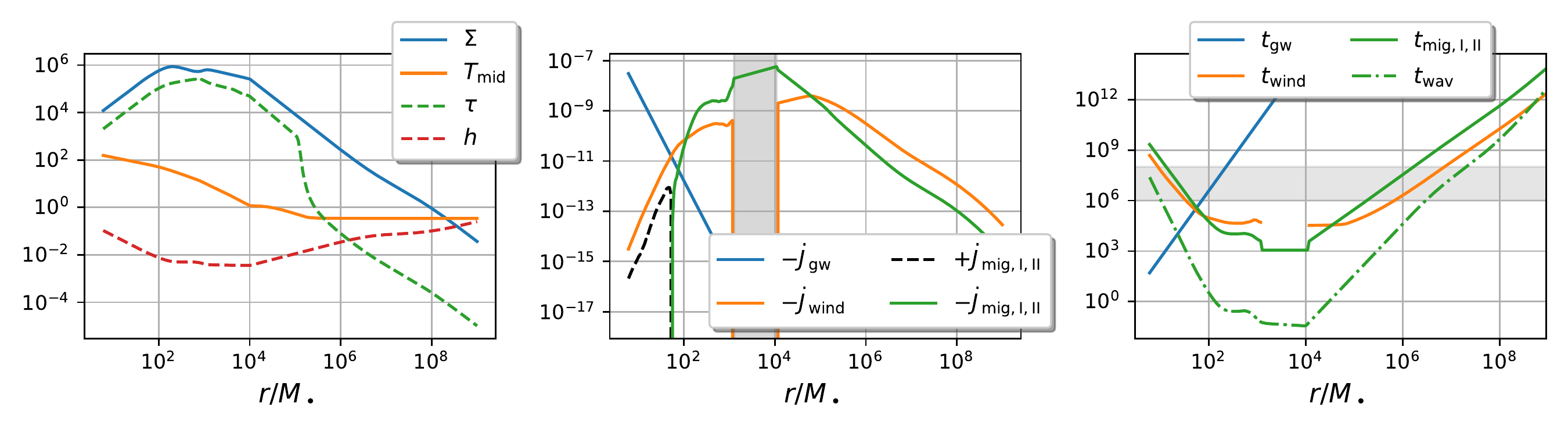}
\includegraphics[scale=0.65]{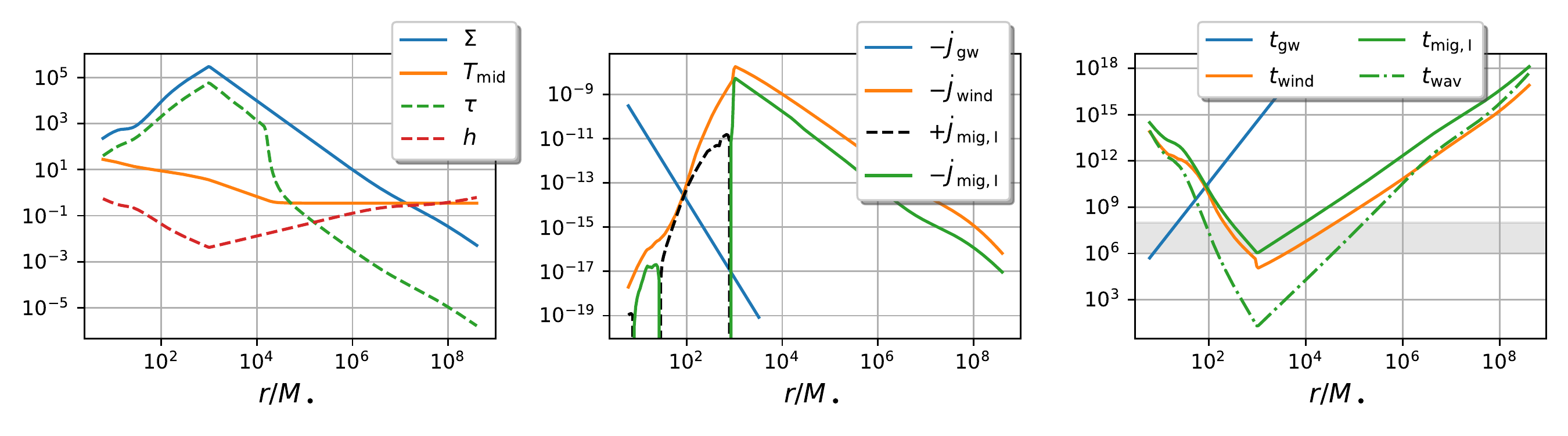}
\caption{\label{fig:alpha2} Upper row:same to Fig.~\ref{fig:alpha} except for a comparison $\alpha$-disk
with low viscosity $\alpha=0.01$, where the gap opening condition is satisfied when an orbiting sBH is located in the range of $\sim (10^3,10^4)M_\bullet$ (vertical gray band in the middle panel).\\
Lower row:same to Fig.~\ref{fig:alpha} except for a comparison $\alpha$-disk
with $M_\bullet=4\times 10^8 M_\odot$ and $\dot M_\bullet = 0.5 \dot M_\bullet^{\rm Edd}$,
where the type-I torque becomes positive in two disconnected regions. }
\end{figure*}

In the three fiducial disk models (Figs.~\ref{fig:alpha} and \ref{fig:TQM}),
the gap opening condition (\ref{eq:gap}) is not satisfied.
In the upper row of Fig.~\ref{fig:alpha2}, we show the structure of an comparison $\alpha$-disk with
low viscosity $\alpha=0.01$ where the gap opening condition  is satisfied if a
sBH orbits around the MBH in the range of $\sim (10^3, 10^4)M_\bullet$.

\subsection{Wind}\label{subsec:wind}
For a sBH embedded in the gas disk, its gravitational attraction influences the surrounding gas meterials, so that they tend to flow towards to the sBH.
If the disk is not rotating and the sBH has no relative motion with repsect to the disk, these gases should flow towards the sBH  in  a nearly-spherical manner. On the other hand, if the disk is rotating and the sBH has nonzero velocity relative to nearby materials, the accretion cannot be spherical. In addition, the infalling materials generally carry nonzero angular momentum relative to the sBH, so that they tend to circularize and form certain  local disk or buldge profile to organize the accretion flow. Depending on the heating processes and magnetic fields, a major part of captured materials may be carried away in the form of outflow
and only the remaining  part is accreted \cite{Yang2014,McKinney2014}.
Because of the circularization process, it is reasonable to expect that
the outflow materials carry minimal net momentum with respect to the sBH.
Therefore all the infalling gas, either finally being accreted by the sBH
or flying away in the outflow, exerts a net torque to the sBH.
As the sBH grows via gas accretion, the processes of sBH capture onto the disk and migration inward
accelerate [Eq.~(\ref{eq:twav})].
Considering the expected strong feedback during accretion,
the calculation of sBH growth in AGN disks needs more sophisticated modelling in separate studies
and we conservatively take it as $\dot m_{\rm bh}=0$ in this paper.

As a result, the ``head wind" with respect to the sBH are captured at places
where the sBH gravity becomes important, and the momentum carried by the wind eventually transfers to the sBH.
The specific torque exerted on the sBH from the wind is written as
\be\label{eq:Jwind}
\dot J_{\rm wind}^{\rm id} = - \frac{r \delta v_\phi \dot m_{\rm wind}}{m_{\rm bh}}\ ,
\ee
where the upper script ``${\rm id}"$ is used to denote quantities of in-disk (id) sBHs
and $\delta v_\phi$ is the relative bulk velocity in the $\phi$ direction.
 The  the head wind strength $\dot m_{\rm wind}$ can be estimated
according to the Bondi-Hoyle-Lyttleton (BHL) rate $\dot m_{\rm BHL}$
with some disk environment corrections \cite{Kocsis2011}
\be
\dot m_{\rm wind} = \dot m_{\rm BHL} \times {\rm min.}\{1, H/r_{\rm BHL}, r_{\rm Hill}/r_{\rm BHL}\}\ .
\ee
The BHL rate and the Bondi radius are well known as
\be
\frac{\dot m_{\rm BHL}}{m_{\rm bh}} = \frac{4\pi\rho m_{\rm bh}}{(v_{\rm rel}^2 +c_s^2)^{3/2}}\ ,
\ee
and $r_{\rm BHL} = m_{\rm bh}/(v_{\rm rel}^2 +c_s^2)$,
where $v_{\rm rel}$ is the relative velocity between the sBH and the local gas,
$v_{\rm rel}^2 = (\delta v_\phi + \delta v_{\rm dr})^2 + \delta v_r^2$,
with $\delta v_r$ being the relative bulk velocity in the  $r$ direction,
and $\delta v_{\rm dr}$ being the relative velocity coming from the differential rotation of the gas
\cite{Kocsis2011}:
\begin{subequations}
  \be
  \delta v_\phi = \frac{3-\gamma}{2} h c_s
  \ee
  \be
  \delta v_r = |v_{\rm gas,r}-v_{\rm bh,r}|
  = \left|-\frac{\dot M}{2\pi r\Sigma}-\frac{\dot J}{dJ/dr}\right|
  \ee
  \be
  \delta v_{\rm dr} = \frac{3}{2}\left(\frac{m_{\rm bh}}{3M}\right)^{1/3} h^{-1} c_s\ ,
  \ee
\end{subequations}
where $\gamma = d\ln\rho/d\ln r$ and $\dot J$ is the change rate of the specific angular momentum of the sBH (or equivalently the specific torque exerted on the sBH) due to sBH-disk interactions and
GW emission, i.e., $\dot J = \dot J_{\rm mig, I,II} + \dot J_{\rm wind} + \dot J_{\rm gw}$,
where
\be\label{eq:Jgw}
\dot J_{\rm gw} = -\frac{32}{5} \frac{m_{\rm bh}}{M}\left(\frac{M}{r}\right)^{7/2}
\ee
is the angular momentum loss rate due to GW emission (assuming a circular orbit),
$\dot J_{\rm wind}$ is the loss rate due to the wind interaction [Eq.~(\ref{eq:Jwind})],
and $\dot J_{\rm mig,I,II} = \dot J_{\rm mig,I}$ or $\dot J_{\rm mig,II}$ depends on
which type of migration is operating.
For sBHs with inclined orbits (so that part of their orbits are outside of the disk (od)), when they hit the accretion disk,  the relative velocity $v_{\rm rel}\sim r\Omega \gg c_s$ is usually much greater than that of the in-disk sBHs. As a result, the wind capture radius is much smaller which greatly reduces  the wind effect,  so that we simply take $\dot J_{\rm wind}^{\rm od} = 0$.

In summary,  migration timescales of in-disk sBHs and those outside are
\be
t_{\rm mig}^{\rm bh,id} = \frac{J}{|\dot J_{\rm mig,I,II} + \dot J_{\rm gw} + \dot J_{\rm wind}^{\rm id}|},\quad
t_{\rm mig}^{\rm bh,od} = \frac{J}{|\dot J_{\rm mig,I} + {\dot J_{\rm gw}}|} \ ,
\ee
where  we take $\dot J_{\rm wind}^{\rm id}=0$ in the case of type-II migration when a gap is open,
or $\dot J_{\rm wind}^{\rm id}$ as in Eq.~(\ref{eq:Jwind}) otherwise.
 Without a detailed model of sBH accretion in the disk environment,
Eq.~(\ref{eq:Jwind}) is merely an order of magnitude estimate of the wind strength.
In the following section, we will see the EMRI rate
in AGN disks is insenstive to the wind strength, and it decreases very mildly even if
we turn off the head wind completely by setting $\dot J_{\rm wind}\equiv0$ (Table~\ref{table}).

Equating the in-disk migration timescale  with the disk life time $T_{\rm disk}$
defines a critical radius within which sBH can successfully migrate into the central MBH,
i.e., the critical radius $r_{\rm crit}(T_{\rm disk})$ is defined by
\be\label{eq:rc}
\int^{r_{\rm crit}}_{r_{\rm min}} t_{\rm mig}^{\rm bh,id}(r) \ d\ln r = T_{\rm disk}\ .
\ee
In right panels of Fig.~\ref{fig:alpha}, we plot the timescales of different processes:
$t_{\rm mig,I}, t_{\rm wav}$ [Eq.~(\ref{eq:twav})],
$t_{\rm gw}:=J/|\dot J_{\rm gw}|$ [Eq.~(\ref{eq:Jgw})], and $t_{\rm wind}:=J/|\dot J_{\rm wind}|$ [Eq.~(\ref{eq:Jwind})], from which we can roughly read the critical radius $r_{\rm crit}(T_{\rm disk})$.

Different from sBHs, the gas accretion onto stars is more complicated considering that both the radiation heating and solar
wind are supposed to alter the local gas environment, by heating up and blowing away surrounding gas. In the presence of an strong isotropic outflow from stars, the star-gas interaction could be completely different from the classical Bondi accretion \cite{Gruzinov2020,Li2020}. Here we simply take $\dot J_{\rm wind}^{\rm star} = 0$ for both in-disk stars and stars outside, i.e.,
\be
  t_{\rm mig}^{\rm star, id} \simeq t_{\rm mig}^{\rm star, od}
   =\frac{m_{\rm bh}}{m_{\rm star}} t_{\rm mig}^{\rm bh, od} \ .
\ee

In this paper, we do not include possible contribution from dynamic friction as sBHs moving through the gas disk,
which is negligible due to subsonic sBH-gas relative motion for sBHs orbiting in the disk \cite{Kocsis2011}.
However the relative gas velocity at different radii ($r_{\rm sBH}\pm H$) is supersonic, and the dynamic friction
with respect to gas in this region may be important. In this case, the supersonic relative velocity
is mostly due to the shear (differential rotation) of the accretion flow instead of the local pressure gradient,
and the standard dynamic friction is automatically incorporated in the migration torque \cite{Ostriker1999}.
For inclined sBHs, dynamic friction is still weaker than the effect of density waves
: consider a sBH on an inclined orbit, penetrating the gas disk with relative velocity
$v_{\rm rel}\sim r\Omega$, the dynamic friction (per unit mass) on the sBH is $f_{\rm df}\sim G^2 m_{\rm bh} \rho/v^2_{\rm rel}$ \cite{Chandrasekhar1943,Ostriker1999},
and the timescale for the dynamic friction to change the sBH's orbit is
$t_{\rm df}\sim (r\Omega)^2/(f_{\rm df} H\Omega)$, which is $\sim t_{\rm mig,I}h^{-2} \gg  t_{\rm mig,I}$.
Another possible contribution which we do not include here is the heating torque  \cite{Masset2017} arising from the
asymmetric distribution of low-density gas around the sBH due to the accretion heating
and the shear of disk flow. As estimated in \cite{Hankla2020}, the heating torque might be comparable with the type-I migration torque assuming a thermal feedback of Eddington luminosity, but neither gravity
nor dynamical feedback from the sBH. In fact, the local gas distribution is sensitive to both the sBH gravity
and the gas outflow which carries away the angular momentum of the sBH as explained in the beginning of the section.
A simple estimate shows that the disk flow shear timescale $1/\Delta\Omega \sim 1/(\lambda_c d\Omega/dr)> 1/(\Omega h)$ for the characteristic size $\lambda_c < H$ of the thermal feedback is much longer than the dynamical timescale $ \lambda_c/c_s < 1/\Omega$ on the same length scale.
Therefore the local gas distribution should be more sensitive to the dynamical processes of gas inflow and outflow, and the heating torque should be much weaker than the estimate assuming neither sBH gravity nor dynamical feedback.

\subsection{Migration traps}

In middel panels of Fig.~\ref{fig:alpha}, we plot the torques exerted on the sBH $\dot J_{\rm gw},
\dot J_{\rm mig,I}$ and $\dot J_{\rm wind}$ for $\alpha$ and $\beta$ type of disks, respectively.
For the $\alpha$ disk, we find that  the torque $J_{\rm mig,I}$ changes its sign around the local density maxima $r\sim 100 M_\bullet$, which is known as the migration trap \cite{Lyra2010,Bellovary2016}.  However, the migration
trap is not present for  the $\beta$-disk, simply because  there is no sign change in $\dot J_{\rm mig,I}$ as the surface density decreases monotonically with $r$. In addition,  although there is a sign change
in $\dot J_{\rm mig,I}$ for the $\alpha$-disk, $\dot J_{\rm wind}$ and $\dot J_{\rm gw}$ dominate in the region where
$\dot J_{\rm mig,I}$ is positive. As a result, the combined torque never changes sign and there is no migration trap in the $\alpha$-disk either.

To compare with previous studies about migration traps \cite[e.g.,][]{Bellovary2016}, we also calculate the disk
structure of an comparison $\alpha$-disk with $M_\bullet=4\times 10^8 M_\odot$, $\dot M_\bullet =0.5 M^{\rm Edd}$
and $\alpha=0.1$, and we show all the disk variables, torques and timescales in Fig.~\ref{fig:alpha2}. From the middle
panel, we see two special radii ($\sim 10 M_\bullet$ and $\sim 10^3M_\bullet$) where $\dot J_{\rm mig,I}$ changes its sign from negative to positive in the decreasing $r$ direction. These two radii are called migration traps by previous studies,
and there have been extensive studies on the consquences of migration traps in AGN disks accumulating compact objects \cite{McKernan2012,McKernan2014,Stone2017,Bartos2017,McKernan2018,Yang2019prl,Yang2019,
Secunda2019,Secunda2020}. As shown in Fig.~\ref{fig:alpha2}, the migration traps are supposed to be
overcome by two counteracting processes: GW emission and wind. We have explored the parameter space $\alpha\in(0.01,0.5)$,
$\dot M_\bullet\in (0.01, 0.5) \dot M_{\bullet}^{\rm Edd}$, $M_\bullet\in(10^5,10^9) M_\odot$, where no migration
trap is found in either $\alpha$-disks or $\beta$-disks.

In the example TQM disk model of Ref.~\cite{Thompson2005}, a salient feature is the presence of a opacity gap
and  consequently a sharp density increase on its inner edge, where the type-I migration torque changes sign
according to Eq.~(\ref{eq:JI}) and has been interpreted as a possible location of migration trap \cite[e.g.,][]{Bellovary2016}. As mentioned in Section~\ref{subsec:disk}, the sharp density increase on the edge of the opacity gap is in fact
resulted by the improper equation of radiation pressure. With a more general  equation of radiation pressure,
we find the density increase is much milder and there is no sign change in the type-I migration torque. We
also explored the parameter space $X\in(0.01,0.1)$, $\dot M_\bullet\in (0.01, 0.5) \dot M_{\bullet}^{\rm Edd}$, $M_\bullet\in(10^5,10^9) M_\odot$, where no migration trap is found in TQM disks.

To summarize, we find no migration trap in the three AGN disk models in a large parameter space we considered.
In $\alpha$-disks, there are locations where the type-I migration torque changes sign,
but the total torque is always negative because of the negative torque from head wind and GW emission.
For EMRI hosts with $M_\bullet < 10^7 M_\odot$, we find no migration trap in their accretion disks even if
 there was no head wind contribution, i.e., $\dot J_{\rm wind}=0$.
In $\beta$-disks, there is no sign change in the type-I migration torque because of the monotonical density and temperature profiles. In TQM disks, there is no sign change in the type-I migration torque either as explained above
\footnote{In fact, Dittmann and Miller \cite{Dittmann2020} also noted that the migration traps in TQM disks no longer stand if a more updated opacity is used in solving the disk structure.}.

In previous studies of hierarchical BBH mergers in migration traps of AGN disks,
the existence of migration traps was established on a fiducial $\alpha$-disk model in
Ref.~\cite{Sirko2003} and a fiducial TQM disk model in Ref.~\cite{Thompson2005}.
As shown above, the migration traps no longer stand after taking account
of the head wind  and/or using a more reasonable equation of radiation pressure.
Therefore, the analysis here raises concerns about the feasibility of hierarchical BBH formation channel in
migration traps of AGN disks (see \cite[e.g.,][]{Leigh2018,McKernan2020b,Tagawa2020} for the impact of migration traps to general BBH mergers in AGN disks).

According to the three disk models considered in this work, we do not expect any migration trap,
but it does not exclude the possibility that these disk models are not good approximations to
the AGN accretion disks in nature. If a migration trap indeed exists in an AGN disk,
sBHs would be trapped and hierarchical BBH mergers would consequently happen
until the remnant BH is so massive that it opens a gap and tears down the trap.
The critical BH mass for gap opening is sensitive to the trap location  and the local disk structure
\cite{Kocsis2011}.  For example, in the scenario considered in \cite{Graham:2020gwr} to associate possible
AGN flares with BBH merger in the disk, the trap is assumed to be $\sim 700 M_\bullet$ away from the MBH with mass $M_\bullet\sim 10^8 M_\odot$ and the the critical BH mass is $\sim O(10^2) M_\odot$ assuming an $\alpha$-disk with $\alpha=0.1$ and accretion rate $\dot M_\bullet = 0.1 \dot M_\bullet^{\rm Edd}$ . This means for typical $10 M_\odot$  sBHs, $\mathcal{O}(10)$-times mergers are expected before a gap opens up and the final BH starts its type II migration.

\section{EMRI formation assisted by AGN accretion disks}\label{sec:emri_disk}

\subsection{Fokker-Planck equation}
As shown in the right panels of Figs.~\ref{fig:alpha}, \ref{fig:TQM}, and in Eq.~(\ref{eq:trlx}),
the timescale of orbit eccentricity/inclination decay  $t_{\rm wav}$,
 as driven by density wave geneartion,  is much shorter than the sBH migration timescale
$t_{\rm mig}$ and the stellar cluster relaxation timescale $t_{\rm rlx}$.
As a result,  one may naively expect that all stars/sBHs are captured onto the disk on the shortest timescale $t_{\rm wav}$, which turns out to be incorrect.
In fact, due to the dense distribution of scatters (dominated by stars) in the disk, a large portion
of  stars/sBHs captured by the disk will be scattered back into the cluster. As demonstrated in previous
studies of star-disk interactions  \cite{Vilkoviskij2002,Kennedy2016,Panamarev2018}, a local equilibrium
is built between the net rate of stars captured onto the disk and the rate of inward migration within the disk.
Assuming a net fraction $\mu_{\rm star}$ of in-cluster stars are captured on to the disk and migrate inward
in a timescale $t_{\rm mig}^{\rm star, id}$, the in-cluster star loss rate due to disk capture can be formulated
as
\be
\left(\frac{\partial f_{\rm star}}{\partial t}\right)_{\rm cap}=-\mu_{\rm star} \frac{f_{\rm star}}{t_{\rm mig}^{\rm star, id} } \ .
\ee
From another aspect, the loss rate should be proportional to the ratio of two timescales,
the inclination damping timescale $t_{\rm wav}^{\rm star}$ and the timescale in which a star is scattered
by in-disk scatters (mainly stars), where the latter is inversely proportional to the local density
of in-disk scatters $n_{\rm star,id}$, i.e., $\left(\frac{\partial f_{\rm star}}{\partial t}\right)_{\rm cap} \propto 1/(t_{\rm wav}^{\rm star}n_{\rm star,id})$.
In the same way, the in-cluster sBH loss rate due to disk capture depends on the
inclination damping timescale $t_{\rm wav}^{\rm bh}$ and the local density of
in-disk scatters (mainly stars) $n_{\rm star, id}$. As a result, we obtain
\be
\frac{1}{ f_{\rm bh}}\left(\frac{\partial f_{\rm bh}}{\partial t}\right)_{\rm cap}
=\frac{t_{\rm wav}^{\rm star}}{t_{\rm wav}^{\rm bh}} \frac{1}{ f_{\rm star}}\left(\frac{\partial f_{\rm star}}{\partial t}\right)_{\rm cap}
=  -\mu_{\rm star} \frac{m_{\rm bh}}{m_{\rm star}}\frac{1}{t_{\rm mig}^{\rm star, id} }\ ,
\ee
The net fraction $\mu_{\rm star}$ should fall in the range  $\sim(h,1)$, with $h$ being the disk aspect ratio.
The exact value of $\mu_{\rm star}$ depends on the detailed balance between the rate of stars captured to the disk,
the fraction of which scattered away from the disk, and the inward migration rate of in-disk scatters,
which require seperate numerical studies to determine.

Due to the interactions with the accretion disk,
stars and sBHs settle as two components: a cluster component and a disk component.
The evolution of disk-component sBHs is relatively simple: their orbits tend to circularize on
the eccentricity damping timescale $t_{\rm wav}^{\rm bh}$, which is much shorter than the migration timescale.
Therefore, the orbital eccentricities of sBHs in the disk have been damped to essentially zero
long before they migrate to the vicinity of the MBH. This is in stark constrast with EMRI formation via
the loss-cone mechanism \cite{Babak2017}.
On the other hand, we expect the distribution functions of cluster-component
stars and sBHs acquire some dependence on the orbital inclination as interacting with the disk.
For convenience, we choose to integrate out the inclination and work with the
inclination-integrated distribution functions $f_i(t, E, R)$ of the
cluster-component stars and sBHs. The orbit-averaged Fokker-Planck equation (\ref{eq:FP}) for
cluster-component sBHs/stars is modified as
\be\label{eq:FP_disk}
  \mathcal C\frac{\partial f}{\partial t}
  = - \frac{\partial}{\partial E} F_E
  - \frac{\partial}{\partial R} F_R + S\ ,
\ee
where $F_E, F_R$ are defined in Eq.~(\ref{eq:flux}), with the advection coefficients modified as
\be\label{eq:coef}
\begin{aligned}
  \mathcal  D_{E,\rm bh} &\rightarrow \mathcal D_{E,\rm bh}  - \mathcal C\frac{E}{t_{\rm mig}^{\rm bh, od}}\ , \\
  \mathcal D_{R,\rm bh}  &\rightarrow \mathcal D_{R,\rm bh}  - \mathcal C\frac{1-R}{t_{\rm wav}^{\rm bh, od}}\ , \\
  \mathcal D_{E,{\rm star}} &\rightarrow \mathcal D_{E,{\rm star}}  -\mathcal C\frac{E}{t_{\rm mig}^{\rm star, od}}\ ,\\
  \mathcal D_{R,{\rm star}} &\rightarrow  \mathcal D_{R,{\rm star}} - \mathcal C\frac{1-R}{t_{\rm wav}^{\rm star, od}}\ ,
\end{aligned}
\ee
where the corrections are due to interactions with the accretion disk and GW emission.
The (negative) source term arises from stars/sBHs capture onto the accretion disk,
\be
S_{\rm bh} = -\mu_{\rm cap}\mathcal C \frac{f_{\rm bh}}{t_{\rm mig}^{\rm star, id}}\ ,\quad
S_{\rm star} = -\mu_{\rm cap}\frac{m_{\rm star}}{m_{\rm bh}}\mathcal C  \frac{f_{\rm star}}{t_{\rm mig}^{\rm star, id}}\ .
\ee
where we have defined $\mu_{\rm cap}=\mu_{\rm star}m_{\rm bh}/m_{\rm star}$, which we expect to be in the range of
$\sim(h,1)m_{\rm bh}/m_{\rm star}$. In this paper, we treat $\mu_{\rm cap}$ as a free parameter and take two representative numbers $\mu_{\rm cap}=1$ (fast disk capture) and $\mu_{\rm cap}=0.1$ (slow disk capture) as working examples.

\begin{figure}
\includegraphics[scale=0.8]{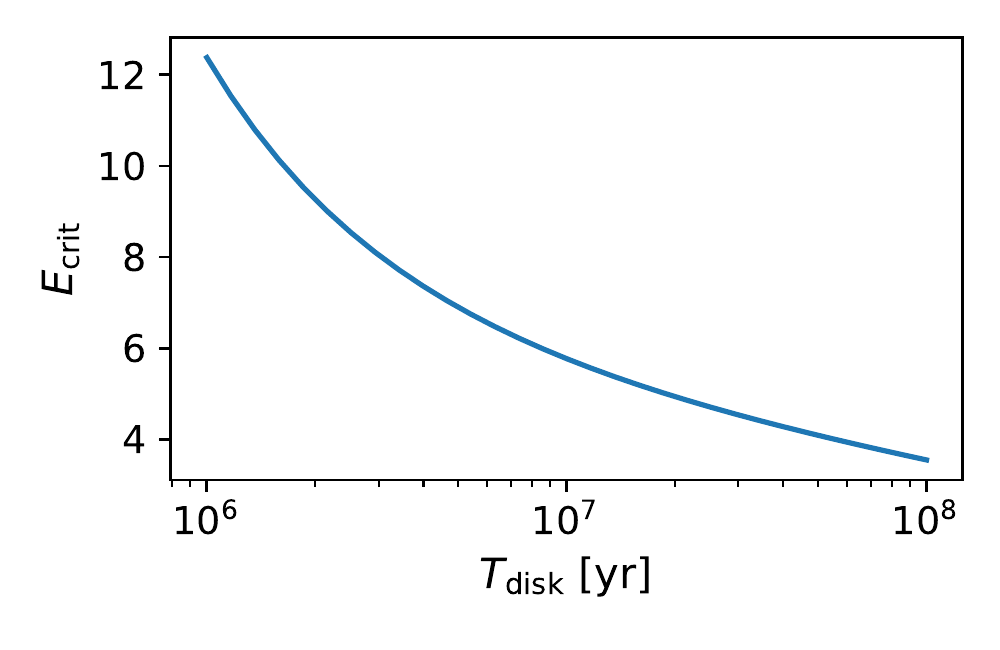}
\caption{\label{fig:Ecrit} The dependence of $E_{\rm crit}\ [\sigma^2]$ on the disk life time $T_{\rm disk}\ [\rm yr]$
for the fiducial $\alpha$-disk shown in Fig.~\ref{fig:alpha}.
}
\end{figure}

Given a disk lifetime $T_{\rm disk}$, only sBHs within some critical radius $r_{\rm crit}(T_{\rm disk})$ have enough time to migrate within the disk to reach the central MBH within $T_{\rm disk}$. Therefore, the EMRI rate is formulated as
\be\label{eq:rcrit}
\Gamma_{\rm emri}(t;T_{\rm disk}) \simeq \int\int_{E>E_{\rm crit}}  -S_{\rm bh}(E,R) dEdR\ ,
\ee
where $E_{\rm crit}(T_{\rm disk}) := \phi(r_{\rm crit}(T_{\rm disk}))/2$ (see Fig.~\ref{fig:Ecrit}).

To compute the EMRI rate per MBH in the presence of an accretion disk,
we evolve the distributions $f_i(t,E,R)$ using Eq.~(\ref{eq:FP_disk}) for a period of time $T_{\rm disk}$.
On the low energy $E\rightarrow0$ (far away from the MBH) boundary, both the sBH/star-disk interactions and
two body scatterings are slow, therefore
\be
f_i(t,E,R)|_{E\rightarrow 0} = f_i(t=0,E,R)|_{E\rightarrow 0}\ .
\ee
On the $R=1$ boundary, the vanishing flux $F_R$ for both sBHs and stars still applies
\be
F_R|_{R\rightarrow1} = 0\ .
\ee
On the $R=R_{\rm lc}$ boundary, we again impose the vanishing flux condition
\be
F_R|_{R= R_{\rm lc}} = 0\ .
\ee
This  boundary condition is different from the one imposed in the no-disk case,
simply because the fast eccentricity damping by density waves drives stars/sBHs away from the loss cone.

\begin{figure*}
\includegraphics[scale=0.6]{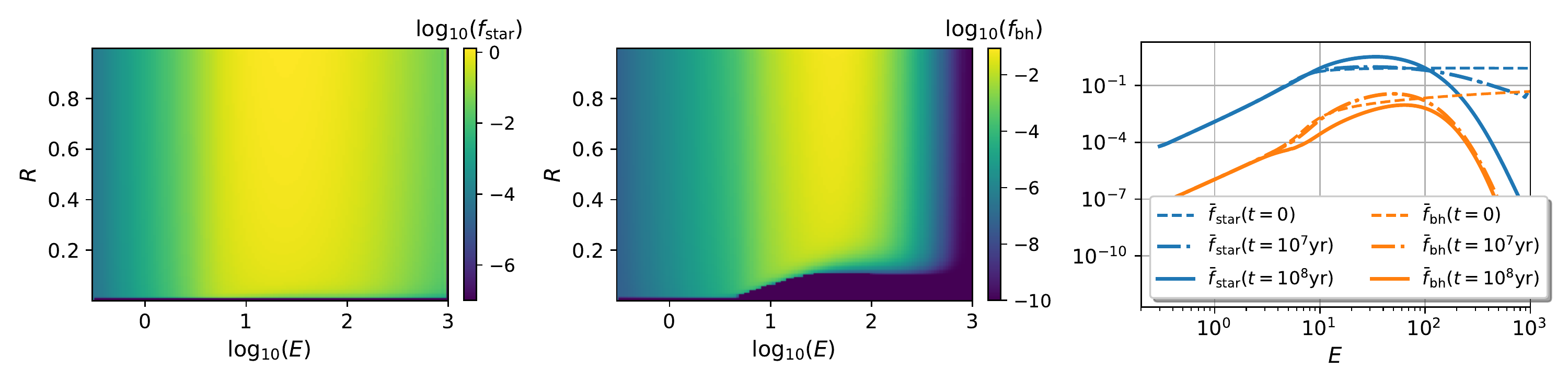}
\includegraphics[scale=0.6]{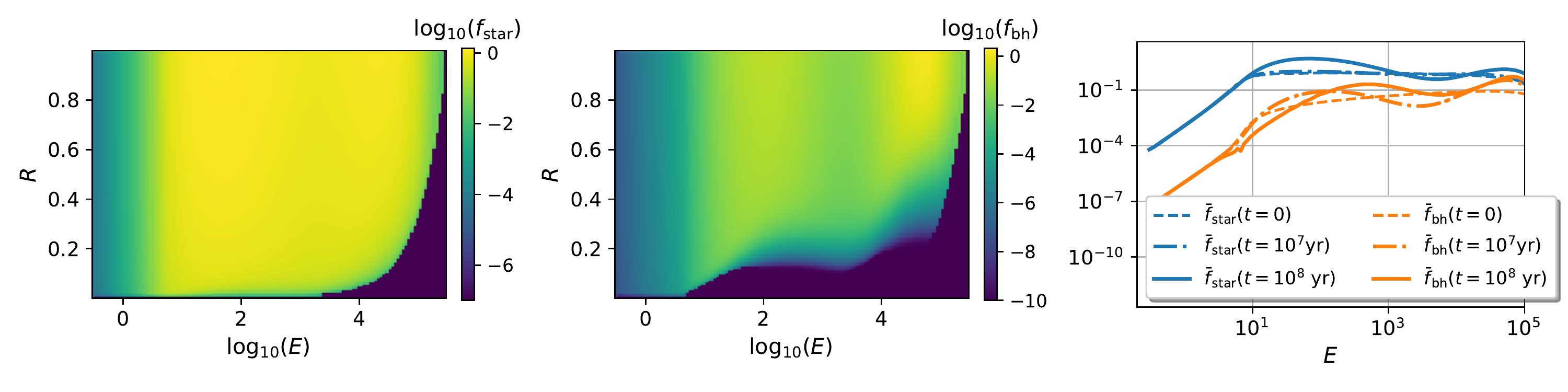}
\caption{\label{fig:f2d_disk} We show the distribution functions for the case of fast disk capture
($\mu_{\rm cap}=1$) in the upper row:$f_{\rm star}(t=10^7 {\rm yr}, E,R)$ (left panel), $f_{\rm bh}(t=10^7 {\rm yr}, E,R)$ (middle panel),
 the time dependence of EMRI rate $\Gamma_{\rm emri}$,
 where $f_i$ are shown in units of $10^5 {\rm pc}^{-3}/(2\pi\sigma^2)^{3/2}$ and
 $E$ is shown in units of $\sigma^2$. The counterparts of the slow disk capture ($\mu_{\rm cap}=0.1$) case are in the lower row.
}
\end{figure*}

\subsection{Numerical method}\label{subsec:num}
The numerical method for solving Eq.~(\ref{eq:FP_disk}) is the same as the one
used for solving Eq.~(\ref{eq:FP}) in Section~\ref{sec:review}.
The only extra numerical subtlety is due to the scale separation: the timescales of
migration and orbit eccentricity decay
$t_{\rm mig},t_{\rm wav}$ are smaller than the cluster relaxation timescale $t_{\rm rlx}$. As a result,
the advection coefficient will be much larger than the diffusion coefficient:
$|\mathcal D_E/E| \gg \mathcal D_{EE}/E^2$ and $|\mathcal D_R| \gg \mathcal D_{RR}$.
To avoid numerical difficulties for resolving the large scale separations,
we choose to regularize the advection coefficients in Eq.~(\ref{eq:coef}) as follows:
\be\label{eq:regularize}
\begin{aligned}
  \mathcal  D_{E,\rm bh} &\rightarrow \mathcal D_{E,\rm bh}
  - \mathcal C\frac{E}{ t_{\rm mig}^{\rm bh, od}+\epsilon T_0}\ , \\
  \mathcal D_{R,\rm bh}  &\rightarrow \mathcal D_{R,\rm bh}  - \eta (1-R)\mathcal D_0\ , \\
  \mathcal D_{E,{\rm star}} &\rightarrow \mathcal D_{E,{\rm star}}
   -\mathcal C\frac{E}{t_{\rm mig}^{\rm bh, od}+\epsilon T_0}\frac{m_{\rm star}}{m_{\rm bh}} \ ,\\
  \mathcal D_{R,{\rm star}} &\rightarrow  \mathcal D_{R,{\rm star}}
  - \eta(1-R)\mathcal D_0\frac{m_{\rm star}}{m_{\rm bh}} \ ,
\end{aligned}
\ee
where $\epsilon$ is a small number ensuring a numerically resolvable scale separation, $T_0=10$ Gyr,
$\eta$ is a large number ensuring a large scale separation between
the regularized advection coefficient $\mathcal D_R$ and the diffusion coefficient $\mathcal D_{RR}$,
and $\mathcal D_0(E)$ is defined as the maximal value of $\mathcal D_{RR,{\rm bh}}(E,R)$ for given $E$.

We choose $\epsilon=10^{-4}$ and $\eta=10^2$ as default regularization parameters. In  Appendix~\ref{apb},
we will show that different choices of these two regularization parameters do not affect the EMRI rate
assisted by accretion disks as long as $\epsilon$ is sufficiently small.

\subsection{Comparison with loss-cone rate}\label{sec:clc}
In order to compare with the canonical EMRI rate associated with the loss cone mechanism, we consider a fiducial model
assuming the same MBH+sBH/star cluster model and parameters as those considered in Section \ref{sec:review}
and  shown in Fig.~\ref{fig:alpha}. We take the final state of Eq.~(\ref{eq:FP}) (Fig.~\ref{fig:lc}) as the initial condition of Eq.~(\ref{eq:FP_disk}), and evolve the equation for a period of time $T_{\rm disk}$.

\begin{figure}
\includegraphics[scale=0.7]{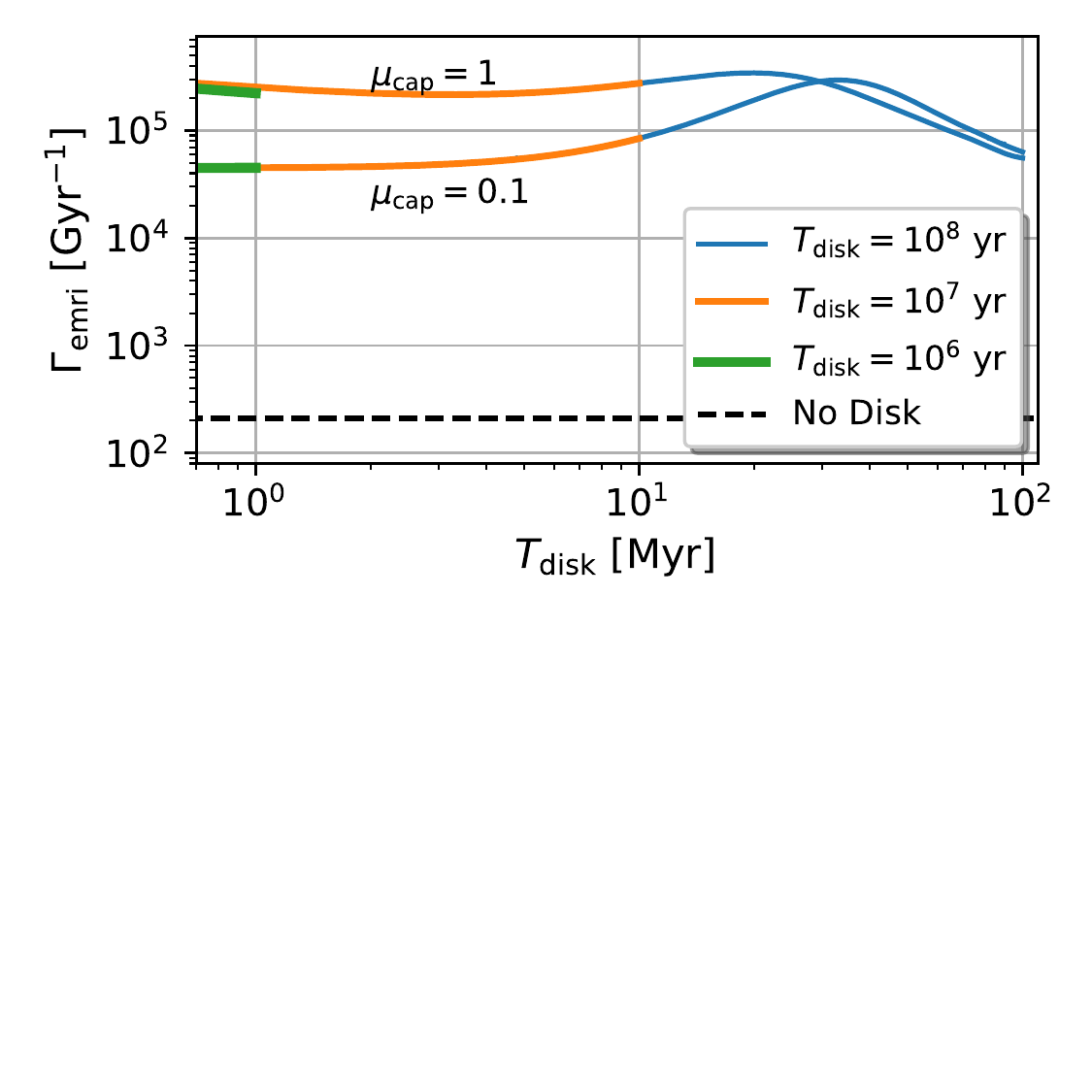}
\caption{\label{fig:Gamma_disk} The time dependence of disk assisted EMRI rate per MBH $\Gamma_{\rm emri}(t;T_{\rm disk})$
for different disk life times $T_{\rm disk}$ and different disk capture parameters $\mu_{\rm cap}$. The dashed line
is the average EMRI rate via loss cone of the fiducial model.}
\end{figure}

\begin{figure}
\includegraphics[scale=0.6]{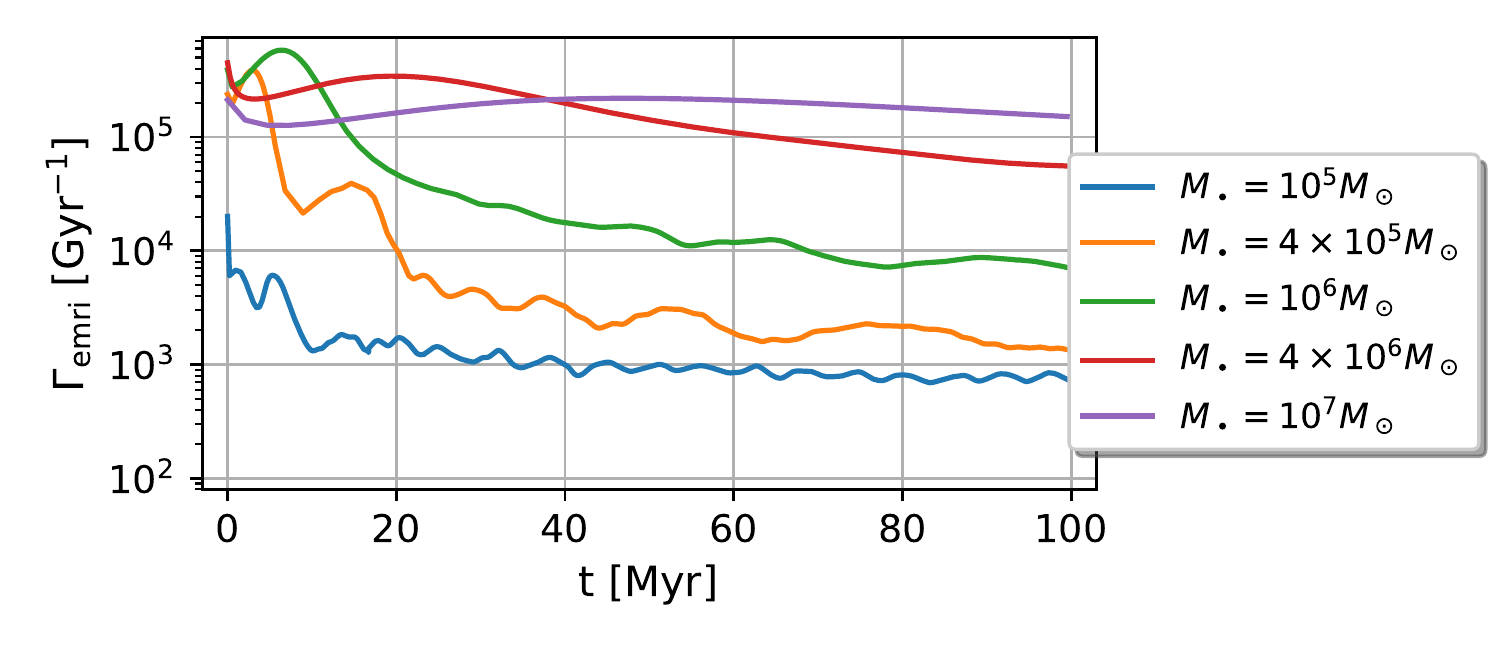}
\caption{\label{fig:M_depend} The disk-assisted EMRI rate $\Gamma_{\rm emri}(t; T_{\rm disk}=10^8\ {\rm yrs})$
for different MBH masses $M_\bullet$, where take $\mu_{\rm cap}=1$.}
\end{figure}

In Fig.~\ref{fig:f2d_disk}, we show the sBH/star distribution functions $f_i(t,E,R)$. In combination with
the initial condition (Fig.~\ref{fig:lc}), we see that sBHs are migrating toward larger $R$ and larger $E$ driven by the density waves.
In Fig.~\ref{fig:Gamma_disk}, we show the time dependence of the disk assisted EMRI rates for two different cases: $\mu_{\rm cap} =1$ or $\mu_{\rm cap} =0.1$.
For both cases, we find that the dependence of $\Gamma_{\rm emri}(t;T_{\rm disk})$ on $T_{\rm disk}$ is relatively weak
(the $\Gamma_{\rm emri}(t;T_{\rm disk})$ curves for different $T_{\rm disk}$ roughly overlap with each other).
This is because of the weak dependence of $E_{\rm crit}$ on $T_{\rm disk}$ (Fig.~\ref{fig:Ecrit}).
For short time $t$, $\Gamma_{\rm emri}$ mainly depends on the initial distributions $f_i(t=0,E,R)$ and is proportional to the parameter $\mu_{\rm cap}$. For long time $t$, we expect
an equilibrium between sBHs migrating inward from $E<E_{\rm crit}$ and sBHs captured
onto the disk in the region $E>E_{\rm crit}$, where the EMRI rate is determined by the sBH supply rate
irrespective of $\mu_{\rm cap}$.
For the case of $\mu_{\rm cap}=1$, we find that the initial EMRI rate
per MBH is $\Gamma_{\rm emri}(t=0;T_{\rm disk})\sim 4\times 10^5\ {\rm Gyr}^{-1}$ (irrespective of disk life time);
after a rapid initial settling (see Fig.~\ref{fig:M_depend}), $\Gamma_{\rm emri}$
then ($t\lesssim 20$ Myr) slowly increases as more sBHs migrate inward from $E<E_{\rm crit}$
than those captured onto the disk; after $t\sim 20$ Myr,  the migration supply and the capture consumption
has reached an equilibrium, where $\Gamma_{\rm emri}$ slowly decreases
due to the decreasing supply arising from the decreasing number of sBHs around $E\sim E_{\rm crit}$.
For $\mu_{\rm cap}=0.1$, the evolution track is similar except starting with a lower initial EMRI rate.

In both examples we see that the disk-assisted rate per MBH is much higher ($\mathcal{O}(10^2 - 10^3)$) than the loss-cone rate, which is due to much more efficient capture and transport mechanisms by the disk. It turns out that this observation is generally true as we vary the parameters for MBH, disk and star cluster. In the following subsection we explore the dependence of disk-assisted EMRI rate on various model parameters.

\subsection{ EMRI rate for different models}
In this subsection, we investigate the dependence of the disk-assisted EMRI rate on different parameters:
the MBH mass $M_\bullet$, the accretion rate $\dot M_\bullet$ and the $\alpha$ parameter of $\alpha$-disk models.
We also explore the EMRI rate for the TQM disk model and different cluster initial condition.
We take the fiducial model $M_\bullet=4\times10^6M_\odot, \dot M_\bullet=0.1 M^{\rm Edd}_\bullet, \alpha=0.1, \mu_{\rm cap}=1$ as reference.

For each MBH with mass $M_\bullet$, we initialize the distributions of surrounding
stars and sBHs according to the Tremaine's cluster model outlined in Section~\ref{subsub:ini},
and evolve the distributions using the Fokker-Planck equation (\ref{eq:FP}) to $t=5$ Gyr.
We then take this finial-state distributions as the initial condition of the disk-modified
Fokker-Planck equation (\ref{eq:FP_disk}), and evolve them to $t=10^8$ yr. In Fig.~\ref{fig:M_depend},
we show the EMRI rate $\Gamma_{\rm emri}(t; T_{\rm disk}=10^8\ {\rm yrs})$ for each $M_\bullet$
(the results of different $T_{\rm disk}$ and $\mu_{\rm cap}$ can be easily inferred from reading Fig.~\ref{fig:Gamma_disk}).
The initial EMRI rates are $\sim \mathcal O(10^5)\ {\rm Gyr}^{-1}$, except for the low mass $M_\bullet=10^5M_\odot$ case,
for which the high EMRI rate via loss cone  has consumed sBHs close to the central MBH.
Their subsequent evolution basically follow the description given in the Sec.~\ref{sec:clc}: an increasing phase where the supply from inward migration dominates and then a decreasing phase where the supply-consumption equilibrium is built. The relevant timescale is longer for larger $M_\bullet$. As a result,
$\Gamma_{\rm emri}$ is roughly a constant in $10^8$ yrs for $M_\bullet=10^7 M_\odot$, while $\Gamma_{\rm emri}$
changes by orders of magnitude for low MBH massses.
In constrast with the loss-cone EMRI channel, the average disk-assisted
EMRI rate (for long disk life time) $\Gamma_{\rm emri}$
increases with the MBH mass $M_\bullet$, because  the capacity of sBH reservior ($\propto M_\bullet$) is larger
for more massive MBHs.

\begin{table*}
\caption{\label{table}Average EMRI rate per MBH $\braket{\Gamma_{\rm emri}(T_{\rm disk})}\ [{\rm Gyr}^{-1}]$ for
different models. In the 1st column are the parameters of initial stellar cluster profiles,
in the 2nd/3rd columns are the MBH mass $M_\bullet$ and the parameter $\mu_{\rm cap}$,
in the 4th column is the $\alpha$ parameter in (default) $\alpha$-disks or $X$ parameter in TQM disks,
in the 5th column is the MBH accretion rate and in the 6th column is the wind state (on or off).
In a few models with TQM disks,
the EMRI rates are nearly zero for short disk  lifetime $T_{\rm disk}$ because the migration timescale in inner parts of the disk is longer than  $T_{\rm disk}$, consequently almost no sBH successfully migrates into the MBH within  $T_{\rm disk}$. }
\begin{tabular}{c c c c c c | c   c  c }
  $(\gamma, \delta)$ & $M_\bullet/M_\odot$ & $\mu_{\rm cap}$ &
  $\alpha\ {\rm or}\ X$ & $\dot M_\bullet/\dot M_\bullet^{\rm Edd}$ & wind &
  $\braket{\Gamma_{\rm emri}(T_{\rm disk}=10^6 {\rm yrs})}$ &
  $\braket{\Gamma_{\rm emri}(T_{\rm disk}=10^7 {\rm yrs})}$ &
  $\braket{\Gamma_{\rm emri}(T_{\rm disk}=10^8 {\rm yrs})}$ \\
  \hline
  \multirow{24}{*}{$(1.5, 0.001)$}
  & $1\times10^7$ & \multirow{5}{*}{$1$} & \multirow{5}{*}{$10^{-1}$} & \multirow{5}{*}{$10^{-1}$} &\multirow{5}{*}{on}
                          & $1.6\times10^5$ & $1.3\times10^5$ & $1.8\times10^5$ \\
  & $4\times10^6$ &  &  &  & & $2.9\times10^5$ & $2.4\times10^5$ & $1.7\times10^5$ \\
  & $1\times10^6$ &  &  &  & & $2.9\times10^5$ & $4.5\times10^5$ & $7.0\times10^4$ \\
  & $4\times10^5$ &  &  &  & & $2.1\times10^5$ & $1.7\times10^5$ & $2.2\times10^4$ \\
  & $1\times10^5$ &  &  &  & & $5.7\times10^3$ & $4.2\times10^3$ & $1.3\times10^3$ \\ \cline{2-9}
  & $1\times10^7$ & \multirow{5}{*}{$10^{-1}$} & \multirow{5}{*}{$10^{-1}$} & \multirow{5}{*}{$10^{-1}$}& \multirow{5}{*}{on}
                          & $1.9\times10^4$ & $2.5\times10^4$ & $1.3\times10^5$\\
  & $4\times10^6$ &  & & & & $4.2\times10^4$ & $5.8\times10^4$ & $1.5\times10^5$ \\
  & $1\times10^6$ &  & & & & $4.3\times10^4$ & $1.9\times10^5$ & $9.3\times10^4$ \\
  & $4\times10^5$ &  & & & & $3.2\times10^4$ & $1.4\times10^5$ & $2.2\times10^4$ \\
  & $1\times10^5$ &  & & & & $2.2\times10^3$ & $5.7\times10^3$ & $1.6\times10^3$\\  \cline{2-9}
  & \multirow{4}{*}{$4\times10^6$} & \multirow{4}{*}{$1$}  &  $10^{-1}$  &$5\times10^{-1}$ & \multirow{4}{*}{on}
                                   & $2.8\times10^5$ & $2.2\times10^5$ & $1.5\times10^5$\\
  &   &   &  $10^{-1}$ & $10^{-2}$ & & $3.4\times10^5$ & $3.4\times10^5$ & $2.5\times10^5$\\
  &   &   &  $3\times10^{-1}$ & $10^{-2}$ & & $3.3\times10^5$ & $2.0\times10^5$ & $2.1\times10^5$  \\
  &   &   &  $10^{-2}$ & $10^{-1}$ & & $2.7\times10^5$ & $1.4\times10^5$ & $1.1\times10^5$ \\ \cline{2-9}
  & $1\times10^7$ & \multirow{5}{*}{$1$} & \multirow{5}{*}{$(10^{-1})_{\rm TQM}$} &\multirow{5}{*}{$10^{-1}$}& \multirow{5}{*}{on}
                         & $\sim 0$ & $\sim 0$ & $2.7\times10^5$\\
  & $4\times10^6$ &  & & & & $\sim 0$ & $1.6\times10^4$ & $2.6\times10^5$ \\
  & $1\times10^6$ &  & & & & $\sim 0$ & $2.5\times10^5$ & $1.7\times10^5$\\
  & $4\times10^5$ &  & & & & $3.4\times 10^2$ & $1.3\times10^5$ & $7.1\times10^4$\\
  & $1\times10^5$ &  & & & & $1.0\times 10^2$ & $1.4\times10^4$ & $5.0\times10^3$\\\cline{2-9}
  & $1\times10^7$ & \multirow{5}{*}{$1$} & \multirow{5}{*}{$10^{-1}$} & \multirow{5}{*}{$10^{-1}$} & \multirow{5}{*}{off}
                          & $1.2\times10^5$ & $1.1\times10^5$ & $1.8\times10^5$ \\
  & $4\times10^6$ &  &  &  & & $2.7\times10^5$ & $2.2\times10^5$ & $1.7\times10^5$ \\
  & $1\times10^6$ &  &  &  & & $2.3\times10^5$ & $4.5\times10^5$ & $6.9\times10^4$ \\
  & $4\times10^5$ &  &  &  & & $1.9\times10^5$ & $1.7\times10^5$ & $2.2\times10^4$ \\
  & $1\times10^5$ &  &  &  & & $5.0\times10^3$ & $4.0\times10^3$ & $1.3\times10^3$ \\
  \hline
  \multirow{5}{*}{$(1.5, 0.002)$}
  & $1\times10^7$  & \multirow{5}{*}{$1$} & \multirow{5}{*}{$10^{-1}$} & \multirow{5}{*}{$10^{-1}$}&\multirow{5}{*}{on}
  & $2.8\times10^5$ & $2.4\times10^5$ & $3.6\times10^5$  \\
  & $4\times10^6$  & & & & & $4.4\times10^5$ & $4.1\times10^5$ & $3.6\times10^5$ \\
  & $1\times10^6$  & & & & & $4.1\times10^5$ & $8.7\times10^5$ & $1.4\times10^5$ \\
  & $4\times10^5$  & & & & & $3.5\times10^5$ & $4.5\times10^5$ & $5.4\times10^4$ \\
  & $1\times10^5$  & & & & & $7.7\times10^4$ & $2.4\times10^4$ & $4.2\times10^3$ \\
  \hline
  \multirow{5}{*}{$(1.8, 0.001)$}
  & $1\times10^7$  & \multirow{5}{*}{$1$} & \multirow{5}{*}{$10^{-1}$} & \multirow{5}{*}{$10^{-1}$}&\multirow{5}{*}{on}
  & $4.2\times10^5$ & $2.8\times10^5$ & $2.4\times10^5$ \\
  & $4\times10^6$  & & & & & $4.2\times10^5$ & $3.0\times10^5$ & $1.5\times10^5$ \\
  & $1\times10^6$  & & & & & $2.1\times10^5$ & $1.9\times10^5$ & $3.3\times10^4$ \\
  & $4\times10^5$  & & & & & $7.2\times10^4$ & $3.8\times10^4$ & $6.2\times10^3$ \\
  & $1\times10^5$  & & & & & $6.4\times10^3$ & $1.9\times10^3$ & $4.6\times10^2$ \\
  \hline
\end{tabular}
\end{table*}

We conclude this section by collecting the average disk-assisted EMRI rate
\be\label{eq:avg}
 \braket{\Gamma_{\rm emri}(T_{\rm disk})}
:= \frac{1}{T_{\rm disk}}\int_0^{T_{\rm disk}} \Gamma_{\rm emri}(t;T_{\rm disk}) dt
\ee
of the representative models in Table~\ref{table}, and briefly outline its parameter dependence as follows.

\emph{Parameter $\mu_{\rm cap}$.} The average EMRI rate $\braket{\Gamma_{\rm emri}(T_{\rm disk}})$ is
should be proportional to the parameter $\mu_{\rm cap}$ for $T_{\rm disk}\rightarrow 0$,
while is independent of $\mu_{\rm cap}$ for long $T_{\rm disk}$ as explained above.
For the fiducial model with $M_\bullet=4\times10^6 M_\odot$,
$\braket{\Gamma_{\rm emri}(T_{\rm disk}=10^8\ {\rm yrs})}\sim \mathcal O(10^5)\ {\rm Gyr}^{-1}$
(irrespective of the parameter $\mu_{\rm cap}$
and the $\alpha$-disk model used), which is higher than the average loss-cone EMRI rate
$\bar\Gamma_{\rm emri}$  by a factor of $\sim \mathcal O(10^3)$ (Fig.~\ref{fig:Gamma_M}),
and $\braket{\Gamma_{\rm emri}(T_{\rm disk}=10^6 \ {\rm yrs}}) \gtrsim \mathcal O(10^5\mu_{\rm cap})\ {\rm Gyr}^{-1}$
is higher by a factor of $\gtrsim 10^3\mu_{\rm cap}$.

\emph{MBH mass $M_\bullet$.} The average EMRI rate $\braket{\Gamma_{\rm emri}(T_{\rm disk})}$ with long $T_{\rm disk}$
is usually higher for a heavier MBH because more sBHs are available in the stellar cluster ($\propto M_\bullet$)
while $\braket{\Gamma_{\rm emri}(T_{\rm disk})}$ with short $T_{\rm disk}$ is more initial condition
dependent.
In a model same to the fiducial model except with a lighter MBH $M_\bullet=10^5 M_\odot$,
$\braket{\Gamma_{\rm emri}(T_{\rm disk}=10^8\ {\rm yrs})} \sim \mathcal O(10^3)\ {\rm Gyr}^{-1}$
(irrespective of $\mu_{\rm cap}$ and the disk model used), which is higher than the average loss-cone EMRI rate
$\bar\Gamma_{\rm emri}$  by a factor of $\sim 15$, and $\braket{\Gamma_{\rm emri}(T_{\rm disk}=10^6 \ {\rm yrs}})
\sim (2-6) \times 10^3\ {\rm Gyr}^{-1}$ (depending on the value of $\mu_{\rm cap}$)
is higher by a factor of $(30-100)$.

\emph{Disk model.}
The dependence of disk-assisted EMRI rate on the $\alpha$ viscosity parameter and on the accretion rate $\dot M_\bullet$ are weak, which can be understood from their impact on the migration timescales of stars and sBHs. Accretion disks with smaller $\alpha$  parameters/higher accretion rates are thicker (higher disk aspect ratio $h$) and therefore the migration timescales arising from density waves are longer [Eq.~(\ref{eq:twav})],
so that the EMRI rate is lower for a disk with a smaller $\alpha$ and a higher $\dot M$.
Note that this trend does not sustain all the way to the regime of extremely low accretion rates
$\lesssim \alpha^2 \dot M^{\rm Edd}_\bullet$ \cite{Rees1982}, where the thin accretion
disk description breaks down. But $\braket{\Gamma_{\rm emri}(T_{\rm disk}})$ is sensitive to which disk
model is assumed ($\alpha$-disk or TQM disk). Due to the large difference in the two disk models, the sBH migration
timescales $t_{\rm mig}^{\rm bh,id}$ are quite different in these two disks, and consequently
the critical radii $r_{\rm crit}(T_{\rm disk})$ [Eq.~(\ref{eq:rcrit})] in these
two disks differ significantly for short $T_{\rm disk}$,
where  $r_{\rm crit}(T_{\rm disk}=10^6\ {\rm yrs}) \sim 10^6 M_\bullet$ for $\alpha$ disks and
$r_{\rm crit}(T_{\rm disk}=10^6\ {\rm yrs}) \sim 10^2 M_\bullet$ for TQM disks
(see Figs.~\ref{fig:alpha} and \ref{fig:TQM}). This explains why
$\braket{\Gamma_{\rm emri}(T_{\rm disk}=10^6 \ {\rm yrs})}$ is much smaller for TQM disks.
However, we notice that the $\alpha$-viscosity prescription is more favored by the current knowledge of turbulence viscosities driven by magnetorotational instability in inner parts of AGN disks. The sharp difference in  $\braket{\Gamma_{\rm emri}(T_{\rm disk}=10^6 \ {\rm yrs})}$ for TQM disks is likely an artifact of too  efficient angular momentum transport assumed in TQM disks.

\emph{Intial density profile of stellar cluster.}
The linear dependence of $\braket{\Gamma_{\rm emri}(T_{\rm disk})}$  on the relative abundance of sBHs $\delta$ is natural for long $T_{\rm disk}$, while the dependence is more complicate for short $T_{\rm disk}$ because the sBH fraction has been substantially changed in the pre-disk evolution. The dependence on the initial stellar density profile (parameterized by the power index $\gamma$) varies
for different MBH masses. For the Galactic nuclear stellar cluster like model ($\gamma=1.8$), sBHs are more concentrated
in the vicinity of the MBH, and therefore $\braket{\Gamma_{\rm emri}(T_{\rm disk})}$ is higher in the case of
heavy MBHs (irrespective of $T_{\rm disk}$). For lighter MBHs, the dependence reverses because more sBHs have been
depleted via the loss cone in the pre-disk phase and therefore less sBHs are available for disk capture. In this model with a steeper intial stellar density profile, the average EMRI rate $\braket{\Gamma_{\rm emri}(T_{\rm disk})}$ changes
only by a factor of few $\lesssim 3$ compared with the rate in the fiducial model.

To summarize,  the average disk-assisted EMRI rate $\braket{\Gamma_{\rm emri}(T_{\rm disk}=10^8\ {\rm yrs})}$ is higher
than the average loss-cone rate by $\mathcal O(10^2-10^3)$ for $M_\bullet=4\times10^6M_\odot$
and by $\mathcal  O(10^1-10^2)$ for $M_\bullet=10^5 M_\odot$  (irrespective of disk models, initial stellar profiles and
the value of parameter $\mu_{\rm cap}$). In the case of short disk lifetime $T_{\rm disk}=10^6\ {\rm yrs}$, $\braket{\Gamma_{\rm emri}(T_{\rm disk})}$ depends on the disk model used and the value of parameter $\mu_{\rm cap}$:
$\braket{\Gamma_{\rm emri}(T_{\rm disk})}$ is higher than the loss-cone rate by $\mathcal O(10^3\mu_{\rm cap})$
for $M_\bullet=4\times10^6M_\odot$ and is higher by $\mathcal O(10^2\mu_{\rm cap})$ for $M_\bullet=10^5M_\odot$
assuming an $\alpha/\beta$-disk, while $\braket{\Gamma_{\rm emri}(T_{\rm disk})}$ is much smaller assuming a TQM disk.

\section{Summary and Discussion}\label{sec:discussion}

\subsection{Model uncertainties}

There are several caveats in this analysis that possibly affect the estimate for disk-assisted EMRI rates.

{\it Disk model}. In this work, we have used the $\alpha$, $\beta$ and TQM model to describe the profiles of accretion disks around MBHs. The difference between $\alpha$-disk model and $\beta$-disk model mainly comes from the prescription of modelling disk viscosities, where the $\beta$-disk viscosity prescription was introduced to avoid the $\alpha$-disk  thermally instability in the radiation pressure dominated region \cite{Lightman1974,Piran1978}.
We also note that these two models yiels similar disk structures for larger distance $r \gtrsim 10^3 M_\bullet$. On the other hand, the TQM model was developed for consistently modelling the star formation in AGN disks and the disk structure
in large distance. In partiuclar, we point out that in Ref.~\cite{Thompson2005}, an equation of radiation pressure which holds only in optically thick
regime was used in both optically thick and thin regimes, which gives rises to an artificially large density variation on the edge of the opacity gap. Although these models are the state-of-the-art tools for describng the AGN accretion disks, there is still large room for improvement before accurately describing the reality,  because of various simplification/approximations taken in the models and the complexity of disk conditions/states in nature. Therefore the real disk profiles may or may not be accurately described by these models, which is a possible source of uncertainty in the rate analysis. From $\alpha$-disks and TQM disks, we find different disk models affect the disk-assisted EMRI rate
because of the diffference in the migration timescales which mildly affect the EMRI rate for long disk lifetime
but change the EMRI rate hugely for short disk lifetime $\sim 10^6$ yrs  because the migration timescale is longer
in inner parts of TQM disks (Fig.~\ref{fig:TQM}) and only nearby sBHs ($r\lesssim 10^2 M_\bullet$) can migrate into the MBH within the disk lifetime. Based on the current understanding of turbulence viscosities driven by magnetorotational instability in
fully ionized accretion disks, the $\alpha$-viscosity is a good approximation.  Therefore the sharp difference
in the disk-assisted EMRI rate for short disk lifetime in TQM disk models is likely the consequence of the artificial  angular momentum transport assumed in TQM disks.

{\it Repeating AGNs}. Disk lifetime characterizes the time duration of the active phase of AGNs. However, it is possible that before the current active phase, there are already a sequence of active phases of AGN, with various lifetimes
\cite{King2015,Schawinski2015}. These active periods may introduce significant change in star cluster distributions due to disk-star/sBH interactions. In fact, if we neglect the evolution of the star cluster distribution during the ``quiet" periods between those active periods,
the presence of previous active periods effectively extends the disk lifetime from $T_{\rm disk}$ to $T_{\rm disk}+t_0$
in Eq.~(\ref{eq:rcrit})
and shifts $t$ to $t+t_0$ in  Fig.~\ref{fig:M_depend}, where $t_0$ is the summation of the lifetimes of all previous active cycles. In other words, the integration upper and lower limit may also need to be shifted by $t_0$ in Eq.~(\ref{eq:avg}). In addition, the distribution of stars and sBHs may still evolve during the quiet phase, which further complicates the picture.
To fully account for these effects, we will need (from observations) information about the fraction of active cycles v.s. quite cycles for AGNs and the total duration of AGN outside which there is no more active phase.
In this work, we use $T_{\rm disk}\in(10^6-10^8)$ yrs as examples, while the total duration of AGN active phases (effective disk lifetime)
should be longer $\sim (10^7-10^9)$ yrs according to Soltan's argument \cite{Soltan1982}.

{\it Initial condition of stellar clusters.} We initialized the stellar cluster following the Tremaine's cluster model,
and explored the dependence of both the loss-cone EMRI rate and the disk-assisted EMRI rate on the sBH fraction $\delta$
and the density profile (parameterized by $\gamma$). We find the loss-cone EMRI rate dependence on $\delta$ is shallower than linear scaling and the disk-assisted EMRI rate dependence is linear (for long disk lifetime). Different initial density profiles affect the loss-cone EMRI rate by changing the total number of stars/sBHs within the influence sphere
$N_{\rm star,bh}(r<r_{\rm h})$.
After a few Gyrs, the distributions within the influence sphere have reached a local equilibrium and the details of initial distributions has been mostly erased except the total number of stars/sBHs. With the accretion disk turned on, the disk-assisted EMRI rate again roughly only depends on the total number instead of other erased details of the initial distributions.
Therefore we do not expect much uncertainty in the EMRI rate estimation arising from unknowns in
the initial condition of stellar clusters except the
total number of stars within the influence $N_{\rm star}(r<r_{\rm h})$ which can be inferred
from the MBH mass as $\sim M_\bullet/M_\odot$.

{\it Torque for inclined orbits}. Based on the studies for planetary systems, a point mass moving along an inclined orbit with respect to a disk excites density waves of various kinds that modify the orbit period, eccentricity and inclination in time \cite{Goldreich1979,Goldreich1980,Ward1989,Tanaka2002,Tanaka2004}. However, we notice that these studies mainly focus on low-inclination orbit. For highly inclined orbits, while the qualitative density wave generation and propagation picture should still apply, the actual torque may deviate from the formulas derived  or fitted for low-inclination orbits.
The migration speed of sBHs is proportional to the magnitude of the migration torque.
The  disk-assisted EMRI rate for short disk lifetime is determined by the capture rate of sBHs within the critical radius $r_{\rm crit}$, so that it is insenstive to the torque; for long disk lifetime   the rate is determined by the migration supply-capture consumption equilibrium, so that it should be proportional to the torque magnitude. The uncertainty in the torque magnitude should proportionally propagate to the EMRI rate estimation.
Though the dependence of the migration torque on the inclination $\iota$ has not been well explored,
we expect the inclination introduces an $\mathcal O(1)$ correction to the migration torque of low-inclination perturbers. As shown in previous studies \cite{Goldreich1979,Goldreich1980,Ward1989}
of  a perturber $m_p$ orbiting at radius $r_p$ in a gas disk,
the perturbation potential can be decomposed into Fourier
components as $\phi^p(r,t)=\sum_{lm}\phi^p_{lm} \cos(m\phi -\omega_{lm}t)$, and the torque $\dot J_{lm}$ arising from density waves of each Fourier component is proportional to $(\phi^p_{lm})^2$. For the most important
Lindblad resonances, their components of perturbation potential are
\be
\begin{aligned}
  \phi^p_{m\pm1,m}(\iota=0)
  &\propto \int_0^{2\pi}\frac{m_p}{|\vec r-\vec r_p|}\cos(m\phi) d\phi \\
  &= \int_0^{2\pi}\frac{m_p\cos(m\phi)}{\sqrt{r^2-2rr_p\cos\phi +r_p^2}}d\phi\ .
\end{aligned}
\ee
Following the same argument, we expect the components in the case of an inclined perturber to be
\be
\begin{aligned}
  \phi^p_{m\pm1,m}(\iota\neq0)
  &\propto \int_0^{2\pi}\frac{m_p}{|\vec r-\vec r_p|}\cos(m\phi) d\phi \\
  &= \int_0^{2\pi}\frac{m_p\cos(m\phi)}{\sqrt{r^2-2rr_p\cos\phi\cos\iota +r_p^2}}d\phi\ .
\end{aligned}
\ee
Comparing the above equations, we find the two components ($\phi^p_{m\pm1,m}(\iota=0)$ v.s. $\phi^p_{m\pm1,m}(\iota\neq 0)$) and therefore
the two migration torques  ($\dot J(\iota=0)$ v.s. $\dot J(\iota\neq 0)$)
are the same within a factor of $\mathcal O(1)$, respectively.

\subsection{Application and future work}

With the EMRI rate computed for different models, the next natural step is to predict the corresponding event rate for space-borne detectors such as LISA and TianQin, based on mass distribution of MBH,  star cluster distribution, disk parameters and detector sensitivity. We will leave this part as future work. It is however evident from the rates listed in
Table~ \ref{table}, the loss-cone rate (Fig.~\ref{fig:Gamma_M}) and the AGN fraction \cite{Galametz2009, Macuga2019}, that the disk-assisted EMRIs should be a good fraction of all EMRIs detected by LISA and TianQin (see the similar estimate in \cite{Tagawa2020}).

It is then important to explore how to distinguish disk-assisted EMRIs from loss-cone EMRIs within future observations. Based on the analysis in \cite{Babak2017}, the eccentricity of loss-cone EMRIs ranges from 0 to $0.2$ with long tail extending to $0.9$ near plunge,  and the inclination distribution should be nearly isotropic. On the other hand, we  expect disk-assisted EMRIs are of effectively zero eccentricity considering that the eccentricity damping timescale is much shorter than the migration timescale, and the inclination with respect to the MBH spin
equatorial plane should be confined by the disk thickness
($\iota \lesssim h$) if the MBH spin direction is aligned with rotation direction of the accretion disk.
 Both parameters of LISA detectable EMRIs are expected to be measured to sub-percent or higher accuracy \cite{Barack2004,Huerta2009}, and therefore can be used for distinguishing the EMRI origins.

As a good fraction of EMRIs detected by LISA should come from systems with AGN, it is possible that the electromagnetic emission from some of these AGNs can be observed. This brings up the opportunity for multi-messenger analysis for these EMRIs. According to Ref.~\cite{Pan:2020dze}, a fraction of low-redshift ($z\lesssim 0.3$) EMRIs can be traced back to their host galaxies with LISA observations alone, and  host galaxies of $\sim 50\%$ EMRIs in low-redshift ($z\lesssim 0.5$) AGNs can be identified with LISA observations alone considering the lower density of AGNs. If the host galaxy of such EMRI can be identified, the distance measurement from gravitaitonal wave observable and redshift measurement from optical observables should allow accurate determination on the Hubble's constant. On the other hand, for those distant EMRIs without host galaxy identification, one may still be able to measure the Hubble's constant using all AGNs in the error volume with the same statistical method introduced in \cite{schutz1986determining}. In addition, for certain disk profiles the EMRI waveform may be significantly modified, so that certain disk properties are able to be constrained with GW observations \cite{Kocsis2011,Barausse:2014tra}. These information can be further compared with electromagnetic observables from the AGN to help reveal the unknowns  about accretion physics.

Lastly, in \cite{Yang:2019iqa} we observed that a pair of sBHs embedded in the accretion disk may be locked into mean motion resonance and then migrate together towards the MBH. The resonance breaks when the pair is close to the MBH, at which stage the inner EMRI should be affected by the gravitational force from the outer sBH, so that the gravitaitonal waveform should be correspondingly satisfied \cite{Bonga:2019ycj}. Such resonance locking for a pair or a chain of objects has been discussed previously for planetary systems \cite{mills2016resonant}, which is also interesting to explore in this disk-assisted EMRI scenario.

\acknowledgements
Z.P. and H.Y. thank B\'{e}atrice Bonga for instructive discussions during early stage of this work.
Z.P. and H.Y. also thank Xinyu Li, Cole Miller, Neal Dalal and Barry McKernan for very helpful discussions and comments.
Z. P. and H. Y. are supported by the Natural Sciences and
Engineering Research Council of Canada and in part by
Perimeter Institute for Theoretical Physics. Research at
Perimeter Institute is supported in part by the Government of
Canada through the Department of Innovation, Science and
Economic Development Canada and by the Province of Ontario through the Ministry of Colleges and Universities.

\appendix
\section{Diffusion and advection coefficients in the Fokker-Planck equation (\ref{eq:FP})}\label{apa}
Following Ref.~\cite{Binney1987}, we extend the calculation of the diffusion and the advection coefficients of a single-component cluster in Ref.~\cite{Shapiro1978,Cohn1978,Cohn1979} to our two-component (stars and sBHs) case. We first define a few auxiliary functions:
\be
\begin{aligned}
  F_0^{(i)}(E,r) &= (4\pi)^2 m_i^2 \ln\Lambda \int_{-\infty}^E dE' \bar f_i(E') \ ,\\
  F_1^{(i)}(E,r) &= (4\pi)^2 m_i^2 \ln\Lambda \int_E^{\phi(r)} dE' \left(\frac{\phi-E'}{\phi-E}\right)^{1/2} \bar f_i(E') \ ,\\
  F_2^{(i)}(E,r) &= (4\pi)^2 m_i^2 \ln\Lambda \int_E^{\phi(r)} dE' \left(\frac{\phi-E'}{\phi-E}\right)^{3/2} \bar f_i(E') \ ,
\end{aligned}
\ee
where $i=\{\rm star,\ bh\}$, $\ln\Lambda$ the Coulomb's logarithm which take as $\ln\Lambda=10$, and
\be
\bar f_i(E) := \int_0^1 f(E,R) dR\ .
\ee
With these auxiliary functions, the coefficients are written as
\be
\begin{aligned}
  \mathcal D_{EE}^{(i)} &= \frac{8\pi^2}{3}J_c^2\int_{r_-}^{r_+} \frac{dr}{v_r} v^2(F_0^{(i)}+F_2^{(i)})
  + (i\leftrightarrow j)\ ,\\
  \mathcal D_{E}^{(i)} &= -8\pi^2J_c^2\int_{r_-}^{r_+} \frac{dr}{v_r} F_1^{(i)}
  +\frac{m_i}{m_j}\times (i\leftrightarrow j)\ ,\\
  \mathcal D_{ER}^{(i)} &= \frac{16\pi^2}{3}J^2\int_{r_-}^{r_+} \frac{dr}{v_r} \left(\frac{v^2}{v_c^2}-1\right)(F_0^{(i)}+F_2^{(i)})
  + (i\leftrightarrow j)\ ,\\
  \mathcal D_{RR}^{(i)} &= \frac{16\pi^2}{3}R\int_{r_-}^{r_+} \frac{dr}{v_r}
  \Bigg\{2\frac{r^2}{v^2} \left[v_t^2\left(\frac{v^2}{v_c^2}-1\right)^2 +v_r^2\right]F_0^{(i)} \\
  &+ 3\frac{r^2}{v^2}v_r^2F_1^{(i)}
  +\frac{r^2}{v^2}\left[2v_t^2\left(\frac{v^2}{v_c^2}-1\right)^2 -v_r^2 \right]F_2^{(i)}\Bigg\}
  + (i\leftrightarrow j)\ ,\\
  \mathcal D_R^{(i)}&= -16\pi^2 R r_c^2 \int_{r_-}^{r_+}  \frac{dr}{v_r} \left(1-\frac{v_c^2}{v^2}\right) F_1^{(i)}
  +\frac{m_i}{m_j}\times (i\leftrightarrow j)\ ,
\end{aligned}
\ee
where $j=\{\rm star, bh\}$, $i\neq j$, and $v_t = J/r$ is the tangential velocity.

\begin{figure}
\includegraphics[scale=0.8]{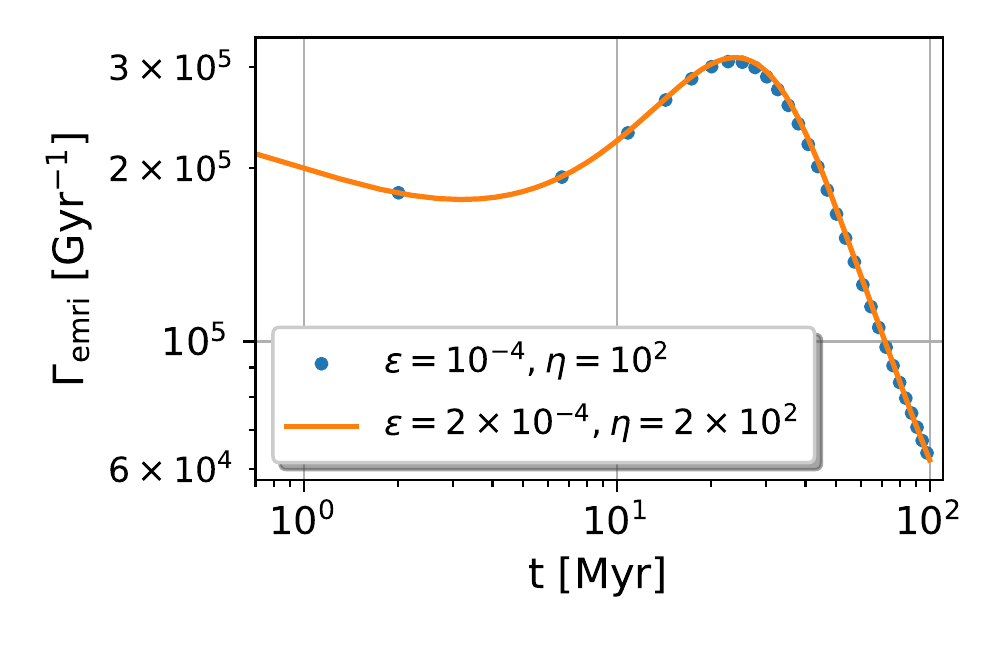}
\caption{\label{fig:rgl} The EMRI rate $\Gamma(t; T_{\rm disk}=10^8{\rm yrs})$ of the fiducial model is independent of the values of regularization parameters
$\epsilon$ and $\eta$ we used in the main text.}
\end{figure}

\section{Sanity check for the regularization algorithm}\label{apb}
As discussed in Section~\ref{subsec:num}, the scale separations between different timescales can be as large
as $10$ orders of magnitude. Therefore we need to the regularize the advection coefficients
ensuring the scale separations are numerically resolvable. There are two parameters our regularization algorithm [Eq.~(\ref{eq:regularize})]: $\epsilon$ and $\eta$, where $\epsilon$ is a small number determining an constant floor of
migrate timescale as $\epsilon T_0$, and $\eta$ is a large number determining a numerically resolvable
scale separation of $\mathcal D_{RR}$
and $\mathcal D_{R}$. In the main text, we choose $\epsilon=10^{-4}$ and $\eta=10^2$
as our default values of the two regularization parameters.
In Fig.~\ref{fig:rgl}, we show the time dependence of EMRI rate of the fiducial fast-disk-capture model with different $\epsilon$ and $\eta$, where we see the EMRI rate has no dependence on the two parameters.

\bibliography{ms}

\begin{thebibliography}{109}%
\makeatletter
\providecommand \@ifxundefined [1]{%
 \@ifx{#1\undefined}
}%
\providecommand \@ifnum [1]{%
 \ifnum #1\expandafter \@firstoftwo
 \else \expandafter \@secondoftwo
 \fi
}%
\providecommand \@ifx [1]{%
 \ifx #1\expandafter \@firstoftwo
 \else \expandafter \@secondoftwo
 \fi
}%
\providecommand \natexlab [1]{#1}%
\providecommand \enquote  [1]{``#1''}%
\providecommand \bibnamefont  [1]{#1}%
\providecommand \bibfnamefont [1]{#1}%
\providecommand \citenamefont [1]{#1}%
\providecommand \href@noop [0]{\@secondoftwo}%
\providecommand \href [0]{\begingroup \@sanitize@url \@href}%
\providecommand \@href[1]{\@@startlink{#1}\@@href}%
\providecommand \@@href[1]{\endgroup#1\@@endlink}%
\providecommand \@sanitize@url [0]{\catcode `\\12\catcode `\$12\catcode
  `\&12\catcode `\#12\catcode `\^12\catcode `\_12\catcode `\%12\relax}%
\providecommand \@@startlink[1]{}%
\providecommand \@@endlink[0]{}%
\providecommand \url  [0]{\begingroup\@sanitize@url \@url }%
\providecommand \@url [1]{\endgroup\@href {#1}{\urlprefix }}%
\providecommand \urlprefix  [0]{URL }%
\providecommand \Eprint [0]{\href }%
\providecommand \doibase [0]{http://dx.doi.org/}%
\providecommand \selectlanguage [0]{\@gobble}%
\providecommand \bibinfo  [0]{\@secondoftwo}%
\providecommand \bibfield  [0]{\@secondoftwo}%
\providecommand \translation [1]{[#1]}%
\providecommand \BibitemOpen [0]{}%
\providecommand \bibitemStop [0]{}%
\providecommand \bibitemNoStop [0]{.\EOS\space}%
\providecommand \EOS [0]{\spacefactor3000\relax}%
\providecommand \BibitemShut  [1]{\csname bibitem#1\endcsname}%
\let\auto@bib@innerbib\@empty
\bibitem [{\citenamefont {{Baker}}\ \emph {et~al.}(2019)\citenamefont
  {{Baker}}, \citenamefont {{Bellovary}}, \citenamefont {{Bender}},
  \citenamefont {{Berti}}, \citenamefont {{Caldwell}}, \citenamefont {{Camp}},
  \citenamefont {{Conklin}}, \citenamefont {{Cornish}}, \citenamefont
  {{Cutler}}, \citenamefont {{DeRosa}}, \citenamefont {{Eracleous}},
  \citenamefont {{Ferrara}}, \citenamefont {{Francis}}, \citenamefont
  {{Hewitson}}, \citenamefont {{Holley-Bockelmann}}, \citenamefont
  {{Hornschemeier}}, \citenamefont {{Hogan}}, \citenamefont {{Kamai}},
  \citenamefont {{Kelly}}, \citenamefont {{Shapiro Key}}, \citenamefont
  {{Larson}}, \citenamefont {{Livas}}, \citenamefont {{Manthripragada}},
  \citenamefont {{McKenzie}}, \citenamefont {{McWilliams}}, \citenamefont
  {{Mueller}}, \citenamefont {{Natarajan}}, \citenamefont {{Numata}},
  \citenamefont {{Rioux}}, \citenamefont {{Sankar}}, \citenamefont
  {{Schnittman}}, \citenamefont {{Shoemaker}}, \citenamefont {{Shoemaker}},
  \citenamefont {{Slutsky}}, \citenamefont {{Spero}}, \citenamefont
  {{Stebbins}}, \citenamefont {{Thorpe}}, \citenamefont {{Vallisneri}},
  \citenamefont {{Ware}}, \citenamefont {{Wass}}, \citenamefont {{Yu}},\ and\
  \citenamefont {{Ziemer}}}]{LISA2019}%
  \BibitemOpen
  \bibfield  {author} {\bibinfo {author} {\bibfnamefont {J.}~\bibnamefont
  {{Baker}}}, \bibinfo {author} {\bibfnamefont {J.}~\bibnamefont
  {{Bellovary}}}, \bibinfo {author} {\bibfnamefont {P.~L.}\ \bibnamefont
  {{Bender}}}, \bibinfo {author} {\bibfnamefont {E.}~\bibnamefont {{Berti}}},
  \bibinfo {author} {\bibfnamefont {R.}~\bibnamefont {{Caldwell}}}, \bibinfo
  {author} {\bibfnamefont {J.}~\bibnamefont {{Camp}}}, \bibinfo {author}
  {\bibfnamefont {J.~W.}\ \bibnamefont {{Conklin}}}, \bibinfo {author}
  {\bibfnamefont {N.}~\bibnamefont {{Cornish}}}, \bibinfo {author}
  {\bibfnamefont {C.}~\bibnamefont {{Cutler}}}, \bibinfo {author}
  {\bibfnamefont {R.}~\bibnamefont {{DeRosa}}}, \bibinfo {author}
  {\bibfnamefont {M.}~\bibnamefont {{Eracleous}}}, \bibinfo {author}
  {\bibfnamefont {E.~C.}\ \bibnamefont {{Ferrara}}}, \bibinfo {author}
  {\bibfnamefont {S.}~\bibnamefont {{Francis}}}, \bibinfo {author}
  {\bibfnamefont {M.}~\bibnamefont {{Hewitson}}}, \bibinfo {author}
  {\bibfnamefont {K.}~\bibnamefont {{Holley-Bockelmann}}}, \bibinfo {author}
  {\bibfnamefont {A.}~\bibnamefont {{Hornschemeier}}}, \bibinfo {author}
  {\bibfnamefont {C.}~\bibnamefont {{Hogan}}}, \bibinfo {author} {\bibfnamefont
  {B.}~\bibnamefont {{Kamai}}}, \bibinfo {author} {\bibfnamefont {B.~J.}\
  \bibnamefont {{Kelly}}}, \bibinfo {author} {\bibfnamefont {J.}~\bibnamefont
  {{Shapiro Key}}}, \bibinfo {author} {\bibfnamefont {S.~L.}\ \bibnamefont
  {{Larson}}}, \bibinfo {author} {\bibfnamefont {J.}~\bibnamefont {{Livas}}},
  \bibinfo {author} {\bibfnamefont {S.}~\bibnamefont {{Manthripragada}}},
  \bibinfo {author} {\bibfnamefont {K.}~\bibnamefont {{McKenzie}}}, \bibinfo
  {author} {\bibfnamefont {S.~T.}\ \bibnamefont {{McWilliams}}}, \bibinfo
  {author} {\bibfnamefont {G.}~\bibnamefont {{Mueller}}}, \bibinfo {author}
  {\bibfnamefont {P.}~\bibnamefont {{Natarajan}}}, \bibinfo {author}
  {\bibfnamefont {K.}~\bibnamefont {{Numata}}}, \bibinfo {author}
  {\bibfnamefont {N.}~\bibnamefont {{Rioux}}}, \bibinfo {author} {\bibfnamefont
  {S.~R.}\ \bibnamefont {{Sankar}}}, \bibinfo {author} {\bibfnamefont
  {J.}~\bibnamefont {{Schnittman}}}, \bibinfo {author} {\bibfnamefont
  {D.}~\bibnamefont {{Shoemaker}}}, \bibinfo {author} {\bibfnamefont
  {D.}~\bibnamefont {{Shoemaker}}}, \bibinfo {author} {\bibfnamefont
  {J.}~\bibnamefont {{Slutsky}}}, \bibinfo {author} {\bibfnamefont
  {R.}~\bibnamefont {{Spero}}}, \bibinfo {author} {\bibfnamefont
  {R.}~\bibnamefont {{Stebbins}}}, \bibinfo {author} {\bibfnamefont
  {I.}~\bibnamefont {{Thorpe}}}, \bibinfo {author} {\bibfnamefont
  {M.}~\bibnamefont {{Vallisneri}}}, \bibinfo {author} {\bibfnamefont
  {B.}~\bibnamefont {{Ware}}}, \bibinfo {author} {\bibfnamefont
  {P.}~\bibnamefont {{Wass}}}, \bibinfo {author} {\bibfnamefont
  {A.}~\bibnamefont {{Yu}}}, \ and\ \bibinfo {author} {\bibfnamefont
  {J.}~\bibnamefont {{Ziemer}}},\ }\href@noop {} {\bibfield  {journal}
  {\bibinfo  {journal} {arXiv e-prints}\ ,\ \bibinfo {eid} {arXiv:1907.06482}}
  (\bibinfo {year} {2019})},\ \Eprint {http://arxiv.org/abs/1907.06482}
  {arXiv:1907.06482 [astro-ph.IM]} \BibitemShut {NoStop}%
\bibitem [{\citenamefont {Mei}\ \emph {et~al.}(2020)\citenamefont {Mei},
  \citenamefont {Bai}, \citenamefont {Bao}, \citenamefont {Barausse},
  \citenamefont {Cai}, \citenamefont {Canuto}, \citenamefont {Cao},
  \citenamefont {Chen}, \citenamefont {Chen}, \citenamefont {Ding},
  \citenamefont {Duan}, \citenamefont {Fan}, \citenamefont {Feng},
  \citenamefont {Fu}, \citenamefont {Gao}, \citenamefont {Gao}, \citenamefont
  {Gong}, \citenamefont {Gou}, \citenamefont {Gu}, \citenamefont {Gu},
  \citenamefont {He}, \citenamefont {Hendry}, \citenamefont {Hong},
  \citenamefont {Hu}, \citenamefont {Hu}, \citenamefont {Hu}, \citenamefont
  {Huang}, \citenamefont {Huang}, \citenamefont {Jiang}, \citenamefont {Jiang},
  \citenamefont {Jiang}, \citenamefont {Jiang}, \citenamefont {Jin},
  \citenamefont {Korol}, \citenamefont {Li}, \citenamefont {Li}, \citenamefont
  {Li}, \citenamefont {Li}, \citenamefont {Li}, \citenamefont {Li},
  \citenamefont {Li}, \citenamefont {Li}, \citenamefont {Li}, \citenamefont
  {Liang}, \citenamefont {Liang}, \citenamefont {Liao}, \citenamefont {Liu},
  \citenamefont {Liu}, \citenamefont {Liu}, \citenamefont {Liu}, \citenamefont
  {Liu}, \citenamefont {Liu}, \citenamefont {Liu}, \citenamefont {Lu},
  \citenamefont {Lu}, \citenamefont {Lu}, \citenamefont {Luo}, \citenamefont
  {Luo}, \citenamefont {Milyukov}, \citenamefont {Ming}, \citenamefont {Pi},
  \citenamefont {Qin}, \citenamefont {Qu}, \citenamefont {Sesana},
  \citenamefont {Shao}, \citenamefont {Shi}, \citenamefont {Su}, \citenamefont
  {Tan}, \citenamefont {Tan}, \citenamefont {Tan}, \citenamefont {Tu},
  \citenamefont {Wang}, \citenamefont {Wang}, \citenamefont {Wang},
  \citenamefont {Wang}, \citenamefont {Wang}, \citenamefont {Wang},
  \citenamefont {Wang}, \citenamefont {Wang}, \citenamefont {Wang},
  \citenamefont {Wang}, \citenamefont {Wang}, \citenamefont {Wei},
  \citenamefont {Wu}, \citenamefont {Xiao}, \citenamefont {Xu}, \citenamefont
  {Xue}, \citenamefont {Yang}, \citenamefont {Yang}, \citenamefont {Yang},
  \citenamefont {Yang}, \citenamefont {Ye}, \citenamefont {Yeh}, \citenamefont
  {Yu}, \citenamefont {Zhai}, \citenamefont {Zhang}, \citenamefont {Zhang},
  \citenamefont {Zhang}, \citenamefont {Zhang}, \citenamefont {Zhang},
  \citenamefont {Zhang}, \citenamefont {Zhang}, \citenamefont {Zhou},
  \citenamefont {Zhou}, \citenamefont {Zhou}, \citenamefont {Zhu},
  \citenamefont {Zi},\ and\ \citenamefont {Luo}}]{TianQin2020}%
  \BibitemOpen
  \bibfield  {author} {\bibinfo {author} {\bibfnamefont {J.}~\bibnamefont
  {Mei}}, \bibinfo {author} {\bibfnamefont {Y.-Z.}\ \bibnamefont {Bai}},
  \bibinfo {author} {\bibfnamefont {J.}~\bibnamefont {Bao}}, \bibinfo {author}
  {\bibfnamefont {E.}~\bibnamefont {Barausse}}, \bibinfo {author}
  {\bibfnamefont {L.}~\bibnamefont {Cai}}, \bibinfo {author} {\bibfnamefont
  {E.}~\bibnamefont {Canuto}}, \bibinfo {author} {\bibfnamefont
  {B.}~\bibnamefont {Cao}}, \bibinfo {author} {\bibfnamefont {W.-M.}\
  \bibnamefont {Chen}}, \bibinfo {author} {\bibfnamefont {Y.}~\bibnamefont
  {Chen}}, \bibinfo {author} {\bibfnamefont {Y.-W.}\ \bibnamefont {Ding}},
  \bibinfo {author} {\bibfnamefont {H.-Z.}\ \bibnamefont {Duan}}, \bibinfo
  {author} {\bibfnamefont {H.}~\bibnamefont {Fan}}, \bibinfo {author}
  {\bibfnamefont {W.-F.}\ \bibnamefont {Feng}}, \bibinfo {author}
  {\bibfnamefont {H.}~\bibnamefont {Fu}}, \bibinfo {author} {\bibfnamefont
  {Q.}~\bibnamefont {Gao}}, \bibinfo {author} {\bibfnamefont {T.}~\bibnamefont
  {Gao}}, \bibinfo {author} {\bibfnamefont {Y.}~\bibnamefont {Gong}}, \bibinfo
  {author} {\bibfnamefont {X.}~\bibnamefont {Gou}}, \bibinfo {author}
  {\bibfnamefont {C.-Z.}\ \bibnamefont {Gu}}, \bibinfo {author} {\bibfnamefont
  {D.-F.}\ \bibnamefont {Gu}}, \bibinfo {author} {\bibfnamefont {Z.-Q.}\
  \bibnamefont {He}}, \bibinfo {author} {\bibfnamefont {M.}~\bibnamefont
  {Hendry}}, \bibinfo {author} {\bibfnamefont {W.}~\bibnamefont {Hong}},
  \bibinfo {author} {\bibfnamefont {X.-C.}\ \bibnamefont {Hu}}, \bibinfo
  {author} {\bibfnamefont {Y.-M.}\ \bibnamefont {Hu}}, \bibinfo {author}
  {\bibfnamefont {Y.}~\bibnamefont {Hu}}, \bibinfo {author} {\bibfnamefont
  {S.-J.}\ \bibnamefont {Huang}}, \bibinfo {author} {\bibfnamefont {X.-Q.}\
  \bibnamefont {Huang}}, \bibinfo {author} {\bibfnamefont {Q.}~\bibnamefont
  {Jiang}}, \bibinfo {author} {\bibfnamefont {Y.-Z.}\ \bibnamefont {Jiang}},
  \bibinfo {author} {\bibfnamefont {Y.}~\bibnamefont {Jiang}}, \bibinfo
  {author} {\bibfnamefont {Z.}~\bibnamefont {Jiang}}, \bibinfo {author}
  {\bibfnamefont {H.-M.}\ \bibnamefont {Jin}}, \bibinfo {author} {\bibfnamefont
  {V.}~\bibnamefont {Korol}}, \bibinfo {author} {\bibfnamefont {H.-Y.}\
  \bibnamefont {Li}}, \bibinfo {author} {\bibfnamefont {M.}~\bibnamefont {Li}},
  \bibinfo {author} {\bibfnamefont {M.}~\bibnamefont {Li}}, \bibinfo {author}
  {\bibfnamefont {P.}~\bibnamefont {Li}}, \bibinfo {author} {\bibfnamefont
  {R.}~\bibnamefont {Li}}, \bibinfo {author} {\bibfnamefont {Y.}~\bibnamefont
  {Li}}, \bibinfo {author} {\bibfnamefont {Z.}~\bibnamefont {Li}}, \bibinfo
  {author} {\bibfnamefont {Z.}~\bibnamefont {Li}}, \bibinfo {author}
  {\bibfnamefont {Z.-X.}\ \bibnamefont {Li}}, \bibinfo {author} {\bibfnamefont
  {Y.-R.}\ \bibnamefont {Liang}}, \bibinfo {author} {\bibfnamefont {Z.-C.}\
  \bibnamefont {Liang}}, \bibinfo {author} {\bibfnamefont {F.-J.}\ \bibnamefont
  {Liao}}, \bibinfo {author} {\bibfnamefont {Q.}~\bibnamefont {Liu}}, \bibinfo
  {author} {\bibfnamefont {S.}~\bibnamefont {Liu}}, \bibinfo {author}
  {\bibfnamefont {Y.-C.}\ \bibnamefont {Liu}}, \bibinfo {author} {\bibfnamefont
  {L.}~\bibnamefont {Liu}}, \bibinfo {author} {\bibfnamefont {P.-B.}\
  \bibnamefont {Liu}}, \bibinfo {author} {\bibfnamefont {X.}~\bibnamefont
  {Liu}}, \bibinfo {author} {\bibfnamefont {Y.}~\bibnamefont {Liu}}, \bibinfo
  {author} {\bibfnamefont {X.-F.}\ \bibnamefont {Lu}}, \bibinfo {author}
  {\bibfnamefont {Y.}~\bibnamefont {Lu}}, \bibinfo {author} {\bibfnamefont
  {Z.-H.}\ \bibnamefont {Lu}}, \bibinfo {author} {\bibfnamefont
  {Y.}~\bibnamefont {Luo}}, \bibinfo {author} {\bibfnamefont {Z.-C.}\
  \bibnamefont {Luo}}, \bibinfo {author} {\bibfnamefont {V.}~\bibnamefont
  {Milyukov}}, \bibinfo {author} {\bibfnamefont {M.}~\bibnamefont {Ming}},
  \bibinfo {author} {\bibfnamefont {X.}~\bibnamefont {Pi}}, \bibinfo {author}
  {\bibfnamefont {C.}~\bibnamefont {Qin}}, \bibinfo {author} {\bibfnamefont
  {S.-B.}\ \bibnamefont {Qu}}, \bibinfo {author} {\bibfnamefont
  {A.}~\bibnamefont {Sesana}}, \bibinfo {author} {\bibfnamefont
  {C.}~\bibnamefont {Shao}}, \bibinfo {author} {\bibfnamefont {C.}~\bibnamefont
  {Shi}}, \bibinfo {author} {\bibfnamefont {W.}~\bibnamefont {Su}}, \bibinfo
  {author} {\bibfnamefont {D.-Y.}\ \bibnamefont {Tan}}, \bibinfo {author}
  {\bibfnamefont {Y.}~\bibnamefont {Tan}}, \bibinfo {author} {\bibfnamefont
  {Z.}~\bibnamefont {Tan}}, \bibinfo {author} {\bibfnamefont {L.-C.}\
  \bibnamefont {Tu}}, \bibinfo {author} {\bibfnamefont {B.}~\bibnamefont
  {Wang}}, \bibinfo {author} {\bibfnamefont {C.-R.}\ \bibnamefont {Wang}},
  \bibinfo {author} {\bibfnamefont {F.}~\bibnamefont {Wang}}, \bibinfo {author}
  {\bibfnamefont {G.-F.}\ \bibnamefont {Wang}}, \bibinfo {author}
  {\bibfnamefont {H.}~\bibnamefont {Wang}}, \bibinfo {author} {\bibfnamefont
  {J.}~\bibnamefont {Wang}}, \bibinfo {author} {\bibfnamefont {L.}~\bibnamefont
  {Wang}}, \bibinfo {author} {\bibfnamefont {P.}~\bibnamefont {Wang}}, \bibinfo
  {author} {\bibfnamefont {X.}~\bibnamefont {Wang}}, \bibinfo {author}
  {\bibfnamefont {Y.}~\bibnamefont {Wang}}, \bibinfo {author} {\bibfnamefont
  {Y.-F.}\ \bibnamefont {Wang}}, \bibinfo {author} {\bibfnamefont
  {R.}~\bibnamefont {Wei}}, \bibinfo {author} {\bibfnamefont {S.-C.}\
  \bibnamefont {Wu}}, \bibinfo {author} {\bibfnamefont {C.-Y.}\ \bibnamefont
  {Xiao}}, \bibinfo {author} {\bibfnamefont {X.-S.}\ \bibnamefont {Xu}},
  \bibinfo {author} {\bibfnamefont {C.}~\bibnamefont {Xue}}, \bibinfo {author}
  {\bibfnamefont {F.-C.}\ \bibnamefont {Yang}}, \bibinfo {author}
  {\bibfnamefont {L.}~\bibnamefont {Yang}}, \bibinfo {author} {\bibfnamefont
  {M.-L.}\ \bibnamefont {Yang}}, \bibinfo {author} {\bibfnamefont {S.-Q.}\
  \bibnamefont {Yang}}, \bibinfo {author} {\bibfnamefont {B.}~\bibnamefont
  {Ye}}, \bibinfo {author} {\bibfnamefont {H.-C.}\ \bibnamefont {Yeh}},
  \bibinfo {author} {\bibfnamefont {S.}~\bibnamefont {Yu}}, \bibinfo {author}
  {\bibfnamefont {D.}~\bibnamefont {Zhai}}, \bibinfo {author} {\bibfnamefont
  {C.}~\bibnamefont {Zhang}}, \bibinfo {author} {\bibfnamefont
  {H.}~\bibnamefont {Zhang}}, \bibinfo {author} {\bibfnamefont {J.-d.}\
  \bibnamefont {Zhang}}, \bibinfo {author} {\bibfnamefont {J.}~\bibnamefont
  {Zhang}}, \bibinfo {author} {\bibfnamefont {L.}~\bibnamefont {Zhang}},
  \bibinfo {author} {\bibfnamefont {X.}~\bibnamefont {Zhang}}, \bibinfo
  {author} {\bibfnamefont {X.}~\bibnamefont {Zhang}}, \bibinfo {author}
  {\bibfnamefont {H.}~\bibnamefont {Zhou}}, \bibinfo {author} {\bibfnamefont
  {M.-Y.}\ \bibnamefont {Zhou}}, \bibinfo {author} {\bibfnamefont {Z.-B.}\
  \bibnamefont {Zhou}}, \bibinfo {author} {\bibfnamefont {D.-D.}\ \bibnamefont
  {Zhu}}, \bibinfo {author} {\bibfnamefont {T.-G.}\ \bibnamefont {Zi}}, \ and\
  \bibinfo {author} {\bibfnamefont {J.}~\bibnamefont {Luo}},\ }\href {\doibase
  10.1093/ptep/ptaa114} {\bibfield  {journal} {\bibinfo  {journal} {Progress of
  Theoretical and Experimental Physics}\ } (\bibinfo {year} {2020}),\
  10.1093/ptep/ptaa114},\ \bibinfo {note} {ptaa114},\ \Eprint
  {http://arxiv.org/abs/https://academic.oup.com/ptep/advance-article-pdf/doi/10.1093/ptep/ptaa114/34068872/ptaa114.pdf}
  {https://academic.oup.com/ptep/advance-article-pdf/doi/10.1093/ptep/ptaa114/34068872/ptaa114.pdf}
  \BibitemShut {NoStop}%
\bibitem [{\citenamefont {{Amaro-Seoane}}\ \emph {et~al.}(2017)\citenamefont
  {{Amaro-Seoane}}, \citenamefont {{Audley}}, \citenamefont {{Babak}},
  \citenamefont {{Baker}}, \citenamefont {{Barausse}}, \citenamefont
  {{Bender}}, \citenamefont {{Berti}}, \citenamefont {{Binetruy}},
  \citenamefont {{Born}}, \citenamefont {{Bortoluzzi}}, \citenamefont {{Camp}},
  \citenamefont {{Caprini}}, \citenamefont {{Cardoso}}, \citenamefont
  {{Colpi}}, \citenamefont {{Conklin}}, \citenamefont {{Cornish}},
  \citenamefont {{Cutler}}, \citenamefont {{Danzmann}}, \citenamefont
  {{Dolesi}}, \citenamefont {{Ferraioli}}, \citenamefont {{Ferroni}},
  \citenamefont {{Fitzsimons}}, \citenamefont {{Gair}}, \citenamefont {{Gesa
  Bote}}, \citenamefont {{Giardini}}, \citenamefont {{Gibert}}, \citenamefont
  {{Grimani}}, \citenamefont {{Halloin}}, \citenamefont {{Heinzel}},
  \citenamefont {{Hertog}}, \citenamefont {{Hewitson}}, \citenamefont
  {{Holley-Bockelmann}}, \citenamefont {{Hollington}}, \citenamefont
  {{Hueller}}, \citenamefont {{Inchauspe}}, \citenamefont {{Jetzer}},
  \citenamefont {{Karnesis}}, \citenamefont {{Killow}}, \citenamefont
  {{Klein}}, \citenamefont {{Klipstein}}, \citenamefont {{Korsakova}},
  \citenamefont {{Larson}}, \citenamefont {{Livas}}, \citenamefont {{Lloro}},
  \citenamefont {{Man}}, \citenamefont {{Mance}}, \citenamefont {{Martino}},
  \citenamefont {{Mateos}}, \citenamefont {{McKenzie}}, \citenamefont
  {{McWilliams}}, \citenamefont {{Miller}}, \citenamefont {{Mueller}},
  \citenamefont {{Nardini}}, \citenamefont {{Nelemans}}, \citenamefont
  {{Nofrarias}}, \citenamefont {{Petiteau}}, \citenamefont {{Pivato}},
  \citenamefont {{Plagnol}}, \citenamefont {{Porter}}, \citenamefont
  {{Reiche}}, \citenamefont {{Robertson}}, \citenamefont {{Robertson}},
  \citenamefont {{Rossi}}, \citenamefont {{Russano}}, \citenamefont {{Schutz}},
  \citenamefont {{Sesana}}, \citenamefont {{Shoemaker}}, \citenamefont
  {{Slutsky}}, \citenamefont {{Sopuerta}}, \citenamefont {{Sumner}},
  \citenamefont {{Tamanini}}, \citenamefont {{Thorpe}}, \citenamefont
  {{Troebs}}, \citenamefont {{Vallisneri}}, \citenamefont {{Vecchio}},
  \citenamefont {{Vetrugno}}, \citenamefont {{Vitale}}, \citenamefont
  {{Volonteri}}, \citenamefont {{Wanner}}, \citenamefont {{Ward}},
  \citenamefont {{Wass}}, \citenamefont {{Weber}}, \citenamefont {{Ziemer}},\
  and\ \citenamefont {{Zweifel}}}]{LISA2017}%
  \BibitemOpen
  \bibfield  {author} {\bibinfo {author} {\bibfnamefont {P.}~\bibnamefont
  {{Amaro-Seoane}}}, \bibinfo {author} {\bibfnamefont {H.}~\bibnamefont
  {{Audley}}}, \bibinfo {author} {\bibfnamefont {S.}~\bibnamefont {{Babak}}},
  \bibinfo {author} {\bibfnamefont {J.}~\bibnamefont {{Baker}}}, \bibinfo
  {author} {\bibfnamefont {E.}~\bibnamefont {{Barausse}}}, \bibinfo {author}
  {\bibfnamefont {P.}~\bibnamefont {{Bender}}}, \bibinfo {author}
  {\bibfnamefont {E.}~\bibnamefont {{Berti}}}, \bibinfo {author} {\bibfnamefont
  {P.}~\bibnamefont {{Binetruy}}}, \bibinfo {author} {\bibfnamefont
  {M.}~\bibnamefont {{Born}}}, \bibinfo {author} {\bibfnamefont
  {D.}~\bibnamefont {{Bortoluzzi}}}, \bibinfo {author} {\bibfnamefont
  {J.}~\bibnamefont {{Camp}}}, \bibinfo {author} {\bibfnamefont
  {C.}~\bibnamefont {{Caprini}}}, \bibinfo {author} {\bibfnamefont
  {V.}~\bibnamefont {{Cardoso}}}, \bibinfo {author} {\bibfnamefont
  {M.}~\bibnamefont {{Colpi}}}, \bibinfo {author} {\bibfnamefont
  {J.}~\bibnamefont {{Conklin}}}, \bibinfo {author} {\bibfnamefont
  {N.}~\bibnamefont {{Cornish}}}, \bibinfo {author} {\bibfnamefont
  {C.}~\bibnamefont {{Cutler}}}, \bibinfo {author} {\bibfnamefont
  {K.}~\bibnamefont {{Danzmann}}}, \bibinfo {author} {\bibfnamefont
  {R.}~\bibnamefont {{Dolesi}}}, \bibinfo {author} {\bibfnamefont
  {L.}~\bibnamefont {{Ferraioli}}}, \bibinfo {author} {\bibfnamefont
  {V.}~\bibnamefont {{Ferroni}}}, \bibinfo {author} {\bibfnamefont
  {E.}~\bibnamefont {{Fitzsimons}}}, \bibinfo {author} {\bibfnamefont
  {J.}~\bibnamefont {{Gair}}}, \bibinfo {author} {\bibfnamefont
  {L.}~\bibnamefont {{Gesa Bote}}}, \bibinfo {author} {\bibfnamefont
  {D.}~\bibnamefont {{Giardini}}}, \bibinfo {author} {\bibfnamefont
  {F.}~\bibnamefont {{Gibert}}}, \bibinfo {author} {\bibfnamefont
  {C.}~\bibnamefont {{Grimani}}}, \bibinfo {author} {\bibfnamefont
  {H.}~\bibnamefont {{Halloin}}}, \bibinfo {author} {\bibfnamefont
  {G.}~\bibnamefont {{Heinzel}}}, \bibinfo {author} {\bibfnamefont
  {T.}~\bibnamefont {{Hertog}}}, \bibinfo {author} {\bibfnamefont
  {M.}~\bibnamefont {{Hewitson}}}, \bibinfo {author} {\bibfnamefont
  {K.}~\bibnamefont {{Holley-Bockelmann}}}, \bibinfo {author} {\bibfnamefont
  {D.}~\bibnamefont {{Hollington}}}, \bibinfo {author} {\bibfnamefont
  {M.}~\bibnamefont {{Hueller}}}, \bibinfo {author} {\bibfnamefont
  {H.}~\bibnamefont {{Inchauspe}}}, \bibinfo {author} {\bibfnamefont
  {P.}~\bibnamefont {{Jetzer}}}, \bibinfo {author} {\bibfnamefont
  {N.}~\bibnamefont {{Karnesis}}}, \bibinfo {author} {\bibfnamefont
  {C.}~\bibnamefont {{Killow}}}, \bibinfo {author} {\bibfnamefont
  {A.}~\bibnamefont {{Klein}}}, \bibinfo {author} {\bibfnamefont
  {B.}~\bibnamefont {{Klipstein}}}, \bibinfo {author} {\bibfnamefont
  {N.}~\bibnamefont {{Korsakova}}}, \bibinfo {author} {\bibfnamefont {S.~L.}\
  \bibnamefont {{Larson}}}, \bibinfo {author} {\bibfnamefont {J.}~\bibnamefont
  {{Livas}}}, \bibinfo {author} {\bibfnamefont {I.}~\bibnamefont {{Lloro}}},
  \bibinfo {author} {\bibfnamefont {N.}~\bibnamefont {{Man}}}, \bibinfo
  {author} {\bibfnamefont {D.}~\bibnamefont {{Mance}}}, \bibinfo {author}
  {\bibfnamefont {J.}~\bibnamefont {{Martino}}}, \bibinfo {author}
  {\bibfnamefont {I.}~\bibnamefont {{Mateos}}}, \bibinfo {author}
  {\bibfnamefont {K.}~\bibnamefont {{McKenzie}}}, \bibinfo {author}
  {\bibfnamefont {S.~T.}\ \bibnamefont {{McWilliams}}}, \bibinfo {author}
  {\bibfnamefont {C.}~\bibnamefont {{Miller}}}, \bibinfo {author}
  {\bibfnamefont {G.}~\bibnamefont {{Mueller}}}, \bibinfo {author}
  {\bibfnamefont {G.}~\bibnamefont {{Nardini}}}, \bibinfo {author}
  {\bibfnamefont {G.}~\bibnamefont {{Nelemans}}}, \bibinfo {author}
  {\bibfnamefont {M.}~\bibnamefont {{Nofrarias}}}, \bibinfo {author}
  {\bibfnamefont {A.}~\bibnamefont {{Petiteau}}}, \bibinfo {author}
  {\bibfnamefont {P.}~\bibnamefont {{Pivato}}}, \bibinfo {author}
  {\bibfnamefont {E.}~\bibnamefont {{Plagnol}}}, \bibinfo {author}
  {\bibfnamefont {E.}~\bibnamefont {{Porter}}}, \bibinfo {author}
  {\bibfnamefont {J.}~\bibnamefont {{Reiche}}}, \bibinfo {author}
  {\bibfnamefont {D.}~\bibnamefont {{Robertson}}}, \bibinfo {author}
  {\bibfnamefont {N.}~\bibnamefont {{Robertson}}}, \bibinfo {author}
  {\bibfnamefont {E.}~\bibnamefont {{Rossi}}}, \bibinfo {author} {\bibfnamefont
  {G.}~\bibnamefont {{Russano}}}, \bibinfo {author} {\bibfnamefont
  {B.}~\bibnamefont {{Schutz}}}, \bibinfo {author} {\bibfnamefont
  {A.}~\bibnamefont {{Sesana}}}, \bibinfo {author} {\bibfnamefont
  {D.}~\bibnamefont {{Shoemaker}}}, \bibinfo {author} {\bibfnamefont
  {J.}~\bibnamefont {{Slutsky}}}, \bibinfo {author} {\bibfnamefont {C.~F.}\
  \bibnamefont {{Sopuerta}}}, \bibinfo {author} {\bibfnamefont
  {T.}~\bibnamefont {{Sumner}}}, \bibinfo {author} {\bibfnamefont
  {N.}~\bibnamefont {{Tamanini}}}, \bibinfo {author} {\bibfnamefont
  {I.}~\bibnamefont {{Thorpe}}}, \bibinfo {author} {\bibfnamefont
  {M.}~\bibnamefont {{Troebs}}}, \bibinfo {author} {\bibfnamefont
  {M.}~\bibnamefont {{Vallisneri}}}, \bibinfo {author} {\bibfnamefont
  {A.}~\bibnamefont {{Vecchio}}}, \bibinfo {author} {\bibfnamefont
  {D.}~\bibnamefont {{Vetrugno}}}, \bibinfo {author} {\bibfnamefont
  {S.}~\bibnamefont {{Vitale}}}, \bibinfo {author} {\bibfnamefont
  {M.}~\bibnamefont {{Volonteri}}}, \bibinfo {author} {\bibfnamefont
  {G.}~\bibnamefont {{Wanner}}}, \bibinfo {author} {\bibfnamefont
  {H.}~\bibnamefont {{Ward}}}, \bibinfo {author} {\bibfnamefont
  {P.}~\bibnamefont {{Wass}}}, \bibinfo {author} {\bibfnamefont
  {W.}~\bibnamefont {{Weber}}}, \bibinfo {author} {\bibfnamefont
  {J.}~\bibnamefont {{Ziemer}}}, \ and\ \bibinfo {author} {\bibfnamefont
  {P.}~\bibnamefont {{Zweifel}}},\ }\href@noop {} {\bibfield  {journal}
  {\bibinfo  {journal} {arXiv e-prints}\ ,\ \bibinfo {eid} {arXiv:1702.00786}}
  (\bibinfo {year} {2017})},\ \Eprint {http://arxiv.org/abs/1702.00786}
  {arXiv:1702.00786 [astro-ph.IM]} \BibitemShut {NoStop}%
\bibitem [{\citenamefont {Babak}\ \emph {et~al.}(2017)\citenamefont {Babak},
  \citenamefont {Gair}, \citenamefont {Sesana}, \citenamefont {Barausse},
  \citenamefont {Sopuerta}, \citenamefont {Berry}, \citenamefont {Berti},
  \citenamefont {Amaro-Seoane}, \citenamefont {Petiteau},\ and\ \citenamefont
  {Klein}}]{Babak:2017tow}%
  \BibitemOpen
  \bibfield  {author} {\bibinfo {author} {\bibfnamefont {S.}~\bibnamefont
  {Babak}}, \bibinfo {author} {\bibfnamefont {J.}~\bibnamefont {Gair}},
  \bibinfo {author} {\bibfnamefont {A.}~\bibnamefont {Sesana}}, \bibinfo
  {author} {\bibfnamefont {E.}~\bibnamefont {Barausse}}, \bibinfo {author}
  {\bibfnamefont {C.~F.}\ \bibnamefont {Sopuerta}}, \bibinfo {author}
  {\bibfnamefont {C.~P.}\ \bibnamefont {Berry}}, \bibinfo {author}
  {\bibfnamefont {E.}~\bibnamefont {Berti}}, \bibinfo {author} {\bibfnamefont
  {P.}~\bibnamefont {Amaro-Seoane}}, \bibinfo {author} {\bibfnamefont
  {A.}~\bibnamefont {Petiteau}}, \ and\ \bibinfo {author} {\bibfnamefont
  {A.}~\bibnamefont {Klein}},\ }\href {\doibase 10.1103/PhysRevD.95.103012}
  {\bibfield  {journal} {\bibinfo  {journal} {Phys. Rev. D}\ }\textbf {\bibinfo
  {volume} {95}},\ \bibinfo {pages} {103012} (\bibinfo {year} {2017})},\
  \Eprint {http://arxiv.org/abs/1703.09722} {arXiv:1703.09722 [gr-qc]}
  \BibitemShut {NoStop}%
\bibitem [{\citenamefont {Bonga}\ \emph {et~al.}(2019)\citenamefont {Bonga},
  \citenamefont {Yang},\ and\ \citenamefont {Hughes}}]{Bonga:2019ycj}%
  \BibitemOpen
  \bibfield  {author} {\bibinfo {author} {\bibfnamefont {B.}~\bibnamefont
  {Bonga}}, \bibinfo {author} {\bibfnamefont {H.}~\bibnamefont {Yang}}, \ and\
  \bibinfo {author} {\bibfnamefont {S.~A.}\ \bibnamefont {Hughes}},\ }\href
  {\doibase 10.1103/PhysRevLett.123.101103} {\bibfield  {journal} {\bibinfo
  {journal} {Phys. Rev. Lett.}\ }\textbf {\bibinfo {volume} {123}},\ \bibinfo
  {pages} {101103} (\bibinfo {year} {2019})},\ \Eprint
  {http://arxiv.org/abs/1905.00030} {arXiv:1905.00030 [gr-qc]} \BibitemShut
  {NoStop}%
\bibitem [{\citenamefont {Yang}\ \emph {et~al.}(2019)\citenamefont {Yang},
  \citenamefont {Bonga}, \citenamefont {Peng},\ and\ \citenamefont
  {Li}}]{Yang:2019iqa}%
  \BibitemOpen
  \bibfield  {author} {\bibinfo {author} {\bibfnamefont {H.}~\bibnamefont
  {Yang}}, \bibinfo {author} {\bibfnamefont {B.}~\bibnamefont {Bonga}},
  \bibinfo {author} {\bibfnamefont {Z.}~\bibnamefont {Peng}}, \ and\ \bibinfo
  {author} {\bibfnamefont {G.}~\bibnamefont {Li}},\ }\href {\doibase
  10.1103/PhysRevD.100.124056} {\bibfield  {journal} {\bibinfo  {journal}
  {Phys. Rev. D}\ }\textbf {\bibinfo {volume} {100}},\ \bibinfo {pages}
  {124056} (\bibinfo {year} {2019})},\ \Eprint
  {http://arxiv.org/abs/1910.07337} {arXiv:1910.07337 [gr-qc]} \BibitemShut
  {NoStop}%
\bibitem [{\citenamefont {Barausse}\ \emph {et~al.}(2014)\citenamefont
  {Barausse}, \citenamefont {Cardoso},\ and\ \citenamefont
  {Pani}}]{Barausse:2014tra}%
  \BibitemOpen
  \bibfield  {author} {\bibinfo {author} {\bibfnamefont {E.}~\bibnamefont
  {Barausse}}, \bibinfo {author} {\bibfnamefont {V.}~\bibnamefont {Cardoso}}, \
  and\ \bibinfo {author} {\bibfnamefont {P.}~\bibnamefont {Pani}},\ }\href
  {\doibase 10.1103/PhysRevD.89.104059} {\bibfield  {journal} {\bibinfo
  {journal} {Phys. Rev. D}\ }\textbf {\bibinfo {volume} {89}},\ \bibinfo
  {pages} {104059} (\bibinfo {year} {2014})},\ \Eprint
  {http://arxiv.org/abs/1404.7149} {arXiv:1404.7149 [gr-qc]} \BibitemShut
  {NoStop}%
\bibitem [{\citenamefont {Berti}\ and\ \citenamefont
  {Volonteri}(2008)}]{Berti:2008af}%
  \BibitemOpen
  \bibfield  {author} {\bibinfo {author} {\bibfnamefont {E.}~\bibnamefont
  {Berti}}\ and\ \bibinfo {author} {\bibfnamefont {M.}~\bibnamefont
  {Volonteri}},\ }\href {\doibase 10.1086/590379} {\bibfield  {journal}
  {\bibinfo  {journal} {Astrophys. J.}\ }\textbf {\bibinfo {volume} {684}},\
  \bibinfo {pages} {822} (\bibinfo {year} {2008})},\ \Eprint
  {http://arxiv.org/abs/0802.0025} {arXiv:0802.0025 [astro-ph]} \BibitemShut
  {NoStop}%
\bibitem [{\citenamefont {Pan}\ and\ \citenamefont {Yang}(2020)}]{Pan:2020dze}%
  \BibitemOpen
  \bibfield  {author} {\bibinfo {author} {\bibfnamefont {Z.}~\bibnamefont
  {Pan}}\ and\ \bibinfo {author} {\bibfnamefont {H.}~\bibnamefont {Yang}},\
  }\href {\doibase 10.3847/1538-4357/abb1b1} {\bibfield  {journal} {\bibinfo
  {journal} {Astrophys. J.}\ }\textbf {\bibinfo {volume} {901}},\ \bibinfo
  {pages} {163} (\bibinfo {year} {2020})},\ \Eprint
  {http://arxiv.org/abs/2007.03783} {arXiv:2007.03783 [astro-ph.CO]}
  \BibitemShut {NoStop}%
\bibitem [{\citenamefont {{Amaro-Seoane}}(2018)}]{Amaro2018}%
  \BibitemOpen
  \bibfield  {author} {\bibinfo {author} {\bibfnamefont {P.}~\bibnamefont
  {{Amaro-Seoane}}},\ }\href {\doibase 10.1007/s41114-018-0013-8} {\bibfield
  {journal} {\bibinfo  {journal} {Living Reviews in Relativity}\ }\textbf
  {\bibinfo {volume} {21}},\ \bibinfo {eid} {4} (\bibinfo {year} {2018})},\
  \Eprint {http://arxiv.org/abs/1205.5240} {arXiv:1205.5240 [astro-ph.CO]}
  \BibitemShut {NoStop}%
\bibitem [{\citenamefont {Gair}\ \emph {et~al.}(2017)\citenamefont {Gair},
  \citenamefont {Babak}, \citenamefont {Sesana}, \citenamefont {Amaro-Seoane},
  \citenamefont {Barausse}, \citenamefont {Berry}, \citenamefont {Berti},\ and\
  \citenamefont {Sopuerta}}]{Gair:2017ynp}%
  \BibitemOpen
  \bibfield  {author} {\bibinfo {author} {\bibfnamefont {J.~R.}\ \bibnamefont
  {Gair}}, \bibinfo {author} {\bibfnamefont {S.}~\bibnamefont {Babak}},
  \bibinfo {author} {\bibfnamefont {A.}~\bibnamefont {Sesana}}, \bibinfo
  {author} {\bibfnamefont {P.}~\bibnamefont {Amaro-Seoane}}, \bibinfo {author}
  {\bibfnamefont {E.}~\bibnamefont {Barausse}}, \bibinfo {author}
  {\bibfnamefont {C.~P.}\ \bibnamefont {Berry}}, \bibinfo {author}
  {\bibfnamefont {E.}~\bibnamefont {Berti}}, \ and\ \bibinfo {author}
  {\bibfnamefont {C.}~\bibnamefont {Sopuerta}},\ }\href {\doibase
  10.1088/1742-6596/840/1/012021} {\bibfield  {journal} {\bibinfo  {journal}
  {J. Phys. Conf. Ser.}\ }\textbf {\bibinfo {volume} {840}},\ \bibinfo {pages}
  {012021} (\bibinfo {year} {2017})},\ \Eprint
  {http://arxiv.org/abs/1704.00009} {arXiv:1704.00009 [astro-ph.GA]}
  \BibitemShut {NoStop}%
\bibitem [{\citenamefont {{Babak}}\ \emph {et~al.}(2017)\citenamefont
  {{Babak}}, \citenamefont {{Gair}}, \citenamefont {{Sesana}}, \citenamefont
  {{Barausse}}, \citenamefont {{Sopuerta}}, \citenamefont {{Berry}},
  \citenamefont {{Berti}}, \citenamefont {{Amaro-Seoane}}, \citenamefont
  {{Petiteau}},\ and\ \citenamefont {{Klein}}}]{Babak2017}%
  \BibitemOpen
  \bibfield  {author} {\bibinfo {author} {\bibfnamefont {S.}~\bibnamefont
  {{Babak}}}, \bibinfo {author} {\bibfnamefont {J.}~\bibnamefont {{Gair}}},
  \bibinfo {author} {\bibfnamefont {A.}~\bibnamefont {{Sesana}}}, \bibinfo
  {author} {\bibfnamefont {E.}~\bibnamefont {{Barausse}}}, \bibinfo {author}
  {\bibfnamefont {C.~F.}\ \bibnamefont {{Sopuerta}}}, \bibinfo {author}
  {\bibfnamefont {C.~P.~L.}\ \bibnamefont {{Berry}}}, \bibinfo {author}
  {\bibfnamefont {E.}~\bibnamefont {{Berti}}}, \bibinfo {author} {\bibfnamefont
  {P.}~\bibnamefont {{Amaro-Seoane}}}, \bibinfo {author} {\bibfnamefont
  {A.}~\bibnamefont {{Petiteau}}}, \ and\ \bibinfo {author} {\bibfnamefont
  {A.}~\bibnamefont {{Klein}}},\ }\href {\doibase 10.1103/PhysRevD.95.103012}
  {\bibfield  {journal} {\bibinfo  {journal} {\prd}\ }\textbf {\bibinfo
  {volume} {95}},\ \bibinfo {eid} {103012} (\bibinfo {year} {2017})},\ \Eprint
  {http://arxiv.org/abs/1703.09722} {arXiv:1703.09722 [gr-qc]} \BibitemShut
  {NoStop}%
\bibitem [{\citenamefont {{Miller}}\ \emph {et~al.}(2005)\citenamefont
  {{Miller}}, \citenamefont {{Freitag}}, \citenamefont {{Hamilton}},\ and\
  \citenamefont {{Lauburg}}}]{Miller2005}%
  \BibitemOpen
  \bibfield  {author} {\bibinfo {author} {\bibfnamefont {M.~C.}\ \bibnamefont
  {{Miller}}}, \bibinfo {author} {\bibfnamefont {M.}~\bibnamefont {{Freitag}}},
  \bibinfo {author} {\bibfnamefont {D.~P.}\ \bibnamefont {{Hamilton}}}, \ and\
  \bibinfo {author} {\bibfnamefont {V.~M.}\ \bibnamefont {{Lauburg}}},\ }\href
  {\doibase 10.1086/497335} {\bibfield  {journal} {\bibinfo  {journal} {\apjl}\
  }\textbf {\bibinfo {volume} {631}},\ \bibinfo {pages} {L117} (\bibinfo {year}
  {2005})},\ \Eprint {http://arxiv.org/abs/astro-ph/0507133}
  {arXiv:astro-ph/0507133 [astro-ph]} \BibitemShut {NoStop}%
\bibitem [{\citenamefont {{Chen}}\ and\ \citenamefont
  {{Han}}(2018)}]{Chen2018}%
  \BibitemOpen
  \bibfield  {author} {\bibinfo {author} {\bibfnamefont {X.}~\bibnamefont
  {{Chen}}}\ and\ \bibinfo {author} {\bibfnamefont {W.-B.}\ \bibnamefont
  {{Han}}},\ }\href {\doibase 10.1038/s42005-018-0053-0} {\bibfield  {journal}
  {\bibinfo  {journal} {Communications Physics}\ }\textbf {\bibinfo {volume}
  {1}},\ \bibinfo {eid} {53} (\bibinfo {year} {2018})},\ \Eprint
  {http://arxiv.org/abs/1801.05780} {arXiv:1801.05780 [astro-ph.HE]}
  \BibitemShut {NoStop}%
\bibitem [{\citenamefont {{Preto}}\ and\ \citenamefont
  {{Amaro-Seoane}}(2010)}]{Preto2010}%
  \BibitemOpen
  \bibfield  {author} {\bibinfo {author} {\bibfnamefont {M.}~\bibnamefont
  {{Preto}}}\ and\ \bibinfo {author} {\bibfnamefont {P.}~\bibnamefont
  {{Amaro-Seoane}}},\ }\href {\doibase 10.1088/2041-8205/708/1/L42} {\bibfield
  {journal} {\bibinfo  {journal} {\apjl}\ }\textbf {\bibinfo {volume} {708}},\
  \bibinfo {pages} {L42} (\bibinfo {year} {2010})},\ \Eprint
  {http://arxiv.org/abs/0910.3206} {arXiv:0910.3206 [astro-ph.GA]} \BibitemShut
  {NoStop}%
\bibitem [{\citenamefont {{Amaro-Seoane}}\ and\ \citenamefont
  {{Preto}}(2011)}]{Amaro2011}%
  \BibitemOpen
  \bibfield  {author} {\bibinfo {author} {\bibfnamefont {P.}~\bibnamefont
  {{Amaro-Seoane}}}\ and\ \bibinfo {author} {\bibfnamefont {M.}~\bibnamefont
  {{Preto}}},\ }\href {\doibase 10.1088/0264-9381/28/9/094017} {\bibfield
  {journal} {\bibinfo  {journal} {Classical and Quantum Gravity}\ }\textbf
  {\bibinfo {volume} {28}},\ \bibinfo {eid} {094017} (\bibinfo {year}
  {2011})},\ \Eprint {http://arxiv.org/abs/1010.5781} {arXiv:1010.5781
  [astro-ph.CO]} \BibitemShut {NoStop}%
\bibitem [{\citenamefont {{Fan}}\ \emph {et~al.}(2020)\citenamefont {{Fan}},
  \citenamefont {{Hu}}, \citenamefont {{Barausse}}, \citenamefont {{Sesana}},
  \citenamefont {{Zhang}}, \citenamefont {{Zhang}}, \citenamefont {{Zi}},\ and\
  \citenamefont {{Mei}}}]{Fan2020}%
  \BibitemOpen
  \bibfield  {author} {\bibinfo {author} {\bibfnamefont {H.-M.}\ \bibnamefont
  {{Fan}}}, \bibinfo {author} {\bibfnamefont {Y.-M.}\ \bibnamefont {{Hu}}},
  \bibinfo {author} {\bibfnamefont {E.}~\bibnamefont {{Barausse}}}, \bibinfo
  {author} {\bibfnamefont {A.}~\bibnamefont {{Sesana}}}, \bibinfo {author}
  {\bibfnamefont {J.-d.}\ \bibnamefont {{Zhang}}}, \bibinfo {author}
  {\bibfnamefont {X.}~\bibnamefont {{Zhang}}}, \bibinfo {author} {\bibfnamefont
  {T.-G.}\ \bibnamefont {{Zi}}}, \ and\ \bibinfo {author} {\bibfnamefont
  {J.}~\bibnamefont {{Mei}}},\ }\href {\doibase 10.1103/PhysRevD.102.063016}
  {\bibfield  {journal} {\bibinfo  {journal} {\prd}\ }\textbf {\bibinfo
  {volume} {102}},\ \bibinfo {eid} {063016} (\bibinfo {year} {2020})},\ \Eprint
  {http://arxiv.org/abs/2005.08212} {arXiv:2005.08212 [astro-ph.HE]}
  \BibitemShut {NoStop}%
\bibitem [{\citenamefont {{Goldreich}}\ and\ \citenamefont
  {{Tremaine}}(1979)}]{Goldreich1979}%
  \BibitemOpen
  \bibfield  {author} {\bibinfo {author} {\bibfnamefont {P.}~\bibnamefont
  {{Goldreich}}}\ and\ \bibinfo {author} {\bibfnamefont {S.}~\bibnamefont
  {{Tremaine}}},\ }\href {\doibase 10.1086/157448} {\bibfield  {journal}
  {\bibinfo  {journal} {\apj}\ }\textbf {\bibinfo {volume} {233}},\ \bibinfo
  {pages} {857} (\bibinfo {year} {1979})}\BibitemShut {NoStop}%
\bibitem [{\citenamefont {{Goldreich}}\ and\ \citenamefont
  {{Tremaine}}(1980)}]{Goldreich1980}%
  \BibitemOpen
  \bibfield  {author} {\bibinfo {author} {\bibfnamefont {P.}~\bibnamefont
  {{Goldreich}}}\ and\ \bibinfo {author} {\bibfnamefont {S.}~\bibnamefont
  {{Tremaine}}},\ }\href {\doibase 10.1086/158356} {\bibfield  {journal}
  {\bibinfo  {journal} {\apj}\ }\textbf {\bibinfo {volume} {241}},\ \bibinfo
  {pages} {425} (\bibinfo {year} {1980})}\BibitemShut {NoStop}%
\bibitem [{\citenamefont {{Ward}}(1989)}]{Ward1989}%
  \BibitemOpen
  \bibfield  {author} {\bibinfo {author} {\bibfnamefont {W.~R.}\ \bibnamefont
  {{Ward}}},\ }\href {\doibase 10.1086/167031} {\bibfield  {journal} {\bibinfo
  {journal} {\apj}\ }\textbf {\bibinfo {volume} {336}},\ \bibinfo {pages} {526}
  (\bibinfo {year} {1989})}\BibitemShut {NoStop}%
\bibitem [{\citenamefont {{Tanaka}}\ \emph {et~al.}(2002)\citenamefont
  {{Tanaka}}, \citenamefont {{Takeuchi}},\ and\ \citenamefont
  {{Ward}}}]{Tanaka2002}%
  \BibitemOpen
  \bibfield  {author} {\bibinfo {author} {\bibfnamefont {H.}~\bibnamefont
  {{Tanaka}}}, \bibinfo {author} {\bibfnamefont {T.}~\bibnamefont
  {{Takeuchi}}}, \ and\ \bibinfo {author} {\bibfnamefont {W.~R.}\ \bibnamefont
  {{Ward}}},\ }\href {\doibase 10.1086/324713} {\bibfield  {journal} {\bibinfo
  {journal} {\apj}\ }\textbf {\bibinfo {volume} {565}},\ \bibinfo {pages}
  {1257} (\bibinfo {year} {2002})}\BibitemShut {NoStop}%
\bibitem [{\citenamefont {{Tanaka}}\ and\ \citenamefont
  {{Ward}}(2004)}]{Tanaka2004}%
  \BibitemOpen
  \bibfield  {author} {\bibinfo {author} {\bibfnamefont {H.}~\bibnamefont
  {{Tanaka}}}\ and\ \bibinfo {author} {\bibfnamefont {W.~R.}\ \bibnamefont
  {{Ward}}},\ }\href {\doibase 10.1086/380992} {\bibfield  {journal} {\bibinfo
  {journal} {\apj}\ }\textbf {\bibinfo {volume} {602}},\ \bibinfo {pages} {388}
  (\bibinfo {year} {2004})}\BibitemShut {NoStop}%
\bibitem [{\citenamefont {{Sirko}}\ and\ \citenamefont
  {{Goodman}}(2003)}]{Sirko2003}%
  \BibitemOpen
  \bibfield  {author} {\bibinfo {author} {\bibfnamefont {E.}~\bibnamefont
  {{Sirko}}}\ and\ \bibinfo {author} {\bibfnamefont {J.}~\bibnamefont
  {{Goodman}}},\ }\href {\doibase 10.1046/j.1365-8711.2003.06431.x} {\bibfield
  {journal} {\bibinfo  {journal} {\mnras}\ }\textbf {\bibinfo {volume} {341}},\
  \bibinfo {pages} {501} (\bibinfo {year} {2003})},\ \Eprint
  {http://arxiv.org/abs/astro-ph/0209469} {arXiv:astro-ph/0209469 [astro-ph]}
  \BibitemShut {NoStop}%
\bibitem [{\citenamefont {{Thompson}}\ \emph {et~al.}(2005)\citenamefont
  {{Thompson}}, \citenamefont {{Quataert}},\ and\ \citenamefont
  {{Murray}}}]{Thompson2005}%
  \BibitemOpen
  \bibfield  {author} {\bibinfo {author} {\bibfnamefont {T.~A.}\ \bibnamefont
  {{Thompson}}}, \bibinfo {author} {\bibfnamefont {E.}~\bibnamefont
  {{Quataert}}}, \ and\ \bibinfo {author} {\bibfnamefont {N.}~\bibnamefont
  {{Murray}}},\ }\href {\doibase 10.1086/431923} {\bibfield  {journal}
  {\bibinfo  {journal} {\apj}\ }\textbf {\bibinfo {volume} {630}},\ \bibinfo
  {pages} {167} (\bibinfo {year} {2005})},\ \Eprint
  {http://arxiv.org/abs/astro-ph/0503027} {arXiv:astro-ph/0503027 [astro-ph]}
  \BibitemShut {NoStop}%
\bibitem [{\citenamefont {{McKernan}}\ \emph {et~al.}(2012)\citenamefont
  {{McKernan}}, \citenamefont {{Ford}}, \citenamefont {{Lyra}},\ and\
  \citenamefont {{Perets}}}]{McKernan2012}%
  \BibitemOpen
  \bibfield  {author} {\bibinfo {author} {\bibfnamefont {B.}~\bibnamefont
  {{McKernan}}}, \bibinfo {author} {\bibfnamefont {K.~E.~S.}\ \bibnamefont
  {{Ford}}}, \bibinfo {author} {\bibfnamefont {W.}~\bibnamefont {{Lyra}}}, \
  and\ \bibinfo {author} {\bibfnamefont {H.~B.}\ \bibnamefont {{Perets}}},\
  }\href {\doibase 10.1111/j.1365-2966.2012.21486.x} {\bibfield  {journal}
  {\bibinfo  {journal} {\mnras}\ }\textbf {\bibinfo {volume} {425}},\ \bibinfo
  {pages} {460} (\bibinfo {year} {2012})},\ \Eprint
  {http://arxiv.org/abs/1206.2309} {arXiv:1206.2309 [astro-ph.GA]} \BibitemShut
  {NoStop}%
\bibitem [{\citenamefont {{McKernan}}\ \emph {et~al.}(2014)\citenamefont
  {{McKernan}}, \citenamefont {{Ford}}, \citenamefont {{Kocsis}}, \citenamefont
  {{Lyra}},\ and\ \citenamefont {{Winter}}}]{McKernan2014}%
  \BibitemOpen
  \bibfield  {author} {\bibinfo {author} {\bibfnamefont {B.}~\bibnamefont
  {{McKernan}}}, \bibinfo {author} {\bibfnamefont {K.~E.~S.}\ \bibnamefont
  {{Ford}}}, \bibinfo {author} {\bibfnamefont {B.}~\bibnamefont {{Kocsis}}},
  \bibinfo {author} {\bibfnamefont {W.}~\bibnamefont {{Lyra}}}, \ and\ \bibinfo
  {author} {\bibfnamefont {L.~M.}\ \bibnamefont {{Winter}}},\ }\href {\doibase
  10.1093/mnras/stu553} {\bibfield  {journal} {\bibinfo  {journal} {\mnras}\
  }\textbf {\bibinfo {volume} {441}},\ \bibinfo {pages} {900} (\bibinfo {year}
  {2014})},\ \Eprint {http://arxiv.org/abs/1403.6433} {arXiv:1403.6433
  [astro-ph.GA]} \BibitemShut {NoStop}%
\bibitem [{\citenamefont {{Stone}}\ \emph {et~al.}(2017)\citenamefont
  {{Stone}}, \citenamefont {{Metzger}},\ and\ \citenamefont
  {{Haiman}}}]{Stone2017}%
  \BibitemOpen
  \bibfield  {author} {\bibinfo {author} {\bibfnamefont {N.~C.}\ \bibnamefont
  {{Stone}}}, \bibinfo {author} {\bibfnamefont {B.~D.}\ \bibnamefont
  {{Metzger}}}, \ and\ \bibinfo {author} {\bibfnamefont {Z.}~\bibnamefont
  {{Haiman}}},\ }\href {\doibase 10.1093/mnras/stw2260} {\bibfield  {journal}
  {\bibinfo  {journal} {\mnras}\ }\textbf {\bibinfo {volume} {464}},\ \bibinfo
  {pages} {946} (\bibinfo {year} {2017})},\ \Eprint
  {http://arxiv.org/abs/1602.04226} {arXiv:1602.04226 [astro-ph.GA]}
  \BibitemShut {NoStop}%
\bibitem [{\citenamefont {{Bartos}}\ \emph {et~al.}(2017)\citenamefont
  {{Bartos}}, \citenamefont {{Kocsis}}, \citenamefont {{Haiman}},\ and\
  \citenamefont {{M{\'a}rka}}}]{Bartos2017}%
  \BibitemOpen
  \bibfield  {author} {\bibinfo {author} {\bibfnamefont {I.}~\bibnamefont
  {{Bartos}}}, \bibinfo {author} {\bibfnamefont {B.}~\bibnamefont {{Kocsis}}},
  \bibinfo {author} {\bibfnamefont {Z.}~\bibnamefont {{Haiman}}}, \ and\
  \bibinfo {author} {\bibfnamefont {S.}~\bibnamefont {{M{\'a}rka}}},\ }\href
  {\doibase 10.3847/1538-4357/835/2/165} {\bibfield  {journal} {\bibinfo
  {journal} {\apj}\ }\textbf {\bibinfo {volume} {835}},\ \bibinfo {eid} {165}
  (\bibinfo {year} {2017})},\ \Eprint {http://arxiv.org/abs/1602.03831}
  {arXiv:1602.03831 [astro-ph.HE]} \BibitemShut {NoStop}%
\bibitem [{\citenamefont {{McKernan}}\ \emph {et~al.}(2018)\citenamefont
  {{McKernan}}, \citenamefont {{Ford}}, \citenamefont {{Bellovary}},
  \citenamefont {{Leigh}}, \citenamefont {{Haiman}}, \citenamefont {{Kocsis}},
  \citenamefont {{Lyra}}, \citenamefont {{Mac Low}}, \citenamefont {{Metzger}},
  \citenamefont {{O'Dowd}}, \citenamefont {{Endlich}},\ and\ \citenamefont
  {{Rosen}}}]{McKernan2018}%
  \BibitemOpen
  \bibfield  {author} {\bibinfo {author} {\bibfnamefont {B.}~\bibnamefont
  {{McKernan}}}, \bibinfo {author} {\bibfnamefont {K.~E.~S.}\ \bibnamefont
  {{Ford}}}, \bibinfo {author} {\bibfnamefont {J.}~\bibnamefont {{Bellovary}}},
  \bibinfo {author} {\bibfnamefont {N.~W.~C.}\ \bibnamefont {{Leigh}}},
  \bibinfo {author} {\bibfnamefont {Z.}~\bibnamefont {{Haiman}}}, \bibinfo
  {author} {\bibfnamefont {B.}~\bibnamefont {{Kocsis}}}, \bibinfo {author}
  {\bibfnamefont {W.}~\bibnamefont {{Lyra}}}, \bibinfo {author} {\bibfnamefont
  {M.~M.}\ \bibnamefont {{Mac Low}}}, \bibinfo {author} {\bibfnamefont
  {B.}~\bibnamefont {{Metzger}}}, \bibinfo {author} {\bibfnamefont
  {M.}~\bibnamefont {{O'Dowd}}}, \bibinfo {author} {\bibfnamefont
  {S.}~\bibnamefont {{Endlich}}}, \ and\ \bibinfo {author} {\bibfnamefont
  {D.~J.}\ \bibnamefont {{Rosen}}},\ }\href {\doibase 10.3847/1538-4357/aadae5}
  {\bibfield  {journal} {\bibinfo  {journal} {\apj}\ }\textbf {\bibinfo
  {volume} {866}},\ \bibinfo {eid} {66} (\bibinfo {year} {2018})},\ \Eprint
  {http://arxiv.org/abs/1702.07818} {arXiv:1702.07818 [astro-ph.HE]}
  \BibitemShut {NoStop}%
\bibitem [{\citenamefont {{Leigh}}\ \emph {et~al.}(2018)\citenamefont
  {{Leigh}}, \citenamefont {{Geller}}, \citenamefont {{McKernan}},
  \citenamefont {{Ford}}, \citenamefont {{Mac Low}}, \citenamefont
  {{Bellovary}}, \citenamefont {{Haiman}}, \citenamefont {{Lyra}},
  \citenamefont {{Samsing}}, \citenamefont {{O'Dowd}}, \citenamefont
  {{Kocsis}},\ and\ \citenamefont {{Endlich}}}]{Leigh2018}%
  \BibitemOpen
  \bibfield  {author} {\bibinfo {author} {\bibfnamefont {N.~W.~C.}\
  \bibnamefont {{Leigh}}}, \bibinfo {author} {\bibfnamefont {A.~M.}\
  \bibnamefont {{Geller}}}, \bibinfo {author} {\bibfnamefont {B.}~\bibnamefont
  {{McKernan}}}, \bibinfo {author} {\bibfnamefont {K.~E.~S.}\ \bibnamefont
  {{Ford}}}, \bibinfo {author} {\bibfnamefont {M.~M.}\ \bibnamefont {{Mac
  Low}}}, \bibinfo {author} {\bibfnamefont {J.}~\bibnamefont {{Bellovary}}},
  \bibinfo {author} {\bibfnamefont {Z.}~\bibnamefont {{Haiman}}}, \bibinfo
  {author} {\bibfnamefont {W.}~\bibnamefont {{Lyra}}}, \bibinfo {author}
  {\bibfnamefont {J.}~\bibnamefont {{Samsing}}}, \bibinfo {author}
  {\bibfnamefont {M.}~\bibnamefont {{O'Dowd}}}, \bibinfo {author}
  {\bibfnamefont {B.}~\bibnamefont {{Kocsis}}}, \ and\ \bibinfo {author}
  {\bibfnamefont {S.}~\bibnamefont {{Endlich}}},\ }\href {\doibase
  10.1093/mnras/stx3134} {\bibfield  {journal} {\bibinfo  {journal} {\mnras}\
  }\textbf {\bibinfo {volume} {474}},\ \bibinfo {pages} {5672} (\bibinfo {year}
  {2018})},\ \Eprint {http://arxiv.org/abs/1711.10494} {arXiv:1711.10494
  [astro-ph.GA]} \BibitemShut {NoStop}%
\bibitem [{\citenamefont {{Yang}}\ \emph
  {et~al.}(2019{\natexlab{a}})\citenamefont {{Yang}}, \citenamefont {{Bartos}},
  \citenamefont {{Gayathri}}, \citenamefont {{Ford}}, \citenamefont {{Haiman}},
  \citenamefont {{Klimenko}}, \citenamefont {{Kocsis}}, \citenamefont
  {{M{\'a}rka}}, \citenamefont {{M{\'a}rka}}, \citenamefont {{McKernan}},\ and\
  \citenamefont {{O'Shaughnessy}}}]{Yang2019prl}%
  \BibitemOpen
  \bibfield  {author} {\bibinfo {author} {\bibfnamefont {Y.}~\bibnamefont
  {{Yang}}}, \bibinfo {author} {\bibfnamefont {I.}~\bibnamefont {{Bartos}}},
  \bibinfo {author} {\bibfnamefont {V.}~\bibnamefont {{Gayathri}}}, \bibinfo
  {author} {\bibfnamefont {K.~E.~S.}\ \bibnamefont {{Ford}}}, \bibinfo {author}
  {\bibfnamefont {Z.}~\bibnamefont {{Haiman}}}, \bibinfo {author}
  {\bibfnamefont {S.}~\bibnamefont {{Klimenko}}}, \bibinfo {author}
  {\bibfnamefont {B.}~\bibnamefont {{Kocsis}}}, \bibinfo {author}
  {\bibfnamefont {S.}~\bibnamefont {{M{\'a}rka}}}, \bibinfo {author}
  {\bibfnamefont {Z.}~\bibnamefont {{M{\'a}rka}}}, \bibinfo {author}
  {\bibfnamefont {B.}~\bibnamefont {{McKernan}}}, \ and\ \bibinfo {author}
  {\bibfnamefont {R.}~\bibnamefont {{O'Shaughnessy}}},\ }\href {\doibase
  10.1103/PhysRevLett.123.181101} {\bibfield  {journal} {\bibinfo  {journal}
  {\prl}\ }\textbf {\bibinfo {volume} {123}},\ \bibinfo {eid} {181101}
  (\bibinfo {year} {2019}{\natexlab{a}})},\ \Eprint
  {http://arxiv.org/abs/1906.09281} {arXiv:1906.09281 [astro-ph.HE]}
  \BibitemShut {NoStop}%
\bibitem [{\citenamefont {{Yang}}\ \emph
  {et~al.}(2019{\natexlab{b}})\citenamefont {{Yang}}, \citenamefont {{Bartos}},
  \citenamefont {{Haiman}}, \citenamefont {{Kocsis}}, \citenamefont
  {{M{\'a}rka}}, \citenamefont {{Stone}},\ and\ \citenamefont
  {{M{\'a}rka}}}]{Yang2019}%
  \BibitemOpen
  \bibfield  {author} {\bibinfo {author} {\bibfnamefont {Y.}~\bibnamefont
  {{Yang}}}, \bibinfo {author} {\bibfnamefont {I.}~\bibnamefont {{Bartos}}},
  \bibinfo {author} {\bibfnamefont {Z.}~\bibnamefont {{Haiman}}}, \bibinfo
  {author} {\bibfnamefont {B.}~\bibnamefont {{Kocsis}}}, \bibinfo {author}
  {\bibfnamefont {Z.}~\bibnamefont {{M{\'a}rka}}}, \bibinfo {author}
  {\bibfnamefont {N.~C.}\ \bibnamefont {{Stone}}}, \ and\ \bibinfo {author}
  {\bibfnamefont {S.}~\bibnamefont {{M{\'a}rka}}},\ }\href {\doibase
  10.3847/1538-4357/ab16e3} {\bibfield  {journal} {\bibinfo  {journal} {\apj}\
  }\textbf {\bibinfo {volume} {876}},\ \bibinfo {eid} {122} (\bibinfo {year}
  {2019}{\natexlab{b}})},\ \Eprint {http://arxiv.org/abs/1903.01405}
  {arXiv:1903.01405 [astro-ph.HE]} \BibitemShut {NoStop}%
\bibitem [{\citenamefont {{Secunda}}\ \emph {et~al.}(2019)\citenamefont
  {{Secunda}}, \citenamefont {{Bellovary}}, \citenamefont {{Mac Low}},
  \citenamefont {{Ford}}, \citenamefont {{McKernan}}, \citenamefont {{Leigh}},
  \citenamefont {{Lyra}},\ and\ \citenamefont {{S{\'a}ndor}}}]{Secunda2019}%
  \BibitemOpen
  \bibfield  {author} {\bibinfo {author} {\bibfnamefont {A.}~\bibnamefont
  {{Secunda}}}, \bibinfo {author} {\bibfnamefont {J.}~\bibnamefont
  {{Bellovary}}}, \bibinfo {author} {\bibfnamefont {M.-M.}\ \bibnamefont {{Mac
  Low}}}, \bibinfo {author} {\bibfnamefont {K.~E.~S.}\ \bibnamefont {{Ford}}},
  \bibinfo {author} {\bibfnamefont {B.}~\bibnamefont {{McKernan}}}, \bibinfo
  {author} {\bibfnamefont {N.~W.~C.}\ \bibnamefont {{Leigh}}}, \bibinfo
  {author} {\bibfnamefont {W.}~\bibnamefont {{Lyra}}}, \ and\ \bibinfo {author}
  {\bibfnamefont {Z.}~\bibnamefont {{S{\'a}ndor}}},\ }\href {\doibase
  10.3847/1538-4357/ab20ca} {\bibfield  {journal} {\bibinfo  {journal} {\apj}\
  }\textbf {\bibinfo {volume} {878}},\ \bibinfo {eid} {85} (\bibinfo {year}
  {2019})},\ \Eprint {http://arxiv.org/abs/1807.02859} {arXiv:1807.02859
  [astro-ph.HE]} \BibitemShut {NoStop}%
\bibitem [{\citenamefont {{Secunda}}\ \emph {et~al.}(2020)\citenamefont
  {{Secunda}}, \citenamefont {{Bellovary}}, \citenamefont {{Mac Low}},
  \citenamefont {{Ford}}, \citenamefont {{McKernan}}, \citenamefont {{Leigh}},
  \citenamefont {{Lyra}}, \citenamefont {{S{\'a}ndor}},\ and\ \citenamefont
  {{Adorno}}}]{Secunda2020}%
  \BibitemOpen
  \bibfield  {author} {\bibinfo {author} {\bibfnamefont {A.}~\bibnamefont
  {{Secunda}}}, \bibinfo {author} {\bibfnamefont {J.}~\bibnamefont
  {{Bellovary}}}, \bibinfo {author} {\bibfnamefont {M.-M.}\ \bibnamefont {{Mac
  Low}}}, \bibinfo {author} {\bibfnamefont {K.~E.~S.}\ \bibnamefont {{Ford}}},
  \bibinfo {author} {\bibfnamefont {B.}~\bibnamefont {{McKernan}}}, \bibinfo
  {author} {\bibfnamefont {N.~W.~C.}\ \bibnamefont {{Leigh}}}, \bibinfo
  {author} {\bibfnamefont {W.}~\bibnamefont {{Lyra}}}, \bibinfo {author}
  {\bibfnamefont {Z.}~\bibnamefont {{S{\'a}ndor}}}, \ and\ \bibinfo {author}
  {\bibfnamefont {J.~I.}\ \bibnamefont {{Adorno}}},\ }\href {\doibase
  10.3847/1538-4357/abbc1d} {\bibfield  {journal} {\bibinfo  {journal} {\apj}\
  }\textbf {\bibinfo {volume} {903}},\ \bibinfo {eid} {133} (\bibinfo {year}
  {2020})},\ \Eprint {http://arxiv.org/abs/2004.11936} {arXiv:2004.11936
  [astro-ph.HE]} \BibitemShut {NoStop}%
\bibitem [{\citenamefont {{McKernan}}\ \emph
  {et~al.}(2020{\natexlab{a}})\citenamefont {{McKernan}}, \citenamefont
  {{Ford}},\ and\ \citenamefont {{O'Shaughnessy}}}]{McKernan2020}%
  \BibitemOpen
  \bibfield  {author} {\bibinfo {author} {\bibfnamefont {B.}~\bibnamefont
  {{McKernan}}}, \bibinfo {author} {\bibfnamefont {K.~E.~S.}\ \bibnamefont
  {{Ford}}}, \ and\ \bibinfo {author} {\bibfnamefont {R.}~\bibnamefont
  {{O'Shaughnessy}}},\ }\href {\doibase 10.1093/mnras/staa2681} {\bibfield
  {journal} {\bibinfo  {journal} {\mnras}\ }\textbf {\bibinfo {volume} {498}},\
  \bibinfo {pages} {4088} (\bibinfo {year} {2020}{\natexlab{a}})},\ \Eprint
  {http://arxiv.org/abs/2002.00046} {arXiv:2002.00046 [astro-ph.HE]}
  \BibitemShut {NoStop}%
\bibitem [{\citenamefont {{McKernan}}\ \emph
  {et~al.}(2020{\natexlab{b}})\citenamefont {{McKernan}}, \citenamefont
  {{Ford}}, \citenamefont {{O'Shaugnessy}},\ and\ \citenamefont
  {{Wysocki}}}]{McKernan2020b}%
  \BibitemOpen
  \bibfield  {author} {\bibinfo {author} {\bibfnamefont {B.}~\bibnamefont
  {{McKernan}}}, \bibinfo {author} {\bibfnamefont {K.~E.~S.}\ \bibnamefont
  {{Ford}}}, \bibinfo {author} {\bibfnamefont {R.}~\bibnamefont
  {{O'Shaugnessy}}}, \ and\ \bibinfo {author} {\bibfnamefont {D.}~\bibnamefont
  {{Wysocki}}},\ }\href {\doibase 10.1093/mnras/staa740} {\bibfield  {journal}
  {\bibinfo  {journal} {\mnras}\ }\textbf {\bibinfo {volume} {494}},\ \bibinfo
  {pages} {1203} (\bibinfo {year} {2020}{\natexlab{b}})},\ \Eprint
  {http://arxiv.org/abs/1907.04356} {arXiv:1907.04356 [astro-ph.HE]}
  \BibitemShut {NoStop}%
\bibitem [{\citenamefont {Graham}\ \emph {et~al.}(2020)\citenamefont {Graham}
  \emph {et~al.}}]{Graham:2020gwr}%
  \BibitemOpen
  \bibfield  {author} {\bibinfo {author} {\bibfnamefont {M.~J.}\ \bibnamefont
  {Graham}} \emph {et~al.},\ }\href {\doibase 10.1103/PhysRevLett.124.251102}
  {\bibfield  {journal} {\bibinfo  {journal} {Phys. Rev. Lett.}\ }\textbf
  {\bibinfo {volume} {124}},\ \bibinfo {pages} {251102} (\bibinfo {year}
  {2020})},\ \Eprint {http://arxiv.org/abs/2006.14122} {arXiv:2006.14122
  [astro-ph.HE]} \BibitemShut {NoStop}%
\bibitem [{\citenamefont {{Tagawa}}\ \emph
  {et~al.}(2020{\natexlab{a}})\citenamefont {{Tagawa}}, \citenamefont
  {{Haiman}},\ and\ \citenamefont {{Kocsis}}}]{Tagawa2020}%
  \BibitemOpen
  \bibfield  {author} {\bibinfo {author} {\bibfnamefont {H.}~\bibnamefont
  {{Tagawa}}}, \bibinfo {author} {\bibfnamefont {Z.}~\bibnamefont {{Haiman}}},
  \ and\ \bibinfo {author} {\bibfnamefont {B.}~\bibnamefont {{Kocsis}}},\
  }\href {\doibase 10.3847/1538-4357/ab9b8c} {\bibfield  {journal} {\bibinfo
  {journal} {\apj}\ }\textbf {\bibinfo {volume} {898}},\ \bibinfo {eid} {25}
  (\bibinfo {year} {2020}{\natexlab{a}})},\ \Eprint
  {http://arxiv.org/abs/1912.08218} {arXiv:1912.08218 [astro-ph.GA]}
  \BibitemShut {NoStop}%
\bibitem [{\citenamefont {{Tagawa}}\ \emph
  {et~al.}(2020{\natexlab{b}})\citenamefont {{Tagawa}}, \citenamefont
  {{Haiman}}, \citenamefont {{Bartos}},\ and\ \citenamefont
  {{Kocsis}}}]{Tagawa2020b}%
  \BibitemOpen
  \bibfield  {author} {\bibinfo {author} {\bibfnamefont {H.}~\bibnamefont
  {{Tagawa}}}, \bibinfo {author} {\bibfnamefont {Z.}~\bibnamefont {{Haiman}}},
  \bibinfo {author} {\bibfnamefont {I.}~\bibnamefont {{Bartos}}}, \ and\
  \bibinfo {author} {\bibfnamefont {B.}~\bibnamefont {{Kocsis}}},\ }\href
  {\doibase 10.3847/1538-4357/aba2cc} {\bibfield  {journal} {\bibinfo
  {journal} {\apj}\ }\textbf {\bibinfo {volume} {899}},\ \bibinfo {eid} {26}
  (\bibinfo {year} {2020}{\natexlab{b}})},\ \Eprint
  {http://arxiv.org/abs/2004.11914} {arXiv:2004.11914 [astro-ph.HE]}
  \BibitemShut {NoStop}%
\bibitem [{\citenamefont {{Tagawa}}\ \emph {et~al.}(2021)\citenamefont
  {{Tagawa}}, \citenamefont {{Kocsis}}, \citenamefont {{Haiman}}, \citenamefont
  {{Bartos}}, \citenamefont {{Omukai}},\ and\ \citenamefont
  {{Samsing}}}]{Tagawa2021}%
  \BibitemOpen
  \bibfield  {author} {\bibinfo {author} {\bibfnamefont {H.}~\bibnamefont
  {{Tagawa}}}, \bibinfo {author} {\bibfnamefont {B.}~\bibnamefont {{Kocsis}}},
  \bibinfo {author} {\bibfnamefont {Z.}~\bibnamefont {{Haiman}}}, \bibinfo
  {author} {\bibfnamefont {I.}~\bibnamefont {{Bartos}}}, \bibinfo {author}
  {\bibfnamefont {K.}~\bibnamefont {{Omukai}}}, \ and\ \bibinfo {author}
  {\bibfnamefont {J.}~\bibnamefont {{Samsing}}},\ }\href {\doibase
  10.3847/2041-8213/abd4d3} {\bibfield  {journal} {\bibinfo  {journal} {\apjl}\
  }\textbf {\bibinfo {volume} {907}},\ \bibinfo {eid} {L20} (\bibinfo {year}
  {2021})},\ \Eprint {http://arxiv.org/abs/2010.10526} {arXiv:2010.10526
  [astro-ph.HE]} \BibitemShut {NoStop}%
\bibitem [{\citenamefont {{Tagawa}}\ \emph
  {et~al.}(2020{\natexlab{c}})\citenamefont {{Tagawa}}, \citenamefont
  {{Kocsis}}, \citenamefont {{Haiman}}, \citenamefont {{Bartos}}, \citenamefont
  {{Omukai}},\ and\ \citenamefont {{Samsing}}}]{Tagawa2020arXiv}%
  \BibitemOpen
  \bibfield  {author} {\bibinfo {author} {\bibfnamefont {H.}~\bibnamefont
  {{Tagawa}}}, \bibinfo {author} {\bibfnamefont {B.}~\bibnamefont {{Kocsis}}},
  \bibinfo {author} {\bibfnamefont {Z.}~\bibnamefont {{Haiman}}}, \bibinfo
  {author} {\bibfnamefont {I.}~\bibnamefont {{Bartos}}}, \bibinfo {author}
  {\bibfnamefont {K.}~\bibnamefont {{Omukai}}}, \ and\ \bibinfo {author}
  {\bibfnamefont {J.}~\bibnamefont {{Samsing}}},\ }\href@noop {} {\bibfield
  {journal} {\bibinfo  {journal} {arXiv e-prints}\ ,\ \bibinfo {eid}
  {arXiv:2012.00011}} (\bibinfo {year} {2020}{\natexlab{c}})},\ \Eprint
  {http://arxiv.org/abs/2012.00011} {arXiv:2012.00011 [astro-ph.HE]}
  \BibitemShut {NoStop}%
\bibitem [{\citenamefont {{Kocsis}}\ \emph {et~al.}(2011)\citenamefont
  {{Kocsis}}, \citenamefont {{Yunes}},\ and\ \citenamefont
  {{Loeb}}}]{Kocsis2011}%
  \BibitemOpen
  \bibfield  {author} {\bibinfo {author} {\bibfnamefont {B.}~\bibnamefont
  {{Kocsis}}}, \bibinfo {author} {\bibfnamefont {N.}~\bibnamefont {{Yunes}}}, \
  and\ \bibinfo {author} {\bibfnamefont {A.}~\bibnamefont {{Loeb}}},\ }\href
  {\doibase 10.1103/PhysRevD.84.024032} {\bibfield  {journal} {\bibinfo
  {journal} {\prd}\ }\textbf {\bibinfo {volume} {84}},\ \bibinfo {eid} {024032}
  (\bibinfo {year} {2011})},\ \Eprint {http://arxiv.org/abs/1104.2322}
  {arXiv:1104.2322 [astro-ph.GA]} \BibitemShut {NoStop}%
\bibitem [{\citenamefont {{Tremaine}}\ \emph {et~al.}(1994)\citenamefont
  {{Tremaine}}, \citenamefont {{Richstone}}, \citenamefont {{Byun}},
  \citenamefont {{Dressler}}, \citenamefont {{Faber}}, \citenamefont
  {{Grillmair}}, \citenamefont {{Kormendy}},\ and\ \citenamefont
  {{Lauer}}}]{Tremaine1994}%
  \BibitemOpen
  \bibfield  {author} {\bibinfo {author} {\bibfnamefont {S.}~\bibnamefont
  {{Tremaine}}}, \bibinfo {author} {\bibfnamefont {D.~O.}\ \bibnamefont
  {{Richstone}}}, \bibinfo {author} {\bibfnamefont {Y.-I.}\ \bibnamefont
  {{Byun}}}, \bibinfo {author} {\bibfnamefont {A.}~\bibnamefont {{Dressler}}},
  \bibinfo {author} {\bibfnamefont {S.~M.}\ \bibnamefont {{Faber}}}, \bibinfo
  {author} {\bibfnamefont {C.}~\bibnamefont {{Grillmair}}}, \bibinfo {author}
  {\bibfnamefont {J.}~\bibnamefont {{Kormendy}}}, \ and\ \bibinfo {author}
  {\bibfnamefont {T.~R.}\ \bibnamefont {{Lauer}}},\ }\href {\doibase
  10.1086/116883} {\bibfield  {journal} {\bibinfo  {journal} {\aj}\ }\textbf
  {\bibinfo {volume} {107}},\ \bibinfo {pages} {634} (\bibinfo {year}
  {1994})},\ \Eprint {http://arxiv.org/abs/astro-ph/9309044}
  {arXiv:astro-ph/9309044 [astro-ph]} \BibitemShut {NoStop}%
\bibitem [{\citenamefont {{Galametz}}\ \emph {et~al.}(2009)\citenamefont
  {{Galametz}}, \citenamefont {{Stern}}, \citenamefont {{Eisenhardt}},
  \citenamefont {{Brodwin}}, \citenamefont {{Brown}}, \citenamefont {{Dey}},
  \citenamefont {{Gonzalez}}, \citenamefont {{Jannuzi}}, \citenamefont
  {{Moustakas}},\ and\ \citenamefont {{Stanford}}}]{Galametz2009}%
  \BibitemOpen
  \bibfield  {author} {\bibinfo {author} {\bibfnamefont {A.}~\bibnamefont
  {{Galametz}}}, \bibinfo {author} {\bibfnamefont {D.}~\bibnamefont {{Stern}}},
  \bibinfo {author} {\bibfnamefont {P.~R.~M.}\ \bibnamefont {{Eisenhardt}}},
  \bibinfo {author} {\bibfnamefont {M.}~\bibnamefont {{Brodwin}}}, \bibinfo
  {author} {\bibfnamefont {M.~J.~I.}\ \bibnamefont {{Brown}}}, \bibinfo
  {author} {\bibfnamefont {A.}~\bibnamefont {{Dey}}}, \bibinfo {author}
  {\bibfnamefont {A.~H.}\ \bibnamefont {{Gonzalez}}}, \bibinfo {author}
  {\bibfnamefont {B.~T.}\ \bibnamefont {{Jannuzi}}}, \bibinfo {author}
  {\bibfnamefont {L.~A.}\ \bibnamefont {{Moustakas}}}, \ and\ \bibinfo {author}
  {\bibfnamefont {S.~A.}\ \bibnamefont {{Stanford}}},\ }\href {\doibase
  10.1088/0004-637X/694/2/1309} {\bibfield  {journal} {\bibinfo  {journal}
  {\apj}\ }\textbf {\bibinfo {volume} {694}},\ \bibinfo {pages} {1309}
  (\bibinfo {year} {2009})},\ \Eprint {http://arxiv.org/abs/0901.1109}
  {arXiv:0901.1109 [astro-ph.CO]} \BibitemShut {NoStop}%
\bibitem [{\citenamefont {{Macuga}}\ \emph {et~al.}(2019)\citenamefont
  {{Macuga}}, \citenamefont {{Martini}}, \citenamefont {{Miller}},
  \citenamefont {{Brodwin}}, \citenamefont {{Hayashi}}, \citenamefont
  {{Kodama}}, \citenamefont {{Koyama}}, \citenamefont {{Overzier}},
  \citenamefont {{Shimakawa}}, \citenamefont {{Tadaki}},\ and\ \citenamefont
  {{Tanaka}}}]{Macuga2019}%
  \BibitemOpen
  \bibfield  {author} {\bibinfo {author} {\bibfnamefont {M.}~\bibnamefont
  {{Macuga}}}, \bibinfo {author} {\bibfnamefont {P.}~\bibnamefont {{Martini}}},
  \bibinfo {author} {\bibfnamefont {E.~D.}\ \bibnamefont {{Miller}}}, \bibinfo
  {author} {\bibfnamefont {M.}~\bibnamefont {{Brodwin}}}, \bibinfo {author}
  {\bibfnamefont {M.}~\bibnamefont {{Hayashi}}}, \bibinfo {author}
  {\bibfnamefont {T.}~\bibnamefont {{Kodama}}}, \bibinfo {author}
  {\bibfnamefont {Y.}~\bibnamefont {{Koyama}}}, \bibinfo {author}
  {\bibfnamefont {R.~A.}\ \bibnamefont {{Overzier}}}, \bibinfo {author}
  {\bibfnamefont {R.}~\bibnamefont {{Shimakawa}}}, \bibinfo {author}
  {\bibfnamefont {K.-i.}\ \bibnamefont {{Tadaki}}}, \ and\ \bibinfo {author}
  {\bibfnamefont {I.}~\bibnamefont {{Tanaka}}},\ }\href {\doibase
  10.3847/1538-4357/ab0746} {\bibfield  {journal} {\bibinfo  {journal} {\apj}\
  }\textbf {\bibinfo {volume} {874}},\ \bibinfo {eid} {54} (\bibinfo {year}
  {2019})},\ \Eprint {http://arxiv.org/abs/1805.06569} {arXiv:1805.06569
  [astro-ph.GA]} \BibitemShut {NoStop}%
\bibitem [{\citenamefont {{Yunes}}\ \emph {et~al.}(2011)\citenamefont
  {{Yunes}}, \citenamefont {{Kocsis}}, \citenamefont {{Loeb}},\ and\
  \citenamefont {{Haiman}}}]{Yunes2011}%
  \BibitemOpen
  \bibfield  {author} {\bibinfo {author} {\bibfnamefont {N.}~\bibnamefont
  {{Yunes}}}, \bibinfo {author} {\bibfnamefont {B.}~\bibnamefont {{Kocsis}}},
  \bibinfo {author} {\bibfnamefont {A.}~\bibnamefont {{Loeb}}}, \ and\ \bibinfo
  {author} {\bibfnamefont {Z.}~\bibnamefont {{Haiman}}},\ }\href {\doibase
  10.1103/PhysRevLett.107.171103} {\bibfield  {journal} {\bibinfo  {journal}
  {\prl}\ }\textbf {\bibinfo {volume} {107}},\ \bibinfo {eid} {171103}
  (\bibinfo {year} {2011})},\ \Eprint {http://arxiv.org/abs/1103.4609}
  {arXiv:1103.4609 [astro-ph.CO]} \BibitemShut {NoStop}%
\bibitem [{\citenamefont {{Derdzinski}}\ \emph {et~al.}(2021)\citenamefont
  {{Derdzinski}}, \citenamefont {{D'Orazio}}, \citenamefont {{Duffell}},
  \citenamefont {{Haiman}},\ and\ \citenamefont
  {{MacFadyen}}}]{Derdzinski2021}%
  \BibitemOpen
  \bibfield  {author} {\bibinfo {author} {\bibfnamefont {A.}~\bibnamefont
  {{Derdzinski}}}, \bibinfo {author} {\bibfnamefont {D.}~\bibnamefont
  {{D'Orazio}}}, \bibinfo {author} {\bibfnamefont {P.}~\bibnamefont
  {{Duffell}}}, \bibinfo {author} {\bibfnamefont {Z.}~\bibnamefont {{Haiman}}},
  \ and\ \bibinfo {author} {\bibfnamefont {A.}~\bibnamefont {{MacFadyen}}},\
  }\href {\doibase 10.1093/mnras/staa3976} {\bibfield  {journal} {\bibinfo
  {journal} {\mnras}\ }\textbf {\bibinfo {volume} {501}},\ \bibinfo {pages}
  {3540} (\bibinfo {year} {2021})},\ \Eprint {http://arxiv.org/abs/2005.11333}
  {arXiv:2005.11333 [astro-ph.HE]} \BibitemShut {NoStop}%
\bibitem [{\citenamefont {{McGee}}\ \emph {et~al.}(2020)\citenamefont
  {{McGee}}, \citenamefont {{Sesana}},\ and\ \citenamefont
  {{Vecchio}}}]{McGee2020}%
  \BibitemOpen
  \bibfield  {author} {\bibinfo {author} {\bibfnamefont {S.}~\bibnamefont
  {{McGee}}}, \bibinfo {author} {\bibfnamefont {A.}~\bibnamefont {{Sesana}}}, \
  and\ \bibinfo {author} {\bibfnamefont {A.}~\bibnamefont {{Vecchio}}},\ }\href
  {\doibase 10.1038/s41550-019-0969-7} {\bibfield  {journal} {\bibinfo
  {journal} {Nature Astronomy}\ }\textbf {\bibinfo {volume} {4}},\ \bibinfo
  {pages} {26} (\bibinfo {year} {2020})},\ \Eprint
  {http://arxiv.org/abs/1811.00050} {arXiv:1811.00050 [astro-ph.HE]}
  \BibitemShut {NoStop}%
\bibitem [{\citenamefont {{Shapiro}}\ and\ \citenamefont
  {{Marchant}}(1978)}]{Shapiro1978}%
  \BibitemOpen
  \bibfield  {author} {\bibinfo {author} {\bibfnamefont {S.~L.}\ \bibnamefont
  {{Shapiro}}}\ and\ \bibinfo {author} {\bibfnamefont {A.~B.}\ \bibnamefont
  {{Marchant}}},\ }\href {\doibase 10.1086/156521} {\bibfield  {journal}
  {\bibinfo  {journal} {\apj}\ }\textbf {\bibinfo {volume} {225}},\ \bibinfo
  {pages} {603} (\bibinfo {year} {1978})}\BibitemShut {NoStop}%
\bibitem [{\citenamefont {{Hopman}}\ and\ \citenamefont
  {{Alexander}}(2005)}]{Hopman2005}%
  \BibitemOpen
  \bibfield  {author} {\bibinfo {author} {\bibfnamefont {C.}~\bibnamefont
  {{Hopman}}}\ and\ \bibinfo {author} {\bibfnamefont {T.}~\bibnamefont
  {{Alexander}}},\ }\href {\doibase 10.1086/431475} {\bibfield  {journal}
  {\bibinfo  {journal} {\apj}\ }\textbf {\bibinfo {volume} {629}},\ \bibinfo
  {pages} {362} (\bibinfo {year} {2005})},\ \Eprint
  {http://arxiv.org/abs/astro-ph/0503672} {arXiv:astro-ph/0503672 [astro-ph]}
  \BibitemShut {NoStop}%
\bibitem [{\citenamefont {{Bar-Or}}\ and\ \citenamefont
  {{Alexander}}(2016)}]{Bar-Or2016}%
  \BibitemOpen
  \bibfield  {author} {\bibinfo {author} {\bibfnamefont {B.}~\bibnamefont
  {{Bar-Or}}}\ and\ \bibinfo {author} {\bibfnamefont {T.}~\bibnamefont
  {{Alexander}}},\ }\href {\doibase 10.3847/0004-637X/820/2/129} {\bibfield
  {journal} {\bibinfo  {journal} {\apj}\ }\textbf {\bibinfo {volume} {820}},\
  \bibinfo {eid} {129} (\bibinfo {year} {2016})},\ \Eprint
  {http://arxiv.org/abs/1508.01390} {arXiv:1508.01390 [astro-ph.GA]}
  \BibitemShut {NoStop}%
\bibitem [{\citenamefont {{Cutler}}\ \emph {et~al.}(1994)\citenamefont
  {{Cutler}}, \citenamefont {{Kennefick}},\ and\ \citenamefont
  {{Poisson}}}]{Cutler1994}%
  \BibitemOpen
  \bibfield  {author} {\bibinfo {author} {\bibfnamefont {C.}~\bibnamefont
  {{Cutler}}}, \bibinfo {author} {\bibfnamefont {D.}~\bibnamefont
  {{Kennefick}}}, \ and\ \bibinfo {author} {\bibfnamefont {E.}~\bibnamefont
  {{Poisson}}},\ }\href {\doibase 10.1103/PhysRevD.50.3816} {\bibfield
  {journal} {\bibinfo  {journal} {\prd}\ }\textbf {\bibinfo {volume} {50}},\
  \bibinfo {pages} {3816} (\bibinfo {year} {1994})}\BibitemShut {NoStop}%
\bibitem [{\citenamefont {{Lightman}}\ and\ \citenamefont
  {{Shapiro}}(1977)}]{Lightman1977}%
  \BibitemOpen
  \bibfield  {author} {\bibinfo {author} {\bibfnamefont {A.~P.}\ \bibnamefont
  {{Lightman}}}\ and\ \bibinfo {author} {\bibfnamefont {S.~L.}\ \bibnamefont
  {{Shapiro}}},\ }\href {\doibase 10.1086/154925} {\bibfield  {journal}
  {\bibinfo  {journal} {\apj}\ }\textbf {\bibinfo {volume} {211}},\ \bibinfo
  {pages} {244} (\bibinfo {year} {1977})}\BibitemShut {NoStop}%
\bibitem [{Note1()}]{Note1}%
  \BibitemOpen
  \bibinfo {note} {A star orbiting around a MBH will be disrupted as long as
  its periapsis is $r_p \lesssim r_{\protect \rm star}(M_\bullet /m_{\protect
  \rm star})^{1/3}$, where $r_{\protect \rm star}\sim 10^6$ km is the typical
  star radius. As a result, we find $J_{\protect \rm lc, star}(E\simeq 0) =
  M_\bullet \protect \sqrt {a(1-e^2)/M_\bullet } =M_\bullet \protect \sqrt
  {r_p(1+e)/M_\bullet } \simeq M_\bullet \protect \sqrt {2r_p/M_\bullet }$,
  where we have used Eq.~(\ref {eq:EJ}) in the first equality, the relation
  $r_p=a(1-e)$ in the second and $e\simeq 1$ in the third.}\BibitemShut {Stop}%
\bibitem [{\citenamefont {{Cohn}}\ and\ \citenamefont
  {{Kulsrud}}(1978)}]{Cohn1978}%
  \BibitemOpen
  \bibfield  {author} {\bibinfo {author} {\bibfnamefont {H.}~\bibnamefont
  {{Cohn}}}\ and\ \bibinfo {author} {\bibfnamefont {R.~M.}\ \bibnamefont
  {{Kulsrud}}},\ }\href {\doibase 10.1086/156685} {\bibfield  {journal}
  {\bibinfo  {journal} {\apj}\ }\textbf {\bibinfo {volume} {226}},\ \bibinfo
  {pages} {1087} (\bibinfo {year} {1978})}\BibitemShut {NoStop}%
\bibitem [{\citenamefont {{Cohn}}(1979)}]{Cohn1979}%
  \BibitemOpen
  \bibfield  {author} {\bibinfo {author} {\bibfnamefont {H.}~\bibnamefont
  {{Cohn}}},\ }\href {\doibase 10.1086/157587} {\bibfield  {journal} {\bibinfo
  {journal} {\apj}\ }\textbf {\bibinfo {volume} {234}},\ \bibinfo {pages}
  {1036} (\bibinfo {year} {1979})}\BibitemShut {NoStop}%
\bibitem [{\citenamefont {{Dehnen}}(1993)}]{Dehnen1993}%
  \BibitemOpen
  \bibfield  {author} {\bibinfo {author} {\bibfnamefont {W.}~\bibnamefont
  {{Dehnen}}},\ }\href {\doibase 10.1093/mnras/265.1.250} {\bibfield  {journal}
  {\bibinfo  {journal} {\mnras}\ }\textbf {\bibinfo {volume} {265}},\ \bibinfo
  {pages} {250} (\bibinfo {year} {1993})}\BibitemShut {NoStop}%
\bibitem [{\citenamefont {{Launhardt}}\ \emph {et~al.}(2002)\citenamefont
  {{Launhardt}}, \citenamefont {{Zylka}},\ and\ \citenamefont
  {{Mezger}}}]{Launhardt2002}%
  \BibitemOpen
  \bibfield  {author} {\bibinfo {author} {\bibfnamefont {R.}~\bibnamefont
  {{Launhardt}}}, \bibinfo {author} {\bibfnamefont {R.}~\bibnamefont
  {{Zylka}}}, \ and\ \bibinfo {author} {\bibfnamefont {P.~G.}\ \bibnamefont
  {{Mezger}}},\ }\href {\doibase 10.1051/0004-6361:20020017} {\bibfield
  {journal} {\bibinfo  {journal} {\aap}\ }\textbf {\bibinfo {volume} {384}},\
  \bibinfo {pages} {112} (\bibinfo {year} {2002})},\ \Eprint
  {http://arxiv.org/abs/astro-ph/0201294} {arXiv:astro-ph/0201294 [astro-ph]}
  \BibitemShut {NoStop}%
\bibitem [{\citenamefont {{Sch{\"o}del}}\ \emph {et~al.}(2014)\citenamefont
  {{Sch{\"o}del}}, \citenamefont {{Feldmeier}}, \citenamefont {{Kunneriath}},
  \citenamefont {{Stolovy}}, \citenamefont {{Neumayer}}, \citenamefont
  {{Amaro-Seoane}},\ and\ \citenamefont {{Nishiyama}}}]{Schodel2014}%
  \BibitemOpen
  \bibfield  {author} {\bibinfo {author} {\bibfnamefont {R.}~\bibnamefont
  {{Sch{\"o}del}}}, \bibinfo {author} {\bibfnamefont {A.}~\bibnamefont
  {{Feldmeier}}}, \bibinfo {author} {\bibfnamefont {D.}~\bibnamefont
  {{Kunneriath}}}, \bibinfo {author} {\bibfnamefont {S.}~\bibnamefont
  {{Stolovy}}}, \bibinfo {author} {\bibfnamefont {N.}~\bibnamefont
  {{Neumayer}}}, \bibinfo {author} {\bibfnamefont {P.}~\bibnamefont
  {{Amaro-Seoane}}}, \ and\ \bibinfo {author} {\bibfnamefont {S.}~\bibnamefont
  {{Nishiyama}}},\ }\href {\doibase 10.1051/0004-6361/201423481} {\bibfield
  {journal} {\bibinfo  {journal} {\aap}\ }\textbf {\bibinfo {volume} {566}},\
  \bibinfo {eid} {A47} (\bibinfo {year} {2014})},\ \Eprint
  {http://arxiv.org/abs/1403.6657} {arXiv:1403.6657 [astro-ph.GA]} \BibitemShut
  {NoStop}%
\bibitem [{\citenamefont {{Chernoff}}\ and\ \citenamefont
  {{Weinberg}}(1990)}]{Chernoff1990}%
  \BibitemOpen
  \bibfield  {author} {\bibinfo {author} {\bibfnamefont {D.~F.}\ \bibnamefont
  {{Chernoff}}}\ and\ \bibinfo {author} {\bibfnamefont {M.~D.}\ \bibnamefont
  {{Weinberg}}},\ }\href {\doibase 10.1086/168451} {\bibfield  {journal}
  {\bibinfo  {journal} {\apj}\ }\textbf {\bibinfo {volume} {351}},\ \bibinfo
  {pages} {121} (\bibinfo {year} {1990})}\BibitemShut {NoStop}%
\bibitem [{\citenamefont {{Binney}}\ and\ \citenamefont
  {{Tremaine}}(1987)}]{Binney1987}%
  \BibitemOpen
  \bibfield  {author} {\bibinfo {author} {\bibfnamefont {J.}~\bibnamefont
  {{Binney}}}\ and\ \bibinfo {author} {\bibfnamefont {S.}~\bibnamefont
  {{Tremaine}}},\ }\href@noop {} {\emph {\bibinfo {title} {{Galactic
  dynamics}}}}\ (\bibinfo {year} {1987})\BibitemShut {NoStop}%
\bibitem [{\citenamefont {{Spitzer}}\ and\ \citenamefont
  {{Hart}}(1971)}]{Spitzer1971}%
  \BibitemOpen
  \bibfield  {author} {\bibinfo {author} {\bibfnamefont {J.}~\bibnamefont
  {{Spitzer}}, \bibfnamefont {Lyman}}\ and\ \bibinfo {author} {\bibfnamefont
  {M.~H.}\ \bibnamefont {{Hart}}},\ }\href {\doibase 10.1086/150855} {\bibfield
   {journal} {\bibinfo  {journal} {\apj}\ }\textbf {\bibinfo {volume} {164}},\
  \bibinfo {pages} {399} (\bibinfo {year} {1971})}\BibitemShut {NoStop}%
\bibitem [{\citenamefont {{Bahcall}}\ and\ \citenamefont
  {{Wolf}}(1976)}]{Bahcall1976}%
  \BibitemOpen
  \bibfield  {author} {\bibinfo {author} {\bibfnamefont {J.~N.}\ \bibnamefont
  {{Bahcall}}}\ and\ \bibinfo {author} {\bibfnamefont {R.~A.}\ \bibnamefont
  {{Wolf}}},\ }\href {\doibase 10.1086/154711} {\bibfield  {journal} {\bibinfo
  {journal} {\apj}\ }\textbf {\bibinfo {volume} {209}},\ \bibinfo {pages} {214}
  (\bibinfo {year} {1976})}\BibitemShut {NoStop}%
\bibitem [{Note2()}]{Note2}%
  \BibitemOpen
  \bibinfo {note} {In principle, we should evolve both the distributions $f_i$
  and the potential field $\phi (r)$ self-consistently. For the problem we are
  discussing, the potential field $\phi (r)$ barely changes \cite {Preto2010}
  because during the evolution time range the distributions evolve mainly
  within the influence radius where the potential field is dominated by the
  MBH.}\BibitemShut {Stop}%
\bibitem [{\citenamefont {{Tremaine}}\ \emph {et~al.}(2002)\citenamefont
  {{Tremaine}}, \citenamefont {{Gebhardt}}, \citenamefont {{Bender}},
  \citenamefont {{Bower}}, \citenamefont {{Dressler}}, \citenamefont {{Faber}},
  \citenamefont {{Filippenko}}, \citenamefont {{Green}}, \citenamefont
  {{Grillmair}}, \citenamefont {{Ho}}, \citenamefont {{Kormendy}},
  \citenamefont {{Lauer}}, \citenamefont {{Magorrian}}, \citenamefont
  {{Pinkney}},\ and\ \citenamefont {{Richstone}}}]{Tremaine2002}%
  \BibitemOpen
  \bibfield  {author} {\bibinfo {author} {\bibfnamefont {S.}~\bibnamefont
  {{Tremaine}}}, \bibinfo {author} {\bibfnamefont {K.}~\bibnamefont
  {{Gebhardt}}}, \bibinfo {author} {\bibfnamefont {R.}~\bibnamefont
  {{Bender}}}, \bibinfo {author} {\bibfnamefont {G.}~\bibnamefont {{Bower}}},
  \bibinfo {author} {\bibfnamefont {A.}~\bibnamefont {{Dressler}}}, \bibinfo
  {author} {\bibfnamefont {S.~M.}\ \bibnamefont {{Faber}}}, \bibinfo {author}
  {\bibfnamefont {A.~V.}\ \bibnamefont {{Filippenko}}}, \bibinfo {author}
  {\bibfnamefont {R.}~\bibnamefont {{Green}}}, \bibinfo {author} {\bibfnamefont
  {C.}~\bibnamefont {{Grillmair}}}, \bibinfo {author} {\bibfnamefont {L.~C.}\
  \bibnamefont {{Ho}}}, \bibinfo {author} {\bibfnamefont {J.}~\bibnamefont
  {{Kormendy}}}, \bibinfo {author} {\bibfnamefont {T.~R.}\ \bibnamefont
  {{Lauer}}}, \bibinfo {author} {\bibfnamefont {J.}~\bibnamefont
  {{Magorrian}}}, \bibinfo {author} {\bibfnamefont {J.}~\bibnamefont
  {{Pinkney}}}, \ and\ \bibinfo {author} {\bibfnamefont {D.}~\bibnamefont
  {{Richstone}}},\ }\href {\doibase 10.1086/341002} {\bibfield  {journal}
  {\bibinfo  {journal} {\apj}\ }\textbf {\bibinfo {volume} {574}},\ \bibinfo
  {pages} {740} (\bibinfo {year} {2002})},\ \Eprint
  {http://arxiv.org/abs/astro-ph/0203468} {arXiv:astro-ph/0203468 [astro-ph]}
  \BibitemShut {NoStop}%
\bibitem [{\citenamefont {{G{\"u}ltekin}}\ \emph {et~al.}(2009)\citenamefont
  {{G{\"u}ltekin}}, \citenamefont {{Richstone}}, \citenamefont {{Gebhardt}},
  \citenamefont {{Lauer}}, \citenamefont {{Tremaine}}, \citenamefont {{Aller}},
  \citenamefont {{Bender}}, \citenamefont {{Dressler}}, \citenamefont
  {{Faber}}, \citenamefont {{Filippenko}}, \citenamefont {{Green}},
  \citenamefont {{Ho}}, \citenamefont {{Kormendy}}, \citenamefont
  {{Magorrian}}, \citenamefont {{Pinkney}},\ and\ \citenamefont
  {{Siopis}}}]{Gultekin2009}%
  \BibitemOpen
  \bibfield  {author} {\bibinfo {author} {\bibfnamefont {K.}~\bibnamefont
  {{G{\"u}ltekin}}}, \bibinfo {author} {\bibfnamefont {D.~O.}\ \bibnamefont
  {{Richstone}}}, \bibinfo {author} {\bibfnamefont {K.}~\bibnamefont
  {{Gebhardt}}}, \bibinfo {author} {\bibfnamefont {T.~R.}\ \bibnamefont
  {{Lauer}}}, \bibinfo {author} {\bibfnamefont {S.}~\bibnamefont {{Tremaine}}},
  \bibinfo {author} {\bibfnamefont {M.~C.}\ \bibnamefont {{Aller}}}, \bibinfo
  {author} {\bibfnamefont {R.}~\bibnamefont {{Bender}}}, \bibinfo {author}
  {\bibfnamefont {A.}~\bibnamefont {{Dressler}}}, \bibinfo {author}
  {\bibfnamefont {S.~M.}\ \bibnamefont {{Faber}}}, \bibinfo {author}
  {\bibfnamefont {A.~V.}\ \bibnamefont {{Filippenko}}}, \bibinfo {author}
  {\bibfnamefont {R.}~\bibnamefont {{Green}}}, \bibinfo {author} {\bibfnamefont
  {L.~C.}\ \bibnamefont {{Ho}}}, \bibinfo {author} {\bibfnamefont
  {J.}~\bibnamefont {{Kormendy}}}, \bibinfo {author} {\bibfnamefont
  {J.}~\bibnamefont {{Magorrian}}}, \bibinfo {author} {\bibfnamefont
  {J.}~\bibnamefont {{Pinkney}}}, \ and\ \bibinfo {author} {\bibfnamefont
  {C.}~\bibnamefont {{Siopis}}},\ }\href {\doibase 10.1088/0004-637X/698/1/198}
  {\bibfield  {journal} {\bibinfo  {journal} {\apj}\ }\textbf {\bibinfo
  {volume} {698}},\ \bibinfo {pages} {198} (\bibinfo {year} {2009})},\ \Eprint
  {http://arxiv.org/abs/0903.4897} {arXiv:0903.4897 [astro-ph.GA]} \BibitemShut
  {NoStop}%
\bibitem [{\citenamefont {{Alexander}}\ and\ \citenamefont
  {{Hopman}}(2009)}]{Alexander2009}%
  \BibitemOpen
  \bibfield  {author} {\bibinfo {author} {\bibfnamefont {T.}~\bibnamefont
  {{Alexander}}}\ and\ \bibinfo {author} {\bibfnamefont {C.}~\bibnamefont
  {{Hopman}}},\ }\href {\doibase 10.1088/0004-637X/697/2/1861} {\bibfield
  {journal} {\bibinfo  {journal} {\apj}\ }\textbf {\bibinfo {volume} {697}},\
  \bibinfo {pages} {1861} (\bibinfo {year} {2009})},\ \Eprint
  {http://arxiv.org/abs/0808.3150} {arXiv:0808.3150 [astro-ph]} \BibitemShut
  {NoStop}%
\bibitem [{\citenamefont {{Chandrasekhar}}(1943)}]{Chandrasekhar1943}%
  \BibitemOpen
  \bibfield  {author} {\bibinfo {author} {\bibfnamefont {S.}~\bibnamefont
  {{Chandrasekhar}}},\ }\href {\doibase 10.1086/144517} {\bibfield  {journal}
  {\bibinfo  {journal} {\apj}\ }\textbf {\bibinfo {volume} {97}},\ \bibinfo
  {pages} {255} (\bibinfo {year} {1943})}\BibitemShut {NoStop}%
\bibitem [{\citenamefont {{Ostriker}}(1999)}]{Ostriker1999}%
  \BibitemOpen
  \bibfield  {author} {\bibinfo {author} {\bibfnamefont {E.~C.}\ \bibnamefont
  {{Ostriker}}},\ }\href {\doibase 10.1086/306858} {\bibfield  {journal}
  {\bibinfo  {journal} {\apj}\ }\textbf {\bibinfo {volume} {513}},\ \bibinfo
  {pages} {252} (\bibinfo {year} {1999})},\ \Eprint
  {http://arxiv.org/abs/astro-ph/9810324} {arXiv:astro-ph/9810324 [astro-ph]}
  \BibitemShut {NoStop}%
\bibitem [{\citenamefont {{Masset}}(2017)}]{Masset2017}%
  \BibitemOpen
  \bibfield  {author} {\bibinfo {author} {\bibfnamefont {F.~S.}\ \bibnamefont
  {{Masset}}},\ }\href {\doibase 10.1093/mnras/stx2271} {\bibfield  {journal}
  {\bibinfo  {journal} {\mnras}\ }\textbf {\bibinfo {volume} {472}},\ \bibinfo
  {pages} {4204} (\bibinfo {year} {2017})},\ \Eprint
  {http://arxiv.org/abs/1708.09807} {arXiv:1708.09807 [astro-ph.EP]}
  \BibitemShut {NoStop}%
\bibitem [{\citenamefont {{Hankla}}\ \emph {et~al.}(2020)\citenamefont
  {{Hankla}}, \citenamefont {{Jiang}},\ and\ \citenamefont
  {{Armitage}}}]{Hankla2020}%
  \BibitemOpen
  \bibfield  {author} {\bibinfo {author} {\bibfnamefont {A.~M.}\ \bibnamefont
  {{Hankla}}}, \bibinfo {author} {\bibfnamefont {Y.-F.}\ \bibnamefont
  {{Jiang}}}, \ and\ \bibinfo {author} {\bibfnamefont {P.~J.}\ \bibnamefont
  {{Armitage}}},\ }\href {\doibase 10.3847/1538-4357/abb4df} {\bibfield
  {journal} {\bibinfo  {journal} {\apj}\ }\textbf {\bibinfo {volume} {902}},\
  \bibinfo {eid} {50} (\bibinfo {year} {2020})},\ \Eprint
  {http://arxiv.org/abs/2005.03785} {arXiv:2005.03785 [astro-ph.EP]}
  \BibitemShut {NoStop}%
\bibitem [{\citenamefont {{Shakura}}\ and\ \citenamefont
  {{Sunyaev}}(1973)}]{Shakura1973}%
  \BibitemOpen
  \bibfield  {author} {\bibinfo {author} {\bibfnamefont {N.~I.}\ \bibnamefont
  {{Shakura}}}\ and\ \bibinfo {author} {\bibfnamefont {R.~A.}\ \bibnamefont
  {{Sunyaev}}},\ }\href@noop {} {\bibfield  {journal} {\bibinfo  {journal}
  {\aap}\ }\textbf {\bibinfo {volume} {500}},\ \bibinfo {pages} {33} (\bibinfo
  {year} {1973})}\BibitemShut {NoStop}%
\bibitem [{\citenamefont {{Alexander}}\ and\ \citenamefont
  {{Ferguson}}(1994)}]{Alexander1994}%
  \BibitemOpen
  \bibfield  {author} {\bibinfo {author} {\bibfnamefont {D.~R.}\ \bibnamefont
  {{Alexander}}}\ and\ \bibinfo {author} {\bibfnamefont {J.~W.}\ \bibnamefont
  {{Ferguson}}},\ }\href {\doibase 10.1086/175039} {\bibfield  {journal}
  {\bibinfo  {journal} {\apj}\ }\textbf {\bibinfo {volume} {437}},\ \bibinfo
  {pages} {879} (\bibinfo {year} {1994})}\BibitemShut {NoStop}%
\bibitem [{\citenamefont {{Iglesias}}\ and\ \citenamefont
  {{Rogers}}(1996)}]{Iglesias1996}%
  \BibitemOpen
  \bibfield  {author} {\bibinfo {author} {\bibfnamefont {C.~A.}\ \bibnamefont
  {{Iglesias}}}\ and\ \bibinfo {author} {\bibfnamefont {F.~J.}\ \bibnamefont
  {{Rogers}}},\ }\href {\doibase 10.1086/177381} {\bibfield  {journal}
  {\bibinfo  {journal} {\apj}\ }\textbf {\bibinfo {volume} {464}},\ \bibinfo
  {pages} {943} (\bibinfo {year} {1996})}\BibitemShut {NoStop}%
\bibitem [{\citenamefont {{Balbus}}\ and\ \citenamefont
  {{Hawley}}(1991)}]{Balbus1991}%
  \BibitemOpen
  \bibfield  {author} {\bibinfo {author} {\bibfnamefont {S.~A.}\ \bibnamefont
  {{Balbus}}}\ and\ \bibinfo {author} {\bibfnamefont {J.~F.}\ \bibnamefont
  {{Hawley}}},\ }\href {\doibase 10.1086/170270} {\bibfield  {journal}
  {\bibinfo  {journal} {\apj}\ }\textbf {\bibinfo {volume} {376}},\ \bibinfo
  {pages} {214} (\bibinfo {year} {1991})}\BibitemShut {NoStop}%
\bibitem [{\citenamefont {{Balbus}}\ and\ \citenamefont
  {{Hawley}}(1998)}]{Balbus1998}%
  \BibitemOpen
  \bibfield  {author} {\bibinfo {author} {\bibfnamefont {S.~A.}\ \bibnamefont
  {{Balbus}}}\ and\ \bibinfo {author} {\bibfnamefont {J.~F.}\ \bibnamefont
  {{Hawley}}},\ }\href {\doibase 10.1103/RevModPhys.70.1} {\bibfield  {journal}
  {\bibinfo  {journal} {Reviews of Modern Physics}\ }\textbf {\bibinfo {volume}
  {70}},\ \bibinfo {pages} {1} (\bibinfo {year} {1998})}\BibitemShut {NoStop}%
\bibitem [{\citenamefont {{Martin}}\ \emph {et~al.}(2019)\citenamefont
  {{Martin}}, \citenamefont {{Nixon}}, \citenamefont {{Pringle}},\ and\
  \citenamefont {{Livio}}}]{Martin2019}%
  \BibitemOpen
  \bibfield  {author} {\bibinfo {author} {\bibfnamefont {R.~G.}\ \bibnamefont
  {{Martin}}}, \bibinfo {author} {\bibfnamefont {C.~J.}\ \bibnamefont
  {{Nixon}}}, \bibinfo {author} {\bibfnamefont {J.~E.}\ \bibnamefont
  {{Pringle}}}, \ and\ \bibinfo {author} {\bibfnamefont {M.}~\bibnamefont
  {{Livio}}},\ }\href {\doibase 10.1016/j.newast.2019.01.001} {\bibfield
  {journal} {\bibinfo  {journal} {New Astronomy}\ }\textbf {\bibinfo {volume}
  {70}},\ \bibinfo {pages} {7} (\bibinfo {year} {2019})},\ \Eprint
  {http://arxiv.org/abs/1901.01580} {arXiv:1901.01580 [astro-ph.HE]}
  \BibitemShut {NoStop}%
\bibitem [{\citenamefont {{Paardekooper}}\ \emph {et~al.}(2010)\citenamefont
  {{Paardekooper}}, \citenamefont {{Baruteau}}, \citenamefont {{Crida}},\ and\
  \citenamefont {{Kley}}}]{Paardekooper2010}%
  \BibitemOpen
  \bibfield  {author} {\bibinfo {author} {\bibfnamefont {S.~J.}\ \bibnamefont
  {{Paardekooper}}}, \bibinfo {author} {\bibfnamefont {C.}~\bibnamefont
  {{Baruteau}}}, \bibinfo {author} {\bibfnamefont {A.}~\bibnamefont {{Crida}}},
  \ and\ \bibinfo {author} {\bibfnamefont {W.}~\bibnamefont {{Kley}}},\ }\href
  {\doibase 10.1111/j.1365-2966.2009.15782.x} {\bibfield  {journal} {\bibinfo
  {journal} {\mnras}\ }\textbf {\bibinfo {volume} {401}},\ \bibinfo {pages}
  {1950} (\bibinfo {year} {2010})},\ \Eprint {http://arxiv.org/abs/0909.4552}
  {arXiv:0909.4552 [astro-ph.EP]} \BibitemShut {NoStop}%
\bibitem [{\citenamefont {{Nelson}}(2005)}]{Nelson2005}%
  \BibitemOpen
  \bibfield  {author} {\bibinfo {author} {\bibfnamefont {R.~P.}\ \bibnamefont
  {{Nelson}}},\ }\href {\doibase 10.1051/0004-6361:20042605} {\bibfield
  {journal} {\bibinfo  {journal} {\aap}\ }\textbf {\bibinfo {volume} {443}},\
  \bibinfo {pages} {1067} (\bibinfo {year} {2005})},\ \Eprint
  {http://arxiv.org/abs/astro-ph/0508486} {arXiv:astro-ph/0508486 [astro-ph]}
  \BibitemShut {NoStop}%
\bibitem [{\citenamefont {{Johnson}}\ \emph {et~al.}(2006)\citenamefont
  {{Johnson}}, \citenamefont {{Goodman}},\ and\ \citenamefont
  {{Menou}}}]{Johnson2006}%
  \BibitemOpen
  \bibfield  {author} {\bibinfo {author} {\bibfnamefont {E.~T.}\ \bibnamefont
  {{Johnson}}}, \bibinfo {author} {\bibfnamefont {J.}~\bibnamefont
  {{Goodman}}}, \ and\ \bibinfo {author} {\bibfnamefont {K.}~\bibnamefont
  {{Menou}}},\ }\href {\doibase 10.1086/505462} {\bibfield  {journal} {\bibinfo
   {journal} {\apj}\ }\textbf {\bibinfo {volume} {647}},\ \bibinfo {pages}
  {1413} (\bibinfo {year} {2006})},\ \Eprint
  {http://arxiv.org/abs/astro-ph/0603235} {arXiv:astro-ph/0603235 [astro-ph]}
  \BibitemShut {NoStop}%
\bibitem [{\citenamefont {{Yang}}\ \emph {et~al.}(2009)\citenamefont {{Yang}},
  \citenamefont {{Mac Low}},\ and\ \citenamefont {{Menou}}}]{Yang2009}%
  \BibitemOpen
  \bibfield  {author} {\bibinfo {author} {\bibfnamefont {C.-C.}\ \bibnamefont
  {{Yang}}}, \bibinfo {author} {\bibfnamefont {M.-M.}\ \bibnamefont {{Mac
  Low}}}, \ and\ \bibinfo {author} {\bibfnamefont {K.}~\bibnamefont
  {{Menou}}},\ }\href {\doibase 10.1088/0004-637X/707/2/1233} {\bibfield
  {journal} {\bibinfo  {journal} {\apj}\ }\textbf {\bibinfo {volume} {707}},\
  \bibinfo {pages} {1233} (\bibinfo {year} {2009})},\ \Eprint
  {http://arxiv.org/abs/0907.1897} {arXiv:0907.1897 [astro-ph.EP]} \BibitemShut
  {NoStop}%
\bibitem [{\citenamefont {{Lin}}\ and\ \citenamefont
  {{Papaloizou}}(1986)}]{Lin1986}%
  \BibitemOpen
  \bibfield  {author} {\bibinfo {author} {\bibfnamefont {D.~N.~C.}\
  \bibnamefont {{Lin}}}\ and\ \bibinfo {author} {\bibfnamefont
  {J.}~\bibnamefont {{Papaloizou}}},\ }\href {\doibase 10.1086/164653}
  {\bibfield  {journal} {\bibinfo  {journal} {\apj}\ }\textbf {\bibinfo
  {volume} {309}},\ \bibinfo {pages} {846} (\bibinfo {year}
  {1986})}\BibitemShut {NoStop}%
\bibitem [{\citenamefont {{Bryden}}\ \emph {et~al.}(1999)\citenamefont
  {{Bryden}}, \citenamefont {{Chen}}, \citenamefont {{Lin}}, \citenamefont
  {{Nelson}},\ and\ \citenamefont {{Papaloizou}}}]{Bryden1999}%
  \BibitemOpen
  \bibfield  {author} {\bibinfo {author} {\bibfnamefont {G.}~\bibnamefont
  {{Bryden}}}, \bibinfo {author} {\bibfnamefont {X.}~\bibnamefont {{Chen}}},
  \bibinfo {author} {\bibfnamefont {D.~N.~C.}\ \bibnamefont {{Lin}}}, \bibinfo
  {author} {\bibfnamefont {R.~P.}\ \bibnamefont {{Nelson}}}, \ and\ \bibinfo
  {author} {\bibfnamefont {J.~C.~B.}\ \bibnamefont {{Papaloizou}}},\ }\href
  {\doibase 10.1086/306917} {\bibfield  {journal} {\bibinfo  {journal} {\apj}\
  }\textbf {\bibinfo {volume} {514}},\ \bibinfo {pages} {344} (\bibinfo {year}
  {1999})}\BibitemShut {NoStop}%
\bibitem [{\citenamefont {{Crida}}\ \emph {et~al.}(2006)\citenamefont
  {{Crida}}, \citenamefont {{Morbidelli}},\ and\ \citenamefont
  {{Masset}}}]{Crida2006}%
  \BibitemOpen
  \bibfield  {author} {\bibinfo {author} {\bibfnamefont {A.}~\bibnamefont
  {{Crida}}}, \bibinfo {author} {\bibfnamefont {A.}~\bibnamefont
  {{Morbidelli}}}, \ and\ \bibinfo {author} {\bibfnamefont {F.}~\bibnamefont
  {{Masset}}},\ }\href {\doibase 10.1016/j.icarus.2005.10.007} {\bibfield
  {journal} {\bibinfo  {journal} {Icarus}\ }\textbf {\bibinfo {volume} {181}},\
  \bibinfo {pages} {587} (\bibinfo {year} {2006})},\ \Eprint
  {http://arxiv.org/abs/astro-ph/0511082} {arXiv:astro-ph/0511082 [astro-ph]}
  \BibitemShut {NoStop}%
\bibitem [{\citenamefont {{Duffell}}\ and\ \citenamefont
  {{MacFadyen}}(2013)}]{Duffell2013}%
  \BibitemOpen
  \bibfield  {author} {\bibinfo {author} {\bibfnamefont {P.~C.}\ \bibnamefont
  {{Duffell}}}\ and\ \bibinfo {author} {\bibfnamefont {A.~I.}\ \bibnamefont
  {{MacFadyen}}},\ }\href {\doibase 10.1088/0004-637X/769/1/41} {\bibfield
  {journal} {\bibinfo  {journal} {\apj}\ }\textbf {\bibinfo {volume} {769}},\
  \bibinfo {eid} {41} (\bibinfo {year} {2013})},\ \Eprint
  {http://arxiv.org/abs/1302.1934} {arXiv:1302.1934 [astro-ph.EP]} \BibitemShut
  {NoStop}%
\bibitem [{\citenamefont {{Syer}}\ and\ \citenamefont
  {{Clarke}}(1995)}]{Syer1995}%
  \BibitemOpen
  \bibfield  {author} {\bibinfo {author} {\bibfnamefont {D.}~\bibnamefont
  {{Syer}}}\ and\ \bibinfo {author} {\bibfnamefont {C.~J.}\ \bibnamefont
  {{Clarke}}},\ }\href {\doibase 10.1093/mnras/277.3.758} {\bibfield  {journal}
  {\bibinfo  {journal} {\mnras}\ }\textbf {\bibinfo {volume} {277}},\ \bibinfo
  {pages} {758} (\bibinfo {year} {1995})},\ \Eprint
  {http://arxiv.org/abs/astro-ph/9505021} {arXiv:astro-ph/9505021 [astro-ph]}
  \BibitemShut {NoStop}%
\bibitem [{\citenamefont {{Duffell}}\ \emph {et~al.}(2014)\citenamefont
  {{Duffell}}, \citenamefont {{Haiman}}, \citenamefont {{MacFadyen}},
  \citenamefont {{D'Orazio}},\ and\ \citenamefont {{Farris}}}]{Duffell2014}%
  \BibitemOpen
  \bibfield  {author} {\bibinfo {author} {\bibfnamefont {P.~C.}\ \bibnamefont
  {{Duffell}}}, \bibinfo {author} {\bibfnamefont {Z.}~\bibnamefont {{Haiman}}},
  \bibinfo {author} {\bibfnamefont {A.~I.}\ \bibnamefont {{MacFadyen}}},
  \bibinfo {author} {\bibfnamefont {D.~J.}\ \bibnamefont {{D'Orazio}}}, \ and\
  \bibinfo {author} {\bibfnamefont {B.~D.}\ \bibnamefont {{Farris}}},\ }\href
  {\doibase 10.1088/2041-8205/792/1/L10} {\bibfield  {journal} {\bibinfo
  {journal} {\apjl}\ }\textbf {\bibinfo {volume} {792}},\ \bibinfo {eid} {L10}
  (\bibinfo {year} {2014})},\ \Eprint {http://arxiv.org/abs/1405.3711}
  {arXiv:1405.3711 [astro-ph.EP]} \BibitemShut {NoStop}%
\bibitem [{\citenamefont {{D{\"u}rmann}}\ and\ \citenamefont
  {{Kley}}(2015)}]{Durmann2015}%
  \BibitemOpen
  \bibfield  {author} {\bibinfo {author} {\bibfnamefont {C.}~\bibnamefont
  {{D{\"u}rmann}}}\ and\ \bibinfo {author} {\bibfnamefont {W.}~\bibnamefont
  {{Kley}}},\ }\href {\doibase 10.1051/0004-6361/201424837} {\bibfield
  {journal} {\bibinfo  {journal} {\aap}\ }\textbf {\bibinfo {volume} {574}},\
  \bibinfo {eid} {A52} (\bibinfo {year} {2015})},\ \Eprint
  {http://arxiv.org/abs/1411.3190} {arXiv:1411.3190 [astro-ph.EP]} \BibitemShut
  {NoStop}%
\bibitem [{\citenamefont {{Yang}}\ \emph {et~al.}(2014)\citenamefont {{Yang}},
  \citenamefont {{Yuan}}, \citenamefont {{Ohsuga}},\ and\ \citenamefont
  {{Bu}}}]{Yang2014}%
  \BibitemOpen
  \bibfield  {author} {\bibinfo {author} {\bibfnamefont {X.-H.}\ \bibnamefont
  {{Yang}}}, \bibinfo {author} {\bibfnamefont {F.}~\bibnamefont {{Yuan}}},
  \bibinfo {author} {\bibfnamefont {K.}~\bibnamefont {{Ohsuga}}}, \ and\
  \bibinfo {author} {\bibfnamefont {D.-F.}\ \bibnamefont {{Bu}}},\ }\href
  {\doibase 10.1088/0004-637X/780/1/79} {\bibfield  {journal} {\bibinfo
  {journal} {\apj}\ }\textbf {\bibinfo {volume} {780}},\ \bibinfo {eid} {79}
  (\bibinfo {year} {2014})},\ \Eprint {http://arxiv.org/abs/1306.1871}
  {arXiv:1306.1871 [astro-ph.HE]} \BibitemShut {NoStop}%
\bibitem [{\citenamefont {{McKinney}}\ \emph {et~al.}(2014)\citenamefont
  {{McKinney}}, \citenamefont {{Tchekhovskoy}}, \citenamefont {{Sadowski}},\
  and\ \citenamefont {{Narayan}}}]{McKinney2014}%
  \BibitemOpen
  \bibfield  {author} {\bibinfo {author} {\bibfnamefont {J.~C.}\ \bibnamefont
  {{McKinney}}}, \bibinfo {author} {\bibfnamefont {A.}~\bibnamefont
  {{Tchekhovskoy}}}, \bibinfo {author} {\bibfnamefont {A.}~\bibnamefont
  {{Sadowski}}}, \ and\ \bibinfo {author} {\bibfnamefont {R.}~\bibnamefont
  {{Narayan}}},\ }\href {\doibase 10.1093/mnras/stu762} {\bibfield  {journal}
  {\bibinfo  {journal} {\mnras}\ }\textbf {\bibinfo {volume} {441}},\ \bibinfo
  {pages} {3177} (\bibinfo {year} {2014})},\ \Eprint
  {http://arxiv.org/abs/1312.6127} {arXiv:1312.6127 [astro-ph.CO]} \BibitemShut
  {NoStop}%
\bibitem [{\citenamefont {{Gruzinov}}\ \emph {et~al.}(2020)\citenamefont
  {{Gruzinov}}, \citenamefont {{Levin}},\ and\ \citenamefont
  {{Matzner}}}]{Gruzinov2020}%
  \BibitemOpen
  \bibfield  {author} {\bibinfo {author} {\bibfnamefont {A.}~\bibnamefont
  {{Gruzinov}}}, \bibinfo {author} {\bibfnamefont {Y.}~\bibnamefont {{Levin}}},
  \ and\ \bibinfo {author} {\bibfnamefont {C.~D.}\ \bibnamefont {{Matzner}}},\
  }\href {\doibase 10.1093/mnras/staa013} {\bibfield  {journal} {\bibinfo
  {journal} {\mnras}\ }\textbf {\bibinfo {volume} {492}},\ \bibinfo {pages}
  {2755} (\bibinfo {year} {2020})},\ \Eprint {http://arxiv.org/abs/1906.01186}
  {arXiv:1906.01186 [astro-ph.HE]} \BibitemShut {NoStop}%
\bibitem [{\citenamefont {{Li}}\ \emph {et~al.}(2020)\citenamefont {{Li}},
  \citenamefont {{Chang}}, \citenamefont {{Levin}}, \citenamefont {{Matzner}},\
  and\ \citenamefont {{Armitage}}}]{Li2020}%
  \BibitemOpen
  \bibfield  {author} {\bibinfo {author} {\bibfnamefont {X.}~\bibnamefont
  {{Li}}}, \bibinfo {author} {\bibfnamefont {P.}~\bibnamefont {{Chang}}},
  \bibinfo {author} {\bibfnamefont {Y.}~\bibnamefont {{Levin}}}, \bibinfo
  {author} {\bibfnamefont {C.~D.}\ \bibnamefont {{Matzner}}}, \ and\ \bibinfo
  {author} {\bibfnamefont {P.~J.}\ \bibnamefont {{Armitage}}},\ }\href
  {\doibase 10.1093/mnras/staa900} {\bibfield  {journal} {\bibinfo  {journal}
  {\mnras}\ }\textbf {\bibinfo {volume} {494}},\ \bibinfo {pages} {2327}
  (\bibinfo {year} {2020})},\ \Eprint {http://arxiv.org/abs/1912.06864}
  {arXiv:1912.06864 [astro-ph.HE]} \BibitemShut {NoStop}%
\bibitem [{\citenamefont {{Lyra}}\ \emph {et~al.}(2010)\citenamefont {{Lyra}},
  \citenamefont {{Paardekooper}},\ and\ \citenamefont {{Mac Low}}}]{Lyra2010}%
  \BibitemOpen
  \bibfield  {author} {\bibinfo {author} {\bibfnamefont {W.}~\bibnamefont
  {{Lyra}}}, \bibinfo {author} {\bibfnamefont {S.-J.}\ \bibnamefont
  {{Paardekooper}}}, \ and\ \bibinfo {author} {\bibfnamefont {M.-M.}\
  \bibnamefont {{Mac Low}}},\ }\href {\doibase 10.1088/2041-8205/715/2/L68}
  {\bibfield  {journal} {\bibinfo  {journal} {\apjl}\ }\textbf {\bibinfo
  {volume} {715}},\ \bibinfo {pages} {L68} (\bibinfo {year} {2010})},\ \Eprint
  {http://arxiv.org/abs/1003.0925} {arXiv:1003.0925 [astro-ph.EP]} \BibitemShut
  {NoStop}%
\bibitem [{\citenamefont {{Bellovary}}\ \emph {et~al.}(2016)\citenamefont
  {{Bellovary}}, \citenamefont {{Mac Low}}, \citenamefont {{McKernan}},\ and\
  \citenamefont {{Ford}}}]{Bellovary2016}%
  \BibitemOpen
  \bibfield  {author} {\bibinfo {author} {\bibfnamefont {J.~M.}\ \bibnamefont
  {{Bellovary}}}, \bibinfo {author} {\bibfnamefont {M.-M.}\ \bibnamefont {{Mac
  Low}}}, \bibinfo {author} {\bibfnamefont {B.}~\bibnamefont {{McKernan}}}, \
  and\ \bibinfo {author} {\bibfnamefont {K.~E.~S.}\ \bibnamefont {{Ford}}},\
  }\href {\doibase 10.3847/2041-8205/819/2/L17} {\bibfield  {journal} {\bibinfo
   {journal} {\apjl}\ }\textbf {\bibinfo {volume} {819}},\ \bibinfo {eid} {L17}
  (\bibinfo {year} {2016})},\ \Eprint {http://arxiv.org/abs/1511.00005}
  {arXiv:1511.00005 [astro-ph.GA]} \BibitemShut {NoStop}%
\bibitem [{Note3()}]{Note3}%
  \BibitemOpen
  \bibinfo {note} {In fact, Dittmann and Miller \cite {Dittmann2020} also noted
  that the migration traps in TQM disks no longer stand if a more updated
  opacity is used in solving the disk structure.}\BibitemShut {Stop}%
\bibitem [{\citenamefont {{Vilkoviskij}}\ and\ \citenamefont
  {{Czerny}}(2002)}]{Vilkoviskij2002}%
  \BibitemOpen
  \bibfield  {author} {\bibinfo {author} {\bibfnamefont {E.~Y.}\ \bibnamefont
  {{Vilkoviskij}}}\ and\ \bibinfo {author} {\bibfnamefont {B.}~\bibnamefont
  {{Czerny}}},\ }\href {\doibase 10.1051/0004-6361:20020255} {\bibfield
  {journal} {\bibinfo  {journal} {\aap}\ }\textbf {\bibinfo {volume} {387}},\
  \bibinfo {pages} {804} (\bibinfo {year} {2002})},\ \Eprint
  {http://arxiv.org/abs/astro-ph/0203226} {arXiv:astro-ph/0203226 [astro-ph]}
  \BibitemShut {NoStop}%
\bibitem [{\citenamefont {{Kennedy}}\ \emph {et~al.}(2016)\citenamefont
  {{Kennedy}}, \citenamefont {{Meiron}}, \citenamefont {{Shukirgaliyev}},
  \citenamefont {{Panamarev}}, \citenamefont {{Berczik}}, \citenamefont
  {{Just}},\ and\ \citenamefont {{Spurzem}}}]{Kennedy2016}%
  \BibitemOpen
  \bibfield  {author} {\bibinfo {author} {\bibfnamefont {G.~F.}\ \bibnamefont
  {{Kennedy}}}, \bibinfo {author} {\bibfnamefont {Y.}~\bibnamefont {{Meiron}}},
  \bibinfo {author} {\bibfnamefont {B.}~\bibnamefont {{Shukirgaliyev}}},
  \bibinfo {author} {\bibfnamefont {T.}~\bibnamefont {{Panamarev}}}, \bibinfo
  {author} {\bibfnamefont {P.}~\bibnamefont {{Berczik}}}, \bibinfo {author}
  {\bibfnamefont {A.}~\bibnamefont {{Just}}}, \ and\ \bibinfo {author}
  {\bibfnamefont {R.}~\bibnamefont {{Spurzem}}},\ }\href {\doibase
  10.1093/mnras/stw908} {\bibfield  {journal} {\bibinfo  {journal} {\mnras}\
  }\textbf {\bibinfo {volume} {460}},\ \bibinfo {pages} {240} (\bibinfo {year}
  {2016})},\ \Eprint {http://arxiv.org/abs/1604.05309} {arXiv:1604.05309
  [astro-ph.GA]} \BibitemShut {NoStop}%
\bibitem [{\citenamefont {{Panamarev}}\ \emph {et~al.}(2018)\citenamefont
  {{Panamarev}}, \citenamefont {{Shukirgaliyev}}, \citenamefont {{Meiron}},
  \citenamefont {{Berczik}}, \citenamefont {{Just}}, \citenamefont {{Spurzem}},
  \citenamefont {{Omarov}},\ and\ \citenamefont
  {{Vilkoviskij}}}]{Panamarev2018}%
  \BibitemOpen
  \bibfield  {author} {\bibinfo {author} {\bibfnamefont {T.}~\bibnamefont
  {{Panamarev}}}, \bibinfo {author} {\bibfnamefont {B.}~\bibnamefont
  {{Shukirgaliyev}}}, \bibinfo {author} {\bibfnamefont {Y.}~\bibnamefont
  {{Meiron}}}, \bibinfo {author} {\bibfnamefont {P.}~\bibnamefont {{Berczik}}},
  \bibinfo {author} {\bibfnamefont {A.}~\bibnamefont {{Just}}}, \bibinfo
  {author} {\bibfnamefont {R.}~\bibnamefont {{Spurzem}}}, \bibinfo {author}
  {\bibfnamefont {C.}~\bibnamefont {{Omarov}}}, \ and\ \bibinfo {author}
  {\bibfnamefont {E.}~\bibnamefont {{Vilkoviskij}}},\ }\href {\doibase
  10.1093/mnras/sty459} {\bibfield  {journal} {\bibinfo  {journal} {\mnras}\
  }\textbf {\bibinfo {volume} {476}},\ \bibinfo {pages} {4224} (\bibinfo {year}
  {2018})},\ \Eprint {http://arxiv.org/abs/1802.03027} {arXiv:1802.03027
  [astro-ph.GA]} \BibitemShut {NoStop}%
\bibitem [{\citenamefont {{Rees}}\ \emph {et~al.}(1982)\citenamefont {{Rees}},
  \citenamefont {{Begelman}}, \citenamefont {{Blandford}},\ and\ \citenamefont
  {{Phinney}}}]{Rees1982}%
  \BibitemOpen
  \bibfield  {author} {\bibinfo {author} {\bibfnamefont {M.~J.}\ \bibnamefont
  {{Rees}}}, \bibinfo {author} {\bibfnamefont {M.~C.}\ \bibnamefont
  {{Begelman}}}, \bibinfo {author} {\bibfnamefont {R.~D.}\ \bibnamefont
  {{Blandford}}}, \ and\ \bibinfo {author} {\bibfnamefont {E.~S.}\ \bibnamefont
  {{Phinney}}},\ }\href {\doibase 10.1038/295017a0} {\bibfield  {journal}
  {\bibinfo  {journal} {\nat}\ }\textbf {\bibinfo {volume} {295}},\ \bibinfo
  {pages} {17} (\bibinfo {year} {1982})}\BibitemShut {NoStop}%
\bibitem [{\citenamefont {{Lightman}}\ and\ \citenamefont
  {{Eardley}}(1974)}]{Lightman1974}%
  \BibitemOpen
  \bibfield  {author} {\bibinfo {author} {\bibfnamefont {A.~P.}\ \bibnamefont
  {{Lightman}}}\ and\ \bibinfo {author} {\bibfnamefont {D.~M.}\ \bibnamefont
  {{Eardley}}},\ }\href {\doibase 10.1086/181377} {\bibfield  {journal}
  {\bibinfo  {journal} {\apjl}\ }\textbf {\bibinfo {volume} {187}},\ \bibinfo
  {pages} {L1} (\bibinfo {year} {1974})}\BibitemShut {NoStop}%
\bibitem [{\citenamefont {{Piran}}(1978)}]{Piran1978}%
  \BibitemOpen
  \bibfield  {author} {\bibinfo {author} {\bibfnamefont {T.}~\bibnamefont
  {{Piran}}},\ }\href {\doibase 10.1086/156069} {\bibfield  {journal} {\bibinfo
   {journal} {\apj}\ }\textbf {\bibinfo {volume} {221}},\ \bibinfo {pages}
  {652} (\bibinfo {year} {1978})}\BibitemShut {NoStop}%
\bibitem [{\citenamefont {{King}}\ and\ \citenamefont
  {{Nixon}}(2015)}]{King2015}%
  \BibitemOpen
  \bibfield  {author} {\bibinfo {author} {\bibfnamefont {A.}~\bibnamefont
  {{King}}}\ and\ \bibinfo {author} {\bibfnamefont {C.}~\bibnamefont
  {{Nixon}}},\ }\href {\doibase 10.1093/mnrasl/slv098} {\bibfield  {journal}
  {\bibinfo  {journal} {\mnras}\ }\textbf {\bibinfo {volume} {453}},\ \bibinfo
  {pages} {L46} (\bibinfo {year} {2015})},\ \Eprint
  {http://arxiv.org/abs/1507.05960} {arXiv:1507.05960 [astro-ph.HE]}
  \BibitemShut {NoStop}%
\bibitem [{\citenamefont {{Schawinski}}\ \emph {et~al.}(2015)\citenamefont
  {{Schawinski}}, \citenamefont {{Koss}}, \citenamefont {{Berney}},\ and\
  \citenamefont {{Sartori}}}]{Schawinski2015}%
  \BibitemOpen
  \bibfield  {author} {\bibinfo {author} {\bibfnamefont {K.}~\bibnamefont
  {{Schawinski}}}, \bibinfo {author} {\bibfnamefont {M.}~\bibnamefont
  {{Koss}}}, \bibinfo {author} {\bibfnamefont {S.}~\bibnamefont {{Berney}}}, \
  and\ \bibinfo {author} {\bibfnamefont {L.~F.}\ \bibnamefont {{Sartori}}},\
  }\href {\doibase 10.1093/mnras/stv1136} {\bibfield  {journal} {\bibinfo
  {journal} {\mnras}\ }\textbf {\bibinfo {volume} {451}},\ \bibinfo {pages}
  {2517} (\bibinfo {year} {2015})},\ \Eprint {http://arxiv.org/abs/1505.06733}
  {arXiv:1505.06733 [astro-ph.GA]} \BibitemShut {NoStop}%
\bibitem [{\citenamefont {{Soltan}}(1982)}]{Soltan1982}%
  \BibitemOpen
  \bibfield  {author} {\bibinfo {author} {\bibfnamefont {A.}~\bibnamefont
  {{Soltan}}},\ }\href {\doibase 10.1093/mnras/200.1.115} {\bibfield  {journal}
  {\bibinfo  {journal} {\mnras}\ }\textbf {\bibinfo {volume} {200}},\ \bibinfo
  {pages} {115} (\bibinfo {year} {1982})}\BibitemShut {NoStop}%
\bibitem [{\citenamefont {{Barack}}\ and\ \citenamefont
  {{Cutler}}(2004)}]{Barack2004}%
  \BibitemOpen
  \bibfield  {author} {\bibinfo {author} {\bibfnamefont {L.}~\bibnamefont
  {{Barack}}}\ and\ \bibinfo {author} {\bibfnamefont {C.}~\bibnamefont
  {{Cutler}}},\ }\href {\doibase 10.1103/PhysRevD.69.082005} {\bibfield
  {journal} {\bibinfo  {journal} {\prd}\ }\textbf {\bibinfo {volume} {69}},\
  \bibinfo {eid} {082005} (\bibinfo {year} {2004})},\ \Eprint
  {http://arxiv.org/abs/gr-qc/0310125} {arXiv:gr-qc/0310125 [gr-qc]}
  \BibitemShut {NoStop}%
\bibitem [{\citenamefont {{Huerta}}\ and\ \citenamefont
  {{Gair}}(2009)}]{Huerta2009}%
  \BibitemOpen
  \bibfield  {author} {\bibinfo {author} {\bibfnamefont {E.~A.}\ \bibnamefont
  {{Huerta}}}\ and\ \bibinfo {author} {\bibfnamefont {J.~R.}\ \bibnamefont
  {{Gair}}},\ }\href {\doibase 10.1103/PhysRevD.79.084021} {\bibfield
  {journal} {\bibinfo  {journal} {\prd}\ }\textbf {\bibinfo {volume} {79}},\
  \bibinfo {eid} {084021} (\bibinfo {year} {2009})},\ \Eprint
  {http://arxiv.org/abs/0812.4208} {arXiv:0812.4208 [gr-qc]} \BibitemShut
  {NoStop}%
\bibitem [{\citenamefont {Schutz}(1986)}]{schutz1986determining}%
  \BibitemOpen
  \bibfield  {author} {\bibinfo {author} {\bibfnamefont {B.~F.}\ \bibnamefont
  {Schutz}},\ }\href@noop {} {\bibfield  {journal} {\bibinfo  {journal}
  {Nature}\ }\textbf {\bibinfo {volume} {323}},\ \bibinfo {pages} {310}
  (\bibinfo {year} {1986})}\BibitemShut {NoStop}%
\bibitem [{\citenamefont {Mills}\ \emph {et~al.}(2016)\citenamefont {Mills},
  \citenamefont {Fabrycky}, \citenamefont {Migaszewski}, \citenamefont {Ford},
  \citenamefont {Petigura},\ and\ \citenamefont
  {Isaacson}}]{mills2016resonant}%
  \BibitemOpen
  \bibfield  {author} {\bibinfo {author} {\bibfnamefont {S.~M.}\ \bibnamefont
  {Mills}}, \bibinfo {author} {\bibfnamefont {D.~C.}\ \bibnamefont {Fabrycky}},
  \bibinfo {author} {\bibfnamefont {C.}~\bibnamefont {Migaszewski}}, \bibinfo
  {author} {\bibfnamefont {E.~B.}\ \bibnamefont {Ford}}, \bibinfo {author}
  {\bibfnamefont {E.}~\bibnamefont {Petigura}}, \ and\ \bibinfo {author}
  {\bibfnamefont {H.}~\bibnamefont {Isaacson}},\ }\href@noop {} {\bibfield
  {journal} {\bibinfo  {journal} {Nature}\ }\textbf {\bibinfo {volume} {533}},\
  \bibinfo {pages} {509} (\bibinfo {year} {2016})}\BibitemShut {NoStop}%
\bibitem [{\citenamefont {{Dittmann}}\ and\ \citenamefont
  {{Miller}}(2020)}]{Dittmann2020}%
  \BibitemOpen
  \bibfield  {author} {\bibinfo {author} {\bibfnamefont {A.~J.}\ \bibnamefont
  {{Dittmann}}}\ and\ \bibinfo {author} {\bibfnamefont {M.~C.}\ \bibnamefont
  {{Miller}}},\ }\href {\doibase 10.1093/mnras/staa463} {\bibfield  {journal}
  {\bibinfo  {journal} {\mnras}\ }\textbf {\bibinfo {volume} {493}},\ \bibinfo
  {pages} {3732} (\bibinfo {year} {2020})},\ \Eprint
  {http://arxiv.org/abs/1911.08685} {arXiv:1911.08685 [astro-ph.HE]}
  \BibitemShut {NoStop}%
\end{thebibliography}%
\end{document}